\pdfoutput=1
\documentclass[a4paper]{article}
\usepackage{jheppub}
\usepackage{dsfont}
\usepackage{amsmath,amssymb,amscd,amsfonts,mathtools}

\DeclareMathOperator{\arcsec}{Arcsec}

\DeclareMathOperator{\arccsc}{Arccsc}

\usepackage[toc,page]{appendix}
\usepackage{epsfig}
\usepackage{epstopdf}
\usepackage{latexsym}
\usepackage{enumerate} 
\usepackage{graphicx}
\usepackage{placeins}
\usepackage{floatrow}
\usepackage{caption}
\usepackage{booktabs}
\usepackage{bbm}
\usepackage{color}
\usepackage{physics}
\usepackage{tensor}
\usepackage{tikz}
\usetikzlibrary{matrix}
\usetikzlibrary{decorations.markings,calc,shapes,decorations.pathmorphing,patterns,decorations.pathreplacing,arrows.meta}
\usetikzlibrary{positioning}
\usepackage{hyperref}
\usepackage{subcaption}

\newcommand{\Nc}{N}
\newcommand{\Nf}{N_{\text{f}}}

\makeatletter
\def\@fpheader{\relax}
\makeatother
\def\R{{\mathcal R}}

\DeclareMathOperator\arcosh{arcosh}
\DeclareMathOperator\vol{Vol}

\newcommand{\be}{\begin{equation}}
\newcommand{\ee}{\end{equation}}
\newcommand{\bea}{\begin{eqnarray}}
\newcommand{\eea}{\end{eqnarray}}
\DeclareMathSizes{10}{10}{7}{7}

\begin{document}

\preprint{$\begin{array}{rr}
	\text{HIP-2024-2/TH}\end{array}$
}

\title{Flavors of entanglement}

\author[a,b]{Niko Jokela,}
\affiliation[a]{Helsinki Institute of Physics, P.O. Box 64, FIN-00014 University of Helsinki, Finland}
\emailAdd{niko.jokela@helsinki.fi}
\author[c]{Jani Kastikainen,}
\affiliation[b]{Department of Physics, P.O. Box 64, FIN-00014 University of Helsinki, Finland}
\affiliation[c]{Institute for Theoretical Physics and Astrophysics and W\"urzburg-Dresden Cluster of Excellence
ct.qmat, Julius-Maximilians-Universit\"at W\"urzburg, Am Hubland, 97074 W\"urzburg, Germany}
\emailAdd{jani.kastikainen@uni-wuerzburg.de}
\author[d]{Jos\'e Manuel Pen\'in,}
\affiliation[d]{INFN, Sezione di Firenze; Via G. Sansone 1; I-50019 Sesto Fiorentino (Firenze), Italy}
\emailAdd{jmanpen@gmail.com}
\author[a,b]{and Helime Ruotsalainen}
\emailAdd{helime.ruotsalainen@helsinki.fi}

\abstract{We employ holography to investigate Liu--Mezei renormalization group monotones in conformal field theories influenced by massive flavor degrees of freedom. We examine the entanglement entropy of a spherical subregion in three holographic field theories -- $\mathcal{N}=1$ Klebanov--Witten theory, $\mathcal{N}=4$ SYM theory, and ABJM theory -- with fundamental flavor. The gravity dual of massive unquenched flavor is described by dynamical D-branes, and we solve their backreaction in the smeared approximation. We compute entanglement entropy using the Ryu--Takayanagi formula in these backreacted geometries. Our findings indicate that the Liu--Mezei A- and F-functions decrease monotonically to leading order in the number of flavors across all examples. Additionally, we calculate the leading flavor contribution to entanglement entropy using an alternative probe brane method that does not require knowledge of backreaction in the bulk geometries. These results consistently match with backreacted calculations in all cases, assuming omission of a specific IR boundary term stemming from a total derivative.}

\maketitle
\flushbottom
\setcounter{page}{2}

\section{Introduction}

Among the intriguing phenomena that arise in the quantum realm, one that lacks a classical analog is entanglement -- an intricate concept whose mechanisms continue to elude complete comprehension. Quarks, the elemental constituents influenced by strong interactions, are bound by the rules of quantum physics. Our understanding of flavor entanglement remains limited. In this study, we take a notable step forward in this direction by employing the principles of string/gauge duality.

Incorporating fundamental components such as quarks within the context of string theory is accomplished by introducing the open string sector. This entails introducing a set of flavor D-branes that intersect with the pre-existing color D-branes, serving as a representation of the gluonic sector \cite{Karch:2002sh}. The method for calculating holographic entanglement entropy (HEE) in the flavor sector follows the conventional Ryu--Takayanagi (RT) prescription in principle \cite{Nishioka:2009un}. However, the intricacies lie in the finer points, and advancement in this regard has encountered obstacles.

Finding a supergravity background that fully incorporates the effects of backreacting localized flavor branes presents a substantial challenge, however, there exist two established approximation strategies. The first involves treating the flavor branes as probes, which corresponds to a quenched treatment of the quarks. In the second approach, the D-branes are smeared (delocalized) \cite{Nunez:2010sf} which allows for an unquenched treatment of the quarks at the expense of breaking the $U(\Nf)$ flavor group to a subgroup. Both approximations are useful in capturing effects of flavors in entanglement.

In this paper, we consider entanglement entropy in three different superconformal field theories that admit deformations by massive flavor degrees of freedom: $\mathcal{N}=1$ Klebanov--Witten (KW) theory \cite{Klebanov:1998hh}, $ \mathcal{N} = 4 $ supersymmetric Yang--Mills (SYM) theory, and Aharony--Bergman--Jafferis--Maldacena (ABJM) theory \cite{Aharony:2008ug}. Four-dimensional KW and SYM theories arise as the low-energy descriptions of a stack of $N$ D3-branes with the gravity dual being type IIB supergravity on $\text{AdS}_5\times \text{T}^{1,1}$ and $\text{AdS}_5\times \text{S}^{5}$, respectively. Both admit deformations by flavor degrees of freedom which are holographically described by including a stack of $\Nf$ D7-branes (the D3-D7 system) \cite{Karch:2002sh}. In KW theory one can consider two types of D7-brane embeddings resulting in two unidentical flavor `profiles' in the bulk called chiral and non-chiral profiles. The backreaction of massless chiral flavor branes to the $\text{AdS}_5\times \text{T}^{1,1}$ background was derived in \cite{Benini:2006hh} and was extended to massive chiral flavor in \cite{Bigazzi:2008zt}. The backreaction of massive non-chiral flavor was solved in \cite{Bigazzi:2008ie}. In the case of $ \text{AdS}_5\times \text{S}^{5}$, the backreaction of massive flavor branes was solved in \cite{Bigazzi:2009bk}.

ABJM theory is a three-dimensional $\text{U}(N)_k\times \text{U}(N)_{-k}$ Chern--Simons matter theory which is holographically dual to 11D supergravity on $\text{AdS}_4\times \text{S}^{7}\slash \mathbb{Z}_k$ in the $N\rightarrow \infty$ limit and to type IIA supergravity on $\text{AdS}_4\times \mathbb{C}\text{P}^3$, in the $N,k\rightarrow \infty$ limit \cite{Aharony:2008ug}. The theory admits deformations by flavor degrees of freedom which are dual to probe flavor D6-branes in the type IIA description \cite{Hohenegger:2009as,Gaiotto:2009tk,Hikida:2009tp,Ammon:2009wc}. Extension to unquenched flavor by accommodating backreaction from smeared D6-branes to the $\text{AdS}_4\times \mathbb{C}\text{P}^3$ background was given in \cite{Conde:2011sw,Bea:2013jxa,Jokela:2012dw,Bea:2014yda} for massless and in \cite{Bea:2013jxa} for massive flavor.

In the first part of this work, we compute vacuum entanglement entropy of a spherical subregion of radius $R$ in flat space in the three theories described above when unquenched massive flavor is present. In three dimensions (ABJM theory), the subregion is a two-dimensional disk, while in four dimensions (KW and SYM theories), it is a three-dimensional solid ball. Instead of working with unquenched massive flavor, we take the quenched flavor limit $\Nf\slash N\rightarrow 0$ and study the leading $\Nf$-correction to the entropy. In ABJM theory, we will consider $N,k,\Nf\rightarrow \infty$ with $\Nf\slash k \rightarrow 0$. We compute the leading correction using two complementary methods: the \textit{backreacted method}, which uses the RT formula in the backreacted geometries, and the \textit{probe brane method}, which uses on-shell probe D-brane actions in the unflavored backgrounds.

In the backreacted method, for HEE one simply evaluates the area of a minimal surface in the backreacted geometry which has been performed in multiple holographic setups in the literature.  More specifically, the leading order in $\Nf$ HEE result was obtained in the perturbed $\text{AdS}_5\times \text{S}^5$ background in \cite{Kontoudi:2013rla}. The paper \cite{Balasubramanian:2018qqx}, on the other hand, used the smeared $\text{AdS}_4\times \mathbb{C}\text{P}^3$ background to compute HEE to all orders in $\Nf$ using numerical methods. We elaborate along this line and extract analytical expressions for the entanglement entropies in the $\Nf\rightarrow 0$ limit in the above setups and also present novel results for the smeared $\text{AdS}_5\times \text{T}^{1,1}$ geometry. See also analogous HEE computations in the few flavor limit using the smeared $\text{AdS}_4\times \mathbb{C}\text{P}^3$ background \cite{Balasubramanian:2018qqx} and that of massive flavors in the Klebanov-Strassler theory \cite{Georgiou:2015pia}, albeit both with strip or slab instead of spherical subregions.

The obvious shortcoming of the backreacted method is that supergravity solutions are available with massive flavor only when the flavor branes are delocalized. This is remedied by the probe brane method which does not require the knowledge of the backreacted geometries, but only of the form of the flavor D-brane action and the embedding of the (localized) brane in the unflavored background. Hence, it allows to compute the contribution of fully localized massive flavor branes to the entanglement entropy, but in the $\Nf\rightarrow 0$ limit. This is, however, enough to compare with the $\Nf\rightarrow 0$ result of the backreacted method and can be viewed as a test of the smeared approximation.

Methods based on probe branes that do not involve backreaction have been developed over a series of papers starting from the seminal works \cite{Chang:2013mca,Jensen:2013lxa,Karch:2014ufa}. The method of \cite{Jensen:2013lxa} applies only to conformal symmetry preserving massless flavors, because it is based on the Casini--Huerta--Myers (CHM) conformal mapping \cite{Casini:2011kv}, but \cite{Chang:2013mca,Karch:2014ufa} also work in the absence of conformal symmetry. These methods have been applied to various conformal symmetry breaking scenarios (such as massive flavor and spontaneous symmetry breaking) in the $\text{AdS}_5\times \text{S}^5$ D3-D7 system in \cite{Jones:2015twa,Vaganov:2015vpq,Kumar:2017vjv,Chalabi:2020tlw}. 

In this work, we will utilize the probe brane method presented in \cite{Karch:2014ufa} instead of the graviton propagator method of \cite{Chang:2013mca}. The method has recently been further developed in \cite{Chalabi:2020tlw} where an additional boundary term argued to vanish in \cite{Karch:2014ufa} was shown to be non-zero. We revisit the analysis and concur with \cite{Chalabi:2020tlw} that the boundary term is generically non-zero. We then apply the method in our $\text{AdS}_5\times \text{S}^5$ and $\text{AdS}_4\times \mathbb{C}\text{P}^3$ cases to compute entanglement entropy of a spherical subregion and compare with results of the backreacted method. We find that the universal terms in the results of both calculations match perfectly with the backreacted results if the non-zero boundary term advocated in \cite{Chalabi:2020tlw} is omitted. In the $\text{AdS}_5\times \text{S}^5$ case, our result also matches with the results of \cite{Jones:2015twa} obtained using the graviton propagator method of \cite{Chang:2013mca} when the boundary term is omitted.

Interestingly, in \cite{Chalabi:2020tlw}, the same boundary term we must omit is retained to achieve compatibility with results derived from the backreacted method. However, our work diverges significantly in two key aspects from \cite{Chalabi:2020tlw}. Firstly, our backreacted supergravity solutions, derived under the smeared approximation, contrast with their exact counterparts. Secondly, while their emphasis lies on the Coulomb branch and the corresponding domain wall solutions, our focus centers on flavor entanglement. Thence our flavor branes reach the conformal boundary, while the domain wall branes of \cite{Chalabi:2020tlw} are localized in the holographic radial coordinate.

The apparent discrepancy raises a twofold puzzle. Either (i) our smeared approximation introduces inaccuracies into our backreacted solutions, particularly as they pertain to massive flavor, or (ii) an unknown mechanism in our case nullifies the boundary term in the probe brane method, a phenomenon left unexplored in \cite{Chalabi:2020tlw}. Although the absence of the boundary term yields perfect matching in both D7- and D6-brane scenarios, supporting (ii), we remain unable to conclusively eliminate the possibility of (i). Notably, the smeared backgrounds, known to reproduce exact observables at leading order in $\Nf$ for massless flavor, exhibit generic discrepancies at higher orders in $\Nf$ as demonstrated in \cite{Conde:2011sw} because the symmetry structure is $U(1)^{\Nf}$ for delocalized flavors. If interpretation (i) holds, it implies a breakdown of the smeared approximation even at the leading order in $\Nf$ when flavor becomes massive but also raises questions regarding the graviton propagator method.

In the second part of this work we study the monotonicity of Liu--Mezei A-functions (in four dimensions) and F-functions (in three dimensions) in field theories deformed by massive flavor \cite{Liu:2013una}. These functions are proposed to measure the number of degrees of freedom of a quantum field theory with a single mass scale and hence are expected to monotonically decrease along renormalization group (RG) flows. We compute the A- and F-functions using the entanglement entropy obtained from both the backreacted and probe brane methods. In all of our examples, the functions obtained from the backreacted method are monotonically decreasing in the dimensionless radius $\R$ (flavor mass times $R$). The same is true for the functions obtained from the probe brane method if we omit the boundary term described above: the A- and F-functions are the same, because all non-universal terms of the entanglement entropy are canceled. If we were to include the boundary term in the probe calculation, both the A- and F-functions would lack monotonicity and continuity.

\subsubsection*{Summary of results}

The paper is organized as follows. In Section~\ref{sec:backreacted} we lay out the results of the HEE computation using RT prescription in the backreacted backgrounds (the backreacted method). We treat the perturbative expansion of the area functional to leading order in $\Nf$ carefully keeping track of total derivatives and explaining how regularity of the RT surface ensures vanishing of the boundary term in the IR. The detailed computation, together with the details of the backreacted geometries in the few flavor limit, have been relegated to Appendices \ref{app:geometries} and \ref{app:RTdetails}. The calculation of the entanglement entropy using the smeared backreacted $\text{AdS}_5\times \text{S}^5$ background was done to leading order in $\Nf$ in \cite{Kontoudi:2013rla} whose result we reproduce up to a sign in one of the terms. In \cite{Bea:2013jxa}, a similar calculation was done using the smeared $\text{AdS}_4\times \mathbb{C}\text{P}^3$ geometry to all orders in $\Nf$ numerically and our calculation gives a new analytic expression for the leading $\Nf$ contribution. The calculation of the entropy in the two types of backreacted $\text{AdS}_5\times \text{T}^{1,1}$ backgrounds are new.

In Section~\ref{sec:probe} we describe the HEE computation using methods that do not require knowledge of the backreacted bulk geometries. First, we compute the entropy of massless flavor by utilizing the CHM map that equates the entropy with the free energy of the theory on a Euclidean sphere. The details of the free energy calculations are relegated to Appendix~\ref{app:spherefreenenergy}. Second, we utilize the probe brane method of \cite{Karch:2014ufa,Chalabi:2020tlw} to compute the entropy of massive flavor in the $\text{AdS}_5\times \text{S}^5$ and $\text{AdS}_4\times \mathbb{C}\text{P}^3$ cases. Details of the calculations are relegated to Appendix \ref{app:probedetails} where we keep track of various boundary terms carefully. The universal terms in the calculations match in both of these cases with the results of the backreacted method of Section~\ref{sec:backreacted} assuming the omission of a certain boundary term localized on the RT surface. The probe $\text{AdS}_4\times \mathbb{C}\text{P}^3$ calculation is new while the $\text{AdS}_5\times \text{S}^5$ entropy was computed in \cite{Karch:2014ufa} and in \cite{Jones:2015twa} which utilizes the graviton propagator method of \cite{Chang:2013mca}. Our calculation matches with theirs up to non-universal terms and our main new contribution is the treatment of boundary terms. The probe calculation in the two $\text{AdS}_5\times \text{T}^{1,1}$ cases and their comparison with the backreacted result is left for future work.

In Section~\ref{subsec:backreactedCFfunctions} we discuss the renormalization group monotones of \cite{Liu:2012eea, Liu:2013una} as constructed using the flavor entanglement entropies and, in particular, illuminate that the small and large mass limits correctly expound the UV and IR fixed points of the dual field theories. We show that there exists redefinitions of the UV cut-off $\epsilon$ and the radius $R$ of the spherical subregion under which the A- and F-functions are invariant. The entropies of the probe and backreacted methods are related by such redefinition implying they have the same A- and F-functions. Section~\ref{sec:discussion} contains our conclusions and outlook.

\section{Holographic entanglement entropy for unquenched theories}\label{sec:backreacted}

In this section we will quickly skim on how to compute the holographic entanglement entropy for spherical subregions that are associated with field theories with unquenched flavors. The computation reduces to the minimization of the Ryu-Takayanagi surface in dual geometries that are obtained through the backreaction of flavor D-branes. We will elaborate on how to obtain the few flavor (Veneziano) limit, $\Nf/\Nc\ll 1$, the results thereof to be compared with the probe calculation in the subsequent section.

\subsection{Backreacted solutions from smeared D-branes}\label{subsec:backreactedsols}

We are interested in solutions of type IIB and IIA supergravity with Euclidean actions, which in string frame read
\begin{equation}
	I_{\text{grav}} = -\frac{1}{2\kappa_{10}^{2}}\int d^{10}x\sqrt{g}\,
	\begin{dcases}
	e^{-2\phi}\,(R_{10}+4\,g^{ab}\,\partial_a\phi\,\partial_b\phi) - \dfrac{1}{4\cdot 5!}\,F_5^{2}-\frac{1}{2}F_1^2\ , \quad &\text{type IIB}\vspace{1mm}\\
	e^{-2\phi}\,(R_{10}+4\,g^{ab}\,\partial_a\phi\,\partial_b\phi) - \dfrac{1}{2\cdot 4!}\,F_4^{2} - \dfrac{1}{2\cdot 2!}\,F_2^{2}\ , \quad &\text{type IIA}\label{eq:typeAaction}
 \end{dcases} \ ,
\end{equation}
where $2\kappa_{10}^2 = (2\pi)^7g_{\text{s}}^2\,\ell_{\text{s}}^8$, $g_{\text{s}}$ is the string coupling, $\ell_{\text{s}}$ is the string length, $\phi$ is the dilaton, $R_{10}$ is the Ricci scalar of $g$, $F_{n} \ (n\in\{1,2,4,5\})$ are RR fluxes of IIB, 
and we use the notation $ F_n^2\equiv F_{a_1\ldots a_n}F^{a_1\ldots a_n} $. They admit (unflavored) solutions with locally $\text{AdS}_{d+1}\times X_{q}$ asymptotics of the type
\begin{equation}
    ds^2_{10} = \ell^2\left(\frac{dr^2}{r^2}+r^{2}ds^{2}_{1,d-1} + ds^2_{X_{q}}\right) \ ,
    \label{eq:nobackreaction}
\end{equation}
where $X_{q}$, with $q = 9-d$, is a Sasaki--Einstein manifold, $ds^{2}_{1,d-1}$ is the flat Minkowski metric on $\mathbb{R}^d$ and $\ell$ is the radius of curvature. In type IIB supergravity, we will consider the D3-brane solution ($d=4$) dual to $SU(N)$ SYM,
\begin{equation}
    \phi = 0\ ,\quad F_5 = 4\ell^4\,(1+*)\,\epsilon_{\text{AdS}_5}\ ,\quad \biggl(\frac{\ell}{\ell_{\text{s}}}\biggr)^{4} = \biggl(\frac{2\pi^{3}}{\vol{(X_5)}}\biggr)\;g^{2}_{\text{YM}}N\ ,\quad g_{\text{s}}= 
\frac{g_{\text{YM}}^{2}}{2\pi} \ ,
\label{eq:D3parameters}
\end{equation}
where $g_{\text{YM}}$ is the Yang--Mills coupling of SYM, and $\epsilon_{\mathrm{AdS}_d}$ is the volume element of $\text{AdS}_d$. Throughout the text, the integral of the volume element $\epsilon_X$ of a space $X$ will be denoted by $\vol{(X)}$. We will consider two choices of $X_5$: a five-sphere $\text{S}^5$ and $\text{T}^{1,1}$ which are $S^1$ fibrations over $\text{S}^4$ and $S^2\times S^2$, respectively.

In type IIA supergravity, we will consider the solution ($d=3$) which is dual to ABJM theory with Chern--Simons gauge group $U(N)_k\times U(N)_{-k}$ \cite{Aharony:2008ug}
\begin{equation}
    \phi= 0\ ,\  F_2 = 2\ell\,J\ ,\ F_4 = 3\ell^3\,\epsilon_{\text{AdS}_4}\ ,\ \biggl(\frac{\ell}{\ell_{\text{s}}}\biggr)^{4} = \biggl(\frac{2^{7}\pi^{5}}{3\vol{(X_6)}}\biggr)\,\frac{N}{k}\ ,\ g_{\text{s}}= \biggl(\frac{2^{11}\pi^{5}}{3\vol{(X_{6})}}\dfrac{N}{k^{5}}\biggr)^{1\slash 4} \ .
    \label{eq:ABJMparameters}
\end{equation}
Here the internal space $X_6$ is equipped with the metric $ds^{2}_{X_6} = 4 ds^2_{\mathbb{C}\text{P}^3}$ where $ds^2_{\mathbb{C}\text{P}^3}$ is the standard Fubini--Study metric on $\mathbb{C}\text{P}^3$ and $J$ is the K\"ahler form of $\mathbb{C}\text{P}^3$ (see Appendix \ref{app:ABJMdetails} for details and our conventions). This solution is the dimensional reduction of the 11-dimensional M2-brane solution $\text{AdS}_4\times \text{S}^{7}\slash \mathbb{Z}_k$ of M-theory to ten dimensions \cite{Aharony:2008ug}. Similarly to the type IIB supergravity solutions, $\mathbb{C}\rm{P}^3$ is a fibration over a four-dimensional base manifold, namely, an $S^2$ fibered over $\text{S}^4$ \cite{Conde:2011sw}.

We will now insert $\Nf$ smeared flavor brane sources to the background geometries. Our new geometries will be determined by the extremization of the supergravity plus branes actions:
\begin{gather}
    I = I_{\text{grav}} + \Nf\ I_{\text{D}p} 
    \label{eq:totalbulkaction}
\end{gather}
with $I_{\text{grav}}=\{I_{\text{IIB}},I_{\text{IIA}}\}$ and $I_{\text{D}p}=\{I_{\text{D7}},I_{\text{D6}}\}$, respectively, and, in string frame  
\begin{gather}
    I_{\text{D7}} = -T_{\text{D7}}\int d^{8}x\sqrt{\hat{g}_8}\,e^{-\phi}\ ,\quad I_{\text{D6}} = -T_{\text{D6}}\int d^{7}x\sqrt{\hat{g}_7}\,e^{-\phi} + T_{\text{D6}}\int  \widehat{C}_7 \ ,
    \label{eq:braneactions}
\end{gather}
with $ F_8 = dC_7 $ and $ F_8 = *F_2 $. Our convention for the brane tension is
\begin{equation}
T_{\text{D}p} = \frac{1}{(2\pi)^{p}g_{\text{s}}\,\ell_{\text{s}}^{p+1}} \ .
\label{eq:tension}
\end{equation}
Here hats denote pullbacks onto the D$p$-brane worldvolumes and $C_7$ is the RR potential for $F_8$. Note also that $p$ denotes the spatial dimensionality of the flavor branes and not of color fluxes that otherwise build up the geometry.

To solve for the backreacted geometry, we assume that the flavor branes are smeared in the transverse directions. This means that the actions \eqref{eq:braneactions} with delta function sources for localized branes are replaced with a smearing form which is integrated over the whole internal space. The smearing form is fixed by symmetries, see \cite{Benini:2006hh,Bigazzi:2009bk} and \cite{Conde:2011sw} for a detailed description in D3-D7 and ABJM cases, respectively (see also Appendix~\ref{app:geometries}). When the dual flavor degrees of freedom are introduced using smearing the configurations preserve $\mathcal{N} = 1$ supersymmetry. When the flavors are massive the brane profile is described by a continuous monotonic function of the holographic radial coordinate, the so-called \textit{profile} function (see below).

After backreacting smeared flavor branes, the geometries will adopt the warped form
\begin{equation}
    ds^2_{10} = h^{-1\slash 2}ds^{2}_{1,d-1} + h^{1\slash 2}e^{2f}d\varrho^2+h^{1\slash 2}ds^2_{\widetilde{X}_{q}}
    \label{eq:smearedbckgs}
\end{equation}
and the metric on the internal space is deformed to
\begin{equation}
    ds^2_{\widetilde{X}_{q}} = 
    \begin{cases}
    e^{2g}\,ds^2_{4}+e^{2f}\,(d\tau + A_1)^2\ , \ &\text{D3-D7} \\
    e^{2g}\,ds^2_{4}+e^{2f}\,ds^2_{S^2}\ , \ &\text{ABJMf}
    \end{cases}\ .
\end{equation}
The holographic radial coordinate $ \varrho \in \mathbb{R} $ with the Poincar\'e horizon located at $\varrho = -\infty$ and the conformal boundary at $ \varrho \rightarrow \infty $. We will define a new radial coordinate $ r = r(\varrho) $ by the equation
\begin{equation}
h =  \frac{1}{\ell^{4}r^{4}} \ ,
\end{equation}
where $ \ell $ is defined for both types of geometries in \eqref{eq:D3parameters} and \eqref{eq:ABJMparameters}. In terms of this coordinate, the coefficient of $ds^{2}_{1,d-1} $ in the metric \eqref{eq:smearedbckgs} is simply $\ell^2r^2$. It takes values $r\in (0,\infty)$ with the Poincar\'e horizon at $ r = 0$ and the conformal boundary at $r = \infty$. In terms of this coordinate, the metric \eqref{eq:smearedbckgs} takes the form
\begin{equation}
ds^{2}_{10} = h^{-1\slash 2}ds^{2}_{1,d-1} + h^{1\slash 2}\,\Sigma^{2}\,dr^{2} + h^{1\slash 2}ds^{2}_{\widetilde{X}_q}\ ,
\label{eq:backreactedmetricr}
\end{equation}
where we have defined the function
\begin{equation}
\Sigma(r) \equiv e^{f(r)}\varrho'(r)\ .
\label{eq:Sigma}
\end{equation}
The functions in the metric $h = h(r)$, $f = f(r)$, $g = g(r)$, the RR fluxes, and the dilaton $\phi=\phi(r)$ are functions of $r$ and are deformed by the backreaction.

We will consider our backreacted geometries in the few flavor limit, and, accordingly, the functions defining our fields will be perturbatively expanded in $\Nf$ as
\begin{gather}
    f = f_0 + \Nf\,f_1 + \mathcal{O}(\Nf^2)\ , \quad g = g_0 + \Nf\,g_1 + \mathcal{O}(\Nf^2)\ \label{eq:funcexps1}\\
    \phi = \phi_0 + \Nf\,\phi_1 + \mathcal{O}(\Nf^2)\ , \quad h = h_0 + \Nf\,h_1 + \mathcal{O}(\Nf^2) \ ,
    \label{eq:funcexps2}
\end{gather}
where for our purposes, only the linear order in $\Nf$ is needed. The expansion parameter in the probe limit with $N,\Nf\rightarrow \infty$ is $\lambda\,\Nf\slash N\ll 1$ \cite{Nunez:2010sf} where the coupling $\lambda = g_{\text{YM}}^2N$ and $\lambda = N\slash k$ in the SYM and ABJM cases, respectively. The fields in the flavorless background are given by
\begin{equation}
    f_0 = g_0 = \log{(\ell^2 r)}\ ,\quad \phi_0 = 0\ , \quad h_0 = \frac{1}{\ell^4r^4}\ .
\end{equation}
Hence, in the no flavor limit the internal space reduces to
\begin{equation}
ds^{2}_{\widetilde{X}_q} = \ell^{4}r^{2}ds^{2}_{X_q}\ .
\end{equation}
We will also expand
\begin{equation}
    \varrho = \varrho_0 + \Nf \,\varrho_1 + \mathcal{O}(\Nf^2)\ ,\quad \Sigma = \Sigma_0\,[1+ \Nf\,\Sigma_1 + \mathcal{O}(\Nf^{2})]\ ,
    \label{eq:sigmaexpansiontext}
\end{equation}
where we require $\Sigma_0(r)= \ell^2 $ so that the metric \eqref{eq:backreactedmetricr} reduces to the $ \text{AdS}_{d+1}\times X_q $ metric \eqref{eq:nobackreaction} of radius $\ell$ in the no flavor limit. From \eqref{eq:Sigma} we see that the condition $\Sigma_0(r)= \ell^2 $ implies $\varrho_0(r) = \log{r}$. The leading corrections $\varrho_1,\Sigma_1$ are then determined by \eqref{eq:funcexps1}--\eqref{eq:funcexps2} after solving for $f_1,g_1,h_1,\phi_1$ from the supergravity equations.

The branes extend in all $\text{AdS}_{d+1}$ directions and their profile function in the holographic radial coordinate is denoted by $p(\varrho)$. In the D3-D7 case, the profiles are (see Appendix \ref{app:T11S5details})
\begin{equation}
p(\varrho) = \Theta(\varrho-\varrho_\text{\tiny{q}})\times
    \begin{dcases}
        1-e^{3(\varrho_\text{\tiny{q}}-\varrho)}\,[1+3(\varrho-\varrho_\text{q})]\,,\quad &\text{T}^{1,1} \text{ chiral}\\
        1-e^{-3(\varrho-\varrho_\text{\tiny{q}})}\,,\quad &\text{T}^{1,1} \text{ non-chiral}\\
        (1-e^{2(\varrho_\text{\tiny{q}}-\varrho)})^2\,,\quad &\text{S}^{5}
    \end{dcases} \ ,
\end{equation}
where $\Theta(\varrho)$ is the Heaviside step function. In the ABJM case, the profile is given by (see Appendix \ref{app:ABJMdetails} for details)
\begin{equation}
    p(\varrho) = (1-e^{-2(\varrho-\varrho_\text{\tiny{q}})})\,\Theta(\varrho-\varrho_\text{\tiny{q}}) \,.
    \label{eq:ABJMp}
\end{equation}
Due to the step function, the branes terminate at the radial location $\varrho =\varrho_\text{\tiny{q}} > 0$ such that the termination point $r_\text{q} = e^{\varrho_\text{\tiny{q}}}$ in the coordinate $r = e^{\varrho}$ is proportional to the flavor mass $m_{\text{q}}$. This defines a region $0 < r < r_\text{q}$ between the Poincar\'e horizon and the tip of the branes which we call the \textit{cavity}, inside of which there are no flavor branes, and where the profile function vanishes. Because of this, the forms of the backreacted solution inside and outside the cavity are different, but they are glued together with continuous derivatives. For massless flavors, the branes extend all the way to the Poincar\'e horizon and the cavity disappears.\footnote{In the massless flavored ABJM theory, one recovers the $AdS_4\times\mathbb{C}\text{P}^3$ geometry but with the radius of curvature depending on the number of flavors, $\ell=\ell(\Nf)$. For the D3-D7 case, however, one recovers \eqref{eq:nobackreaction} only in the probe limit.} The detailed form of the corrections $f_1,g_1,h_1,\phi_1$ for each of the cases studied are listed in the Appendices~\ref{app:T11S5details}~and~\ref{app:ABJMdetails}.

\paragraph{UV asymptotics of the backreacted geometries.} It is well known that at finite $\Nf$ the backreacted D3-D7 geometries are not asymptotically locally AdS in the UV $r\rightarrow \infty$, but contain a Landau pole due to a diverging dilaton. Therefore the geometries should only be trusted up to a large but finite value of the holographic radial coordinate $r = r_* < \infty$ determined in \cite{Benini:2006hh,Bigazzi:2008zt}. As a result, holographic quantities, such as on-shell actions or areas of minimal surfaces, are to be integrated only up $r_*$. However, at linear order in $\Nf$ the Landau pole is no longer visible and does not pose problems, because the geometries turn out to be asymptotically locally AdS.

To see this, consider the $\Nf$-expansions \eqref{eq:sigmaT11chiral}, \eqref{eq:sigmaT11nonchiral}, and \eqref{eq:sigmaS5} of the function $\Sigma(r)$ derived in Appendix~\ref{app:T11S5details} for the three D3-D7 geometries. In all cases, they have the UV $r\rightarrow \infty$ expansion
\begin{equation}
    \Sigma(r) = \ell^2\,\biggl(1+\Nf\,\frac{g_{\text{s}}\vol{(X_3)}}{32\vol{(X_5)}} +\mathcal{O}(\Nf^2)\biggr)+\mathcal{O}(r^{-2})\,,\quad r\rightarrow \infty\,.
\end{equation}
Therefore in the UV, the backreacted metric \eqref{eq:backreactedmetricr} takes the form
\begin{equation}
    ds^2 = \ell^2\biggl[(1+2\Nf\, \delta\ell)\,\frac{dr^2}{r^2}+r^{2}ds^{2}_{1,d-1} + ds^2_{X_{q}}\biggr] + \ldots\,,
\end{equation}
where we have defined $\delta\ell \equiv \frac{g_{\text{s}}\vol{(X_3)}}{32\vol{(X_5)}}$. By the coordinate change $r^2 = (1 + 2\Nf\, \delta\ell)\,\hat{r}^2  $, this can be written as
\begin{equation}
    ds^2 = \hat{\ell}^2\biggl(\frac{d\hat{r}^2}{\hat{r}^2}+\hat{r}^{2}ds^{2}_{1,d-1}\biggr) + \ell^2ds^2_{X_{q}} + \ldots\,,
    \label{eq:asympAdS}
\end{equation}
where $\hat{\ell}^2\equiv \ell^2(1+2\Nf\, \delta\ell)$ is the $\Nf$-corrected AdS radius squared. Thus, the geometry is asymptotically locally AdS (for massless flavor the geometry is exact) and the internal space factor has a different radius than the AdS factor.\footnote{The relative squashing of AdS and internal space factors at asymptotic infinity is generic and also a property of the backreacted ABJM geometries (see for example \cite{Bea:2013jxa}) which do not contain a Landau pole at finite $\Nf$.} This implies that we can take the UV limit $r_* \rightarrow \infty$ and perform all calculations as usual in AdS/CFT at linear order in $\Nf$ (as done for example in \cite{Kontoudi:2013rla}).



\subsection{Calculation of the entanglement entropy for spherical regions\label{subsec:perturbativeRT}}

We want to compute the entanglement entropy of a spherical subregion $B_R$ of radius $R$ embedded on a Cauchy slice $\mathbb{R}^{d-1}$ of flat $d$-dimensional Minkowski space. Holographically, the entanglement entropy is given by
\begin{equation}
S(R) = \frac{\text{Area}\,(\Sigma_8)}{4G_{\text{N}}} = \frac{1}{4G_{\text{N}}}\int_{\Sigma_8} d^{8}x\,e^{-2\phi}\sqrt{\hat{g}} \ ,
\label{eq:RTformula}
\end{equation}
where $ \Sigma_8 $ is a spacelike codimension-two RT surface embedded in the constant time-slice of the full ten-dimensional bulk geometry, $\hat{g}$ is the induced metric of $\Sigma_8$ in the string frame, $\phi$ is the dilaton, and the Newton's constant is given by $16\pi G_{\text{N}} = (2\pi)^7g_{\text{s}}^2\ell_{\text{s}}^8 $. The RT surface is the minimal area surface (extremum of the entropy functional \eqref{eq:RTformula}) which is anchored to the entangling surface $\partial\Sigma_8 = \partial B_R = S^{d-2}$ on the conformal boundary and which is homologous to $B_R$. Asymptotically, the bulk geometry is locally $\text{AdS}_{d+1}\times X_q$ and the RT surface is required to wrap the whole internal space $X_q$ \cite{Jones:2016iwx}. Note that the dilaton factor appears explicitly in \eqref{eq:RTformula}, because it is written in the string frame.

We will introduce spherical coordinates $(\rho,\Omega_{d-2})$ on the Cauchy slice $\mathbb{R}^{d-1}$ whose (Euclidean) metric then takes the form
\begin{equation}
    ds^2_{d-1} = d\rho^2 + \rho^2d\Omega^2_{d-2}\ .
\end{equation}
The entangling surface $\partial B_R$ is located at $\rho = R$. We parametrize the embedding of the RT surface $ \Sigma_8 $ in the bulk as $ \rho = \rho(r) $ and impose the boundary conditions
\begin{equation}
\rho(r_*) = 0 \ ,\quad \rho(\infty) = R\ , \quad \rho'(r_*)^{-1} = 0\ .
\label{eq:rhoboundaryconditions}
\end{equation}
The first condition determines the turning point $r_*$ of the RT surface, the second condition anchores the surface to $\partial B_R$ at the conformal boundary and the third condition ensures that there is no conical singularity at the tip of the surface. The induced metric $\hat{g}$ on $ \Sigma_8 $ for the backgrounds in \eqref{eq:smearedbckgs} is given by
\begin{equation}
ds^2 = h(r)^{-1\slash 2}\left[ \rho'(r)^{2} + h(r)\,\Sigma(r)^2 \right] dr^{2}  + h(r)^{-1\slash 2}\rho(r)^{2}d\Omega^2_{d-2}+h(r)^{1\slash 2}ds^2_{\widetilde{X}_q}\ , 
\end{equation}
so that the entropy functional becomes
\begin{equation}
S(R) = \mathcal{N}_S\int_{r_*}^{1\slash \epsilon} dr\,H(r)\,\rho(r)^{d-2}\sqrt{h(r)\,\Sigma(r)^2+\rho'(r)^{2}} \ ,
\label{eq:firstEE}
\end{equation}
where $\Sigma(r)$ is defined in \eqref{eq:Sigma}, we have defined
\begin{equation}
H(r) = \ell^{-8}\,e^{-2\phi+4g}(h^\frac{1}{2}e^{f})^{5-d}\ ,\quad \mathcal{N}_S = \frac{2\vol{(X_{q})}\vol{(S^{d-2})}}{(2\pi)^{6}\,g_{\text{s}}^{2}}\biggl(\frac{\ell}{\ell_{\text{s}}}\biggr)^{8}
\end{equation}
and $ \epsilon $ is the UV cut-off corresponding to $ z = \epsilon $ in the Fefferham--Graham (FG) coordinate $ z = 1\slash r $. In these expressions, the functions $f,g,h,\phi$ for each dimension $d$ have to be understood as the ones corresponding to each geometry and $\vol{(S^{d-2})} $ in $\mathcal{N}_S$ comes from integration over the boundary angular coordinates $\Omega_{d-2}$. Explicitly, we have
\begin{equation}
    \mathcal{N}_S =
    \begin{dcases}
        \frac{2\pi\sqrt{2}}{3}\,N^{3\slash 2}k^{1\slash 2}\ ,\quad &\text{ABJM}\\
        2\Nc^{2}\ ,\quad &\text{SYM } \text{S}^5\\
        \frac{27}{8}\Nc^{2}\ ,\quad &\text{SYM } \text{T}^{1,1}
    \end{dcases} \, ,
\end{equation}
where we have used $\vol{(\text{S}^5)} = \pi^3 $, $\vol{(\text{T}^{1,1})} = \frac{16}{27}\,\pi^3$ and $\vol{(X_6)} = 2^6\vol{(\mathbb{C}\text{P}^3)} = \frac{64}{3}\,\pi^3 $.

The RT surface is a minimal area surface and its embedding $\rho(r)$ extremizes $S(R)$. In the variation of \eqref{eq:firstEE} with respect to $\rho(r)$, the boundary terms at the end-points $r = r_*$ and $r = 1\slash\epsilon \rightarrow \infty$ vanish under the Dirichlet boundary conditions $\rho(r_*) = 0$,  $\rho(\infty) = R$. Thence setting the variation to zero amounts to solve the Euler--Lagrange equation for $\rho(r)$, which is a second-order ordinary differential equation whose solution is the embedding of the RT surface. This solution contains two integration constants, which are fixed in terms of $(r_*,R)$ by Dirichlet boundary conditions at the two end-points, and requiring that the embedding does not have a conical singularity at its tip, $\rho'(r_*)^{-1} = 0$, fixes $r_* $ in terms of $R$. Thus the RT surface is completely fixed in terms of $R$.

\paragraph{Few flavor expansion.} We will now evaluate \eqref{eq:firstEE} at the extremum to leading order in the small flavor expansion. We will accordingly expand the embedding and the location $r_*$ of the tip of the RT surface as
\begin{equation}
\rho(r) = \rho_0(r) + \Nf\,\rho_1(r)+ \mathcal{O}(\Nf^{2}) \ ,\quad r_* = r_{*,0} + \Nf\,r_{*,1}+ \mathcal{O}(\Nf^{2})\,,
\label{eq:rhoexp}
\end{equation}
where we assume the boundary conditions
\begin{equation}
    \rho_0(r_{*,0}) = 0,\quad \rho_0(\infty) = R,\quad \rho_0'(r_{*,0})^{-1} = 0\,,
    \label{eq:rho0conditions}
\end{equation}
such that $r_{*,0}$ is the location of the tip of the zeroth-order embedding $\rho_0$ where it is regular by the last condition. The boundary conditions for $\rho_1(r)$ follow by combining \eqref{eq:rho0conditions} with the boundary conditions \eqref{eq:rhoboundaryconditions} of the full embedding $\rho(r)$. The result is
\begin{equation}
    \rho_1(r_{*,0}) = -\rho_0'(r_{*,0})\,r_{*,1},\quad \rho_1(\infty) = 0,\quad \frac{\rho_1'(r_{*,0})}{\rho_0'(r_{*,0})^2} = -\frac{\rho_0''(r_{*,0})}{\rho_0'(r_{*,0})^2}\,r_{*,1}\,,
    \label{eq:rho1boundaryconditions}
\end{equation}
where the first and third equations should be understood as a limit since $\rho_0'(r_{*,0}) = \infty$ by imposing the absence of a conical singularity. Expanding the Euler--Lagrange equation for $\rho(r)$ to linear order in $\Nf$ gives second-order equations for $\rho_0(r)$ and $\rho_1(r)$ whose solutions are fixed in terms of $R$ by the boundary conditions \eqref{eq:rho0conditions} and \eqref{eq:rho1boundaryconditions} respectively (as explained above). This also fixes $r_{*,0}$ and $r_{*,1}$ in terms of $R$.

In the few flavor expansion \eqref{eq:funcexps1}--\eqref{eq:funcexps2}, we get
\begin{equation}
 H(r) =  H_0(r)\,[1+\Nf\, H_1(r)+ \mathcal{O}(\Nf^{2})] \ , 
\end{equation}
where explicitly $H_0(r) = r^{d-1}$ in ABJM $d = 3$ and D3-brane $d = 4$ geometries. The entanglement entropy \eqref{eq:firstEE} has the expansion
\begin{equation}
 \frac{S(R)}{\mathcal{N}_S} = \int_{r_*}^{1\slash \epsilon} dr\,\mathcal{S}^{(0)}(\rho(r),\rho'(r),r) + \Nf \int_{r_*}^{1\slash \epsilon} dr\,\mathcal{S}^{(1)}(\rho(r),\rho'(r),r) + \mathcal{O}(\Nf^{2}) \ ,
\label{eq:entexp}
\end{equation}
where we have defined
\begin{align}\label{eq:S0andS1functionals}
\mathcal{S}^{(0)}(\rho,\rho',r) &= H_0(r)\rho(r)^{d-2}\sqrt{h_0(r)\,\Sigma_0(r)^2+\rho'(r)^2}\nonumber\\
\mathcal{S}^{(1)}(\rho,\rho',r) &=\frac{\rho(r)^{d-2}H_0(r) h_0(r)\Sigma_0(r)^2}{\sqrt{h_0(r)\Sigma_0(r)^2+\rho'(r)^2}}\bigg[\frac{h_1(r)}{2h_0(r)}+\Sigma_1(r)+\frac{h_0(r)\Sigma_0(r)^2+\rho'(r)^2}{h_0(r) \Sigma_0(r)^2}H_1(r)\bigg] \ .
\end{align}
Substituting the expansion \eqref{eq:rhoexp} gives to zeroth-order in $\Nf$
\begin{equation}
\frac{S(R)}{\mathcal{N}_S} = \int_{r_{*,0}}^{1\slash \epsilon} dr\,\mathcal{S}^{(0)}(\rho_0(r),\rho'_0(r),r) + \mathcal{O}(\Nf) \ .
\end{equation}
Extremizing the functional above gives an equation for $\rho_0(r)$ whose solution involves two integration constants that are fixed in terms of $(r_{*,0},R)$ by the first two conditions in \eqref{eq:rho0conditions}. The third regularity condition fixes $r_{*,0} = 1\slash R$, giving the solution
\begin{equation}
\rho_0(r) = \sqrt{R^{2} - \frac{1}{r^{2}}}
\label{eq:rho0}
\end{equation}
which is a hemisphere of radius $ R $ in the Poincar\'e coordinate $ z = 1\slash r $. This is the smooth minimal surface in the non-backreacted geometry.

Substituting the expansions \eqref{eq:rhoexp} into the first term of \eqref{eq:entexp} produces three boundary terms at order $\Nf$: one from the variation of the lower limit $ r_* $ of integration and two from the variation of $ \mathcal{S}^{(0)} $ with respect to $ \rho(r) $ (since $\rho_0(r)$ is an extremum there is no bulk term at this order). In the second term, we can replace $ \rho \rightarrow \rho_0 $, since the difference is of order $ \mathcal{O}(\Nf^{2}) $. The result is
\begin{equation}\label{eq:entexp-order2}
S(R) = S^{(0)}(R) + \Nf\,S^{(1)}(R)+ \mathcal{O}(\Nf^{2}) \ ,
\end{equation}
where
\begin{equation}
S^{(0)}(R) = \mathcal{N}_S\int_{1\slash R}^{1\slash \epsilon} dr\, \mathcal{S}^{(0)}(\rho_0,\rho_0',r) \ , \quad S^{(1)}(R) = \mathcal{N}_S\,\biggl(B_1 - B_2+ \int_{1\slash R}^{1\slash \epsilon} dr\,\mathcal{S}^{(1)}(\rho_0,\rho'_0,r)\biggr)
\label{eq:S0andS1}
\end{equation}
and the boundary terms are
\begin{align}
B_1 &= \frac{\partial \mathcal{S}^{(0)}(\rho_0,\rho'_0,r)}{\partial \rho'_0(r)}\,\rho_1(r)\bigg\lvert_{r = 1\slash \epsilon}\\
B_2 &= \biggr(\frac{\partial \mathcal{S}^{(0)}(\rho_0,\rho'_0,r)}{\partial \rho'_0(r)}\,\rho_1(r)+\mathcal{S}^{(0)}(\rho_0,\rho_0',r)\, r_{*,1}\biggl)\bigg\lvert_{r=1\slash R}.
\label{eq:RTboundaryterms}
\end{align}
Since there is no physical boundary at the tip $r = 1\slash R$ of the RT surface, the boundary term $B_2$ is expected to vanish. Indeed by using the first boundary condition in \eqref{eq:rho1boundaryconditions}, we get
\begin{equation}
    B_2 = \lim_{r\rightarrow 1\slash R}\biggr(\frac{\partial \mathcal{S}^{(0)}(\rho_0,\rho'_0,r)}{\partial \rho'_0(r)}\,\rho_0'(r)+\mathcal{S}^{(0)}(\rho_0,\rho_0',r)\biggl)\,r_{*,1} = 0\,,
    \label{eq:vanishingB2}
\end{equation}
where the last equality follows after substituting the regular zeroth-order solution \eqref{eq:rho0}. The exact reason for the vanishing of $B_2$ is the regularity condition imposed on the zeroth-order embedding $\rho_0(r)$ (the third condition in \eqref{eq:rho0conditions}). If one does not impose this condition (which amounts to keeping $r_{*,0}$ in the $\rho_0(r)$ solution free) then $B_2\neq 0$.\footnote{The general zeroth-order embedding is given by $\rho_0(r) = \sqrt{c_+^2-r^{-2}}+c_-$ with two integration constants which are fixed by the first two conditions in \eqref{eq:rho0conditions} as $c_{\pm} = \frac{R}{2}\pm\frac{1}{2Rr_{*,0}^2}$. This solution has a conical singularity at $r_{*,0}$ and the boundary term \eqref{eq:vanishingB2} evaluated in the limit $r\rightarrow r_{*,0}$ gives $B_2 = -\left\lvert\frac{1-R^2r_{*,0}^2}{1 + R^2r_{*,0}^2} \right\lvert\,\frac{r_{*,1}}{r_{*,0}}$.} The first boundary term $B_1$ is generically non-vanishing and it is determined by the UV asymptotics of the solution $\rho_1(r)$ (see Appendix \ref{app:RTdetails}). Thus the computation of entanglement entropy to linear order in $\Nf$ requires to solve the correction $\rho_1(r)$ to the embedding only near the conformal boundary with the remaining data being determined by the zeroth-order embedding \eqref{eq:rho0}.

\begin{figure}[t]
    \centering
   \includegraphics[width=0.75\textwidth]{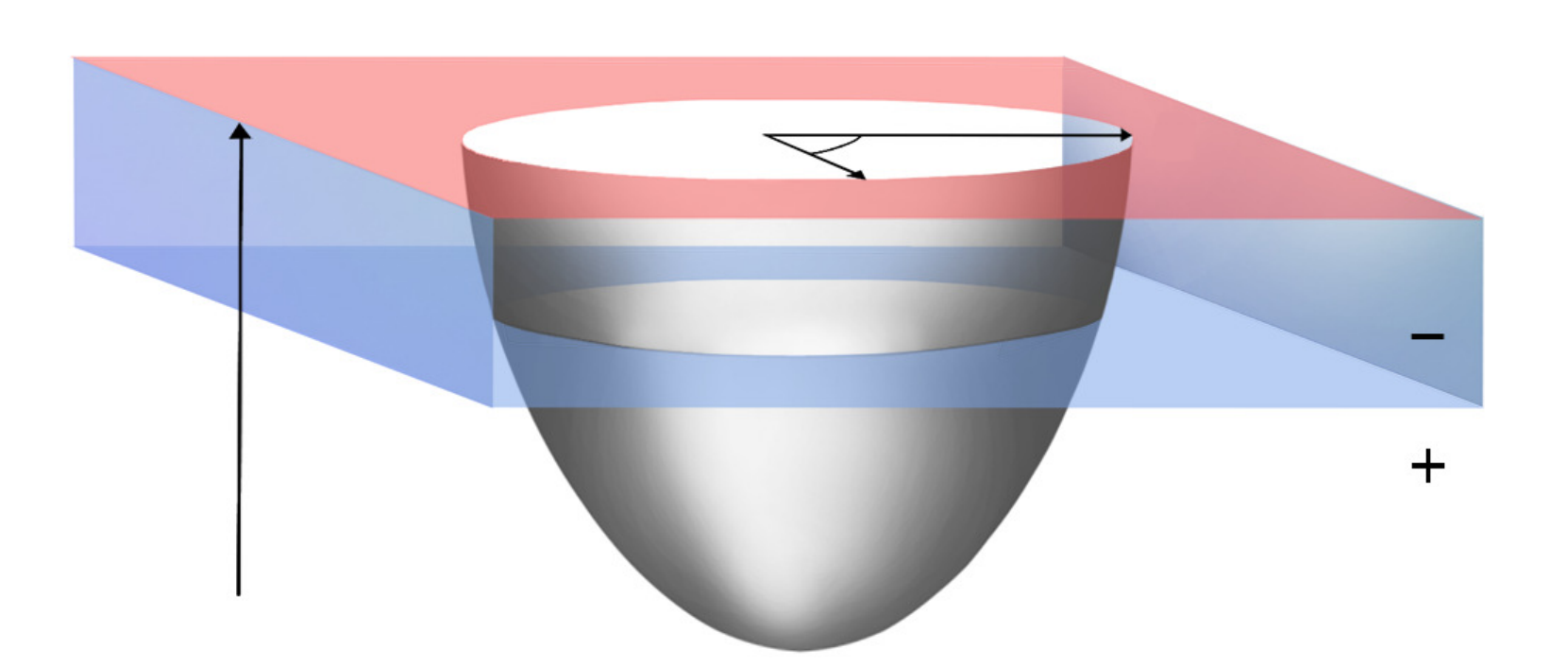}
   \put(-120,20){$\rho(r)$}
   \put(-86,105){$R$}
   \put(-286,68){$r_\text{q}$}
   \put(-285,16){$r$}
    \caption{The sketch of the RT surface in the backreacted geometry anchored to a sphere subregion of radius $R$ in the field theory residing at the boundary (red layer). Here we have chosen to depict a large subregion such that the corresponding minimal surface enters the cavity region (unshaded area), $r<r_\text{q}$. Note that the RT surface is deformed away from a pure hemisphere embedding in the presence of massive flavors because the external spacetime departs from anti de Sitter.}
    \label{fig:cavity}
\end{figure}
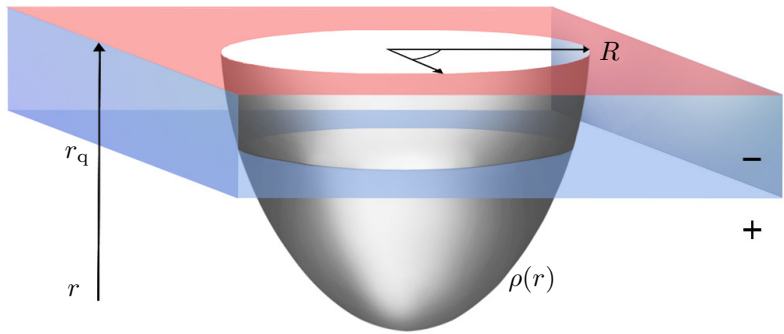

The entanglement entropy probes the cavity region of the backreacted geometry as determined by the RT surface \eqref{eq:rho0} in the non-backreacted geometry. For $ 0 <r_\text{q} R < 1$, the RT surface \eqref{eq:rho0} remains completely outside of the cavity while for $r_\text{q} R>1$ it is partially inside. See Fig.~\ref{fig:cavity} for illustration. Recall that in our geometries, the cavity is an $\mathcal{O}(\Nf)$ effect, thus it is completely encoded by the flavor correction to the entropy $S^{(1)}(R)$ as:
\begin{equation}
    S(R) =S^{(0)}(R)+\Nf\times
    \begin{dcases}
        S_-^{(1)}(R)\ ,\quad & r_\text{q} R <1\\
        S_+^{(1)}(R)\ ,\quad & r_\text{q} R >1
    \end{dcases} \ ,
\end{equation}
with $S_-=S_+$ at $r_\text{q} R  = 1$ so that the transition in the entanglement entropy between the two regimes is continuous. To simplify the notation, we will define the dimensionless radius
\begin{equation}
    \R = r_\text{q} R \ .
    \label{eq:dimensionlessradius}
\end{equation}

\subsubsection{Smeared D7-branes on the conifold}

We will now present the main result for the entanglement entropy of the geometry given by the D3-D7 intersection on the conifold, whose explicit form is detailed in Appendix~\ref{app:T11S5details}. We will discuss two types of solutions, the \textit{chiral} \cite{Benini:2006hh,Bigazzi:2008zt} and \textit{non-chiral} profiles \cite{Ouyang:2003df,Kuperstein:2004hy,Bigazzi:2008ie} for the flavor configurations -- the unflavored case is the well-known ${\cal N}=1$ $SU(\Nc)\times SU(\Nc)$ Klebanov--Witten quiver gauge theory \cite{Klebanov:1998hh}. The detailed computation of the entanglement entropy is relegated to Appendix~\ref{app:EET11} and is obtained by application of the formulas derived in the previous section. We will distinguish two configurations: those embeddings in which the RT surface penetrates the cavity and those in which the RT surface lies completely outside the cavity. 

We find the entanglement entropy of the unflavored SYM theory to be
\begin{align}
S^{(0)}(R)=\frac{27}{16}\Nc^{2}\left(\frac{R^2}{\epsilon^2}-\log{\frac{2R}{\epsilon}}-\frac{1}{2}\right),
\label{eq:S0conifold}
\end{align}
and, to first order in $\Nf$, the flavor correction outside ($-$) and inside ($+$) the cavity reads as follows.

\paragraph{Chiral profile.}
\begin{align}
S_-^{(1)}(R) &= \frac{81g_{\text{YM}}^{2}\Nc^{2}}{512\pi^{2}}\,\biggl[\frac{R^{2}}{\epsilon^{2}}-2\log{\frac{2R}{\epsilon}}-\frac{1}{2}+\frac{3\pi}{4}\,\R^3\,\biggl(\log{\frac{\R }{2}}-\frac{13}{12}\biggr)\biggr]\label{eq:S1conifoldchiral1}\\
S_+^{(1)}(R) &= \frac{81g_{\text{YM}}^{2}\Nc^{2}}{512\pi^{2}}\,\biggl[\frac{R^{2}}{\epsilon^{2}}-2\log{\frac{2R}{\epsilon}}-\frac{1}{2}+2\arcosh{\R}-\frac{6}{7\R}\sqrt{\R^2-1}\,\biggl(1+\frac{19}{6}\,\R^2\biggr)\biggr.\nonumber\\
&\qquad \qquad  +\frac{4}{21}\,{}_3F_2\left(\begin{matrix} -\frac{1}{2}, \frac{3}{2}, \frac{3}{2}\\\frac{5}{2}, \frac{5}{2}\end{matrix};\frac{1}{\R^2}\right)-\frac{12}{7}\,\R^2\,{}_3F_2\left(\begin{matrix} \frac{1}{2}, \frac{1}{2}, \frac{1}{2}\\\frac{3}{2}, \frac{3}{2}\end{matrix};\frac{1}{\R^2}\right)-\frac{11}{7}\R^3\arcsin{\R^{-1}}\biggr]\label{eq:S1conifoldchiral2} \ ,
\end{align}

\paragraph{Non-chiral profile.}
\begin{align}
S_-^{(1)}(R) &= \frac{81g_{\text{YM}}^{2}\Nc^{2}}{512\pi^{2}}\,\biggl[\frac{R^{2}}{\epsilon^{2}}-2\log{\frac{2R}{\epsilon}}-\frac{1}{2}-\frac{\pi}{4}\,\R^3\biggr]\label{eq:S1conifoldnonchiral1}\\
S_+^{(1)}(R) &= \frac{81g_{\text{YM}}^{2}\Nc^{2}}{512\pi^{2}}\,\biggl[\frac{R^{2}}{\epsilon^{2}}-2\log{\frac{2R}{\epsilon}}-\frac{1}{2}-\frac{\sqrt{\R^2-1}}{\R}\,\biggl(1+\frac{3}{2}\,\R^2\biggr)\biggr.\nonumber \\
&\qquad\qquad\quad  -\R^3\arctan{\left(\R -\sqrt{\R^2-1}\right)}+ 2\arcosh{\R}\biggr]\label{eq:S1conifoldnonchiral2} \ .
\end{align}
We will defer the discussion of the physics of these results until after we have defined, in particular, the entanglement A-functions, which neatly capture the number of degrees of freedom at a given energy scale.

\subsubsection{Smeared D7-branes on the five-sphere}
In this subsection we review the entanglement entropy computation for flavor corrections for D7-branes on $\text{S}^5$ \cite{Kontoudi:2013rla} for completeness. The results for the entanglement entropy in our analysis for the $\text{S}^5$, which we display below, agree with the ones found in \cite{Kontoudi:2013rla} up to trifles.\footnote{Our result for the entropy matches the one obtained in \cite{Kontoudi:2013rla} in equation (3.36), up to an overall sign difference in the second $\epsilon_q$ correction enclosed by square brackets in (3.36).}

For the leading order (unflavored) term we find
\begin{equation}
    S^{(0)}(R)=\Nc^{2}\left(\frac{R^2}{\epsilon^2}-\log{\frac{2R}{\epsilon}}-\frac{1}{2}\right) 
\end{equation}
and for the $\Nf$ correction we obtain
\begin{align}
S_-^{(1)}(R) &= \frac{g_{\text{YM}}^{2}\Nc^{2}}{16\pi^{2}}\,\biggl[\frac{R^{2}}{\epsilon^{2}}-\biggl(2+\frac{8}{3}\,\R^2\biggr)\log{\frac{2R}{\epsilon}}-\frac{1}{2}+\frac{28}{9}\,\R^2+\frac{4}{15}\,\R^4\biggr]\label{eq:SD7S5minus}\\
S_+^{(1)}(R) &= \frac{g_{\text{YM}}^{2}\Nc^{2}}{16\pi^{2}}\,\biggl[\frac{R^{2}}{\epsilon^{2}}-\biggl(2+\frac{8}{3}\,\R^2\biggr)\log{\frac{2R}{\epsilon}}-\frac{1}{2}+\frac{28}{9}\,\R^2+\frac{4}{15}\,\R^4\biggr.\label{eq:SD7S5plus}\\
&\biggl.\qquad\qquad\quad-\frac{32}{45\R}\sqrt{\R^2-1}\,\biggl(1+\frac{83}{16}\,\R^2+\frac{3}{8}\,\R^4\biggr) + \biggl(2+\frac{3}{2}\,\R^2\biggr)\arcosh{\R}\biggr]\nonumber \ .
\end{align}

\subsubsection{Smeared D6-branes on the complex projective three-space}
Let us continue to present the main result for the entanglement entropy for a disk in the ABJM geometry with massive fundamental flavors \cite{Conde:2011sw,Bea:2013jxa}, but in the limit of small number of flavors \cite{Bea:2016ekp,Balasubramanian:2018qqx}. We defer the discussion about the background geometry to Appendix~\ref{app:ABJMdetails}, while the entanglement entropy calculation is found in Appendix~\ref{app:EEABJM}. As in the D3-D7 geometries, we will distinguish two configurations depending on whether the entangling surface enters the cavity. 

We find for the entropy of unflavored ABJM theory the result \cite{Bea:2013jxa}
\begin{equation}
    S^{(0)}(R) = \frac{\pi\sqrt{2}}{3}\,N^{3\slash 2}k^{1\slash 2}\left(\frac{R}{\epsilon}-1\right)\ ,
\end{equation}
and for the flavor correction we obtain
\begin{equation}
S^{(1)}(R) = \frac{\pi}{4}N\sqrt{2\lambda}\times
\begin{dcases}
\frac{5}{4}\frac{R}{\epsilon}-1-\frac{1}{3}\,\R^2 \ , \quad &\R <1\ , \quad \text{outside cavity}\\
\frac{5}{4}\frac{R}{\epsilon}-\R -\frac{1}{3\R} \ , \quad &\R >1 \ , \quad \text{inside cavity} 
\end{dcases} \ .
\label{eq:S1backreactedABJM}
\end{equation}
At leading order in the mass $r_\text{q} \rightarrow 0$, the correction is quadratic in mass.
For generic number of flavors the result is numerically available in \cite{Bea:2013jxa}.

\section{Flavored entanglement entropy using probe branes}\label{sec:probe}
In the previous section, we studied corrections to the entanglement entropy of a spherical subregion caused by the addition of massive flavor degrees of freedom in the theory. We derived the corrections holographically by utilizing the RT formula: we first constructed the backreacted bulk geometry sourced by smeared flavor branes and then computed the area of the minimal surface in this geometry. The backreacted solution is constructed with a special distribution of branes: one in which the branes are smeared, but we expect it to give the correct entanglement entropy to leading order in the number of flavor fields $\Nf$.

In this section, we derive the massive flavor corrections to the entanglement entropy using methods that do not require the knowledge of the backreacted bulk geometry. In the massless flavor limit, we do this by using a conformal map \cite{Casini:2011kv} to compute the entanglement entropy correction from the free energy on a sphere. For massive flavor, we use the method developed in \cite{Karch:2014ufa,Chalabi:2020tlw} (see also \cite{Chang:2013mca} for earlier work). In both cases, the result is valid in the probe limit and the flavor correction to the entanglement entropy is computed simply by the on-shell flavor D-brane action in the unflavored background. The culmination is that we precisely match with the results in the backreacted computations presented in the previous section at all values of the flavor masses. It is worth noting, however, that this matching is contingent upon addressing the boundary term challenge and accounting for non-universal terms.

\subsection{Entropy of massless flavor from the sphere free energy}\label{subsec:masslessprobe}

When the flavor is massless, $r_\text{q} = 0$, the conformal symmetry of the flavored field theory is restored. In a CFT$_d$ living on Minkowski space, the entanglement entropy $S(R)$ of a spherical spatial subregion of radius $R$ is equal to the (Euclidean) free energy $F_{\text{S}^{d}} = -\log{Z_{\text{S}^{d}}}$ of the theory on a unit sphere $\text{S}^{d}$,
\begin{equation}
    S(R) = F_{\text{S}^{d}} \ ,
    \label{eq:entropyequalsF}
\end{equation}
which follows from the Casini--Huerta--Myers conformal map \cite{Casini:2011kv}. Here $Z_{\text{S}^{d}}$ denotes the Euclidean path integral of the CFT on $\text{S}^{d}$. In $d=3$, this implies that $S(R)$ is independent of $R$, and in $d=4$, that the coefficient of the divergence $\log{(2R\slash \epsilon)}$ is independent of $R$.

We can use the equality \eqref{eq:entropyequalsF} to compute the massless value of the entanglement entropy of a spherical subregion in the flavored theory. In the probe limit, the free energy has the expansion
\begin{equation}
    F_{\text{S}^{d}} = F^{(0)}_{\text{S}^{d}} + \Nf\, F^{(1)}_{\text{S}^{d}} + \mathcal{O}(\Nf)^2 \ ,
\end{equation}
where holographically the coefficients are computed by bulk on-shell actions as
\begin{equation}
        F^{(0)}_{\text{S}^{d}} = I^{\text{on-shell}}_{\text{grav}}\ ,\quad F^{(1)}_{\text{S}^{d}} = I^{\text{on-shell}}_{\text{D}p} \ .
\end{equation}
Here $I^{\text{on-shell}}_{\text{grav}}$ is the type IIA or IIB supergravity action \eqref{eq:typeAaction} evaluated on the solution that does not include brane backreaction, the unflavored solution, and $I^{\text{on-shell}}_{\text{D}p}$ is the D-brane action evaluated on the embedding that extremizes $I_{\text{D}p}$ in the unflavored solution. These on-shell actions can be computed using the $(d+1)$-dimensional effective actions of type IIB and 11-dimensional supergravities in the D3-D7 and ABJM systems, respectively. The computations are done using lower-dimensional effective actions, because evaluation of the supergravity actions directly gives zero and the prescription for obtaining the correct value is still an open problem (see~\cite{Kurlyand:2022vzv,Mkrtchyan:2022xrm,Chakrabarti:2022jcb} for recent advances). Details of the calculations are relegated to Appendix~\ref{app:spherefreenenergy}.

\paragraph{SYM theory.} In Appendix~\ref{app:spherefreenenergySYM}, we show that the unflavored free energy and the flavor contribution are given by
\begin{equation}
F^{(0)}_{\text{S}^{4}} = \frac{1}{16\pi^7g_{\text{s}}^{2}}\,\biggl(\frac{\ell}{\ell_{\text{s}}}\biggr)^{8}\vol{(X_{5})}\vol{(\text{AdS}_{5})}\ ,\quad F^{(1)}_{\text{S}^{4}} =\frac{1}{(2\pi)^{7}g_{\text{s}}}\,\biggl(\frac{\ell}{\ell_{\text{s}}}\biggr)^{8}\vol{(M_3)}\vol{(\text{AdS}_{5})} \ ,
\end{equation}
where $M_3\subset X_5$ is the part of the internal space wrapped by the probe D7-brane. Substituting parameters \eqref{eq:D3parameters} gives
\begin{equation}
F_{\text{S}^{4}} = \frac{\pi N^{2}}{\vol{(X_5)}}\biggl(1 + \frac{g_{\text{YM}}^2\Nf\vol{(M_3)}}{16\pi\vol{(X_5)}}+ \mathcal{O}(\Nf^{2})\biggr)\vol{(\text{AdS}_5)} \ .
\end{equation}
In these expressions, $\vol{(\text{AdS}_{5})}$ denotes the regularized volume of $\text{AdS}_5$ which has a logarithmic divergence whose coefficient is universal and independent of the regularization prescription \cite{Graham:1999jg}
\begin{equation}
    \vol{(\text{AdS}_{5})} = \pi^2\log{\epsilon} + \ldots \ .
\end{equation}
In the two cases, the volumes are given by
\begin{equation*}
\left\{
\begin{alignedat}{3}
    \vol{(M_3)} &= 2\pi^2\ ,&\quad\vol{(X_5)} &= \pi^3\ ,&\quad X_5 &= \text{S}^5\\
    \vol{(M_3)} &= \frac{16}{9}\,\pi^2\ ,&\quad \vol{(X_5)} &= \frac{16}{27}\,\pi^3\ ,&\quad X_5 &= \text{T}^{1,1}
\end{alignedat}
\right.
\end{equation*}
so that we get a prediction for the coefficient of the logarithmic divergence in the entropy
\begin{equation}
S(R)\big\lvert_{r_\text{\tiny{q}} = 0}\, = 
\begin{dcases}
N^{2}\,\biggl(1 + \frac{g_{\text{YM}}^2\Nf}{8\pi^2}+ \mathcal{O}(\Nf^{2})\biggr)\log{\epsilon}\\
\frac{27}{16}\,N^{2}\,\biggl(1 + \frac{3g_{\text{YM}}^2\Nf}{16\pi^2}+ \mathcal{O}(\Nf^{2})\biggr)\log{\epsilon} 
\end{dcases}\ .
\label{eq:D7entropiesfromF}
\end{equation}
\paragraph{ABJM theory.} The on-shell action of the type IIA supergravity ABJM solution is computed by first uplifting to 11-dimensional supergravity and then dimensionally reducing to a four-dimensional effective action. The on-shell D6-brane action on the other hand is directly computed in ten dimensions. Details can be found in Appendix~\ref{app:spherefreenenergyABJM} and the results are
\begin{equation}
    F_{\text{S}^{3}}^{(0)} = \frac{3}{\pi^{8}}\,\biggl(\frac{L}{\ell_{\text{s}}}\biggr)^{9}\vol{(\text{S}^{7}\slash \,\mathbb{Z}_k)}\vol{(\text{AdS}_4)}\ ,\quad F^{(1)}_{\text{S}^{3}} =\frac{12}{(2\pi)^{6}g_{\text{s}}}\biggl(\frac{\ell}{\ell_{\text{s}}}\biggr)^{7} \vol{(\mathbb{R}\rm{P}^{3})}\vol{(\text{AdS}_{4})} \ ,
    \label{eq:ABJMFs}
\end{equation}
where $L$ is the radius of curvature of the 11-dimensional solution and it is related to the 10-dimensional one via $ \ell^3 = g_{\text{s}}L^3 $. Due to the contribution of the WZ action, the D6-brane on-shell action differs from the on-shell value of the DBI action alone by a factor of $3\slash 2$ (see also \cite{Jokela:2021knd}). This same factor was discussed in \cite{Jensen:2013lxa}.

Substituting the parameters \eqref{eq:ABJMparameters} gives
\begin{equation}
    F_{\text{S}^3} =\sqrt{\frac{\pi}{24\vol{(\mathbb{C}\rm{P}^3)}}}\,N^{3\slash 2}k^{1\slash 2}\,\biggl(1 + \frac{\pi\vol{(\mathbb{R}\rm{P}^3)}}{4\vol{(\mathbb{C}\rm{P}^3})}\frac{\Nf}{k}+ \mathcal{O}(\Nf^2)\biggr)\vol{(\text{AdS}_4)} \ ,
\end{equation}
where we used that $\vol{(\text{S}^{7}\slash \,\mathbb{Z}_k)} = \pi \vol{(\mathbb{C}\rm{P}^3 )}\slash k = \pi^4\slash (3k)$ and $\vol{(\mathbb{R}\rm{P}^3)} = \pi^2$. The renormalized volume of $\text{AdS}_4$ is \cite{Graham:1999jg}
\begin{equation}
    \vol{(\text{AdS}_{4})} = \frac{(2\pi)^2}{3}\,.
\end{equation}
Thus we get the following $\Nf$-expansion for the finite term in the entanglement entropy
\begin{equation}
    S(R)\big\lvert_{r_\text{\tiny{q}} = 0}\, = \frac{\pi\sqrt{2}}{3}N^{3\slash 2}k^{1\slash 2}\,\biggl(1 + \frac{3}{4}\frac{\Nf}{k}+ \mathcal{O}(\Nf^2)\biggr) \ ,
    \label{eq:D6entropyfromF}
\end{equation}
when the flavor is massless.

\subsection{Entropy of massive flavor from probe brane actions}\label{sec:probeentropy}

We will now move on to consider flavor with non-zero mass and compute the flavor correction to the entanglement entropy of a spherical subregion in the probe limit using the method of \cite{Karch:2014ufa,Chalabi:2020tlw}. We will first briefly review the method and then present the results of the computations whose details are relegated to Appendix \ref{app:probedetails}. The advantage of the method is that it does not require the knowledge of the backreacted bulk geometry.

The idea of \cite{Karch:2014ufa,Chalabi:2020tlw} is to redo the derivation of the RT formula \cite{Lewkowycz:2013nqa} in the case when the bulk theory contains flavor branes. Recall that the total bulk action \eqref{eq:totalbulkaction} is given by
\begin{equation}
    I[g,\xi] = I_{\text{grav}}[g] + \Nf\,I_{\text{D}p}[\xi,g] \ ,
    \label{eq:sugrawithbranes}
\end{equation}
where $g$ is the background metric $g$ and $\xi$ is the embedding of the flavor brane. The field theory replica trick computes the entanglement entropy of a spherical subregion of radius $R$ as an Euclidean partition function on a space where the periodicity of the Euclidean time $\tau$ is $2\pi n$ and $n$ is the number of replicas \cite{Calabrese:2004eu,Calabrese:2009qy}. In the $N,\Nf \rightarrow \infty $ limit with $\Nf\slash N$ finite, the so-called Veneziano limit, this partition function is computed holographically by the on-shell action of a solution $\{g_n,\xi_n\}$ of the coupled equations of motion of the action \eqref{eq:sugrawithbranes}. The absence of conical singularities in the bulk requires that the metric $g_n$ takes the form
\begin{equation}
    ds^2 = du^2 + \frac{u^2}{n^2}\,d\tau^2 + \sigma_{nAB}\,d\hat{x}^Ad\hat{x}^B + \ldots \ ,\quad u\rightarrow 0 
    \label{eq:shrinkingcondition}
\end{equation}
near the codimension-two surface $u = 0$ (with induced metric $\sigma_n$) where the Euclidean time circle shrinks to zero size \cite{Lewkowycz:2013nqa}. Here $u$ is a holographic radial coordinate, $\hat{x}^A$ and $A = 1,\ldots, 8$ are the rest of the eight coordinates of the ten-dimensional geometry and ellipsis denote subleading terms in $u\rightarrow 0$. The resulting entanglement entropy is given by the derivative \cite{Lewkowycz:2013nqa}
\begin{equation}
S(R) = \biggl(n\frac{d}{dn}-1\biggr)\,I_n[g_n,\xi_n]\bigg\lvert_{n=1} 
\end{equation}
of the on-shell action $I_n$ where the subscript $n$ indicates that $\tau$ is integrated from zero to $2\pi n$. Assuming that the solution $\{g_n,\xi_n\}$ respects replica symmetry $I_n[g_n,\xi_n] = nI_1[g_n,\xi_n]$, this can be simplified to
\begin{equation}
S(R) = -\frac{d}{dn}I[g_n,\xi_n]\bigg\lvert_{n=1} \ ,
\label{eq:LMentropy}
\end{equation}
where $\tau$ is integrated from $0$ to $2\pi$ and we denote $I_1\equiv I$ from now on. One can show that this formula simplifies to the area \eqref{eq:RTformula} of the RT surface (but in the Einstein frame) in the backreacted metric $g_n$ assuming certain IR boundary term coming from the brane action vanishes \cite{Karch:2014ufa} (as also mentioned in \cite{Chalabi:2020tlw}).

The computation of \eqref{eq:LMentropy} to all orders in $\Nf\slash N$ requires the knowledge of the fully backreacted solution $g_n$ due to the brane sources. However, in the probe limit, where $\Nf\slash N \ll 1$, the situation simplifies as first shown in \cite{Karch:2014ufa,Chalabi:2020tlw}. In this limit the solution can be expanded as
\begin{equation}
    g_n = g_{n,0} + \Nf\,g_{n,1} + \mathcal{O}(\Nf^2)\ ,\quad \xi_n = \xi_{n,0} + \Nf\,\xi_{n,1} + \mathcal{O}(\Nf^2) \ ,
    \label{eq:probelimit}
\end{equation}
where $g_{n,0}$ extremizes the supergravity action $I_{\text{grav}}[g]$ without brane sources and the embedding $\xi_{n,0}$ extremizes the brane action $I_{\text{D}p}[\xi,g_{n,0}]$ with respect to $\xi$ in the unflavored metric $g_{n,0}$. The terms $g_{n,1}$ and $\xi_{n,1}$ denote the leading order backreaction of flavor to the metric and the embedding, respectively. 

The observation in \cite{Karch:2014ufa,Chalabi:2020tlw} was that \eqref{eq:LMentropy} has an expansion in the probe limit
\begin{equation}\label{eq:entropy-expansion}
S(R) = S^{(0)}(R) + \Nf\, S^{(1)}(R) + \mathcal{O}(\Nf^{2}) \ ,
\end{equation}
where the coefficients are
\begin{equation}
S^{(0)}(R) = -\frac{d}{dn}I_{\text{grav}}[g_{n,0}]\bigg\lvert_{n=1} ,\quad S^{(1)}(R) = -\frac{d}{dn} I_{\text{D}p}[\xi_{n,0},g_{n,0}]\bigg\lvert_{n=1} 
\label{eq:bannonentropy}
\end{equation}
with all $\tau$ integrals from zero to $2\pi$. In other words, the flavor correction $S^{(1)}(R)$ is computed simply by the on-shell brane action of the unflavored solution $\{g_{n,0},\xi_{n,0}\}$ without the need to compute any backreaction. The on-shell actions in the expression \eqref{eq:bannonentropy} have to be renormalized appropriately such that the UV divergence of entanglement entropy is preserved which will be done below.

\paragraph{Entanglement entropy of spherical regions.} The above discussion is general and also applies to other subregions beyond spherical regions. However, in our case of a spherical entangling surface, we can write down the ten-dimensional solution $g_{n,0}$ of supergravity (without brane sources) explicitly. It is given by (see \cite{Chalabi:2020tlw})
\begin{equation}
ds^{2}_{10} = \ell^{2}\left(ds^{2}_{\text{BH}_{d+1}} + ds^{2}_{X_{q}}\right) \ .
\label{eq:10gn0}
\end{equation}
where the $(d+1)$-dimensional hyperbolic Euclidean black hole geometry (of unit radius) is given by
\begin{equation}
ds^{2}_{\text{BH}_{d+1}} = f_n(\zeta)\,d\tau^{2}+\frac{d\zeta^{2}}{f_n(\zeta)}+\zeta^{2}(dv^{2}+\sinh^{2}{v}\,d\Omega^{2}_{d-2})  \ ,
\label{eq:hypbh}
\end{equation}
where $ \tau \sim \tau + 2\pi n $ is the Euclidean time, $v>0$ and \cite{Rangamani:2016dms,Chalabi:2020tlw}
\begin{equation}
f_n(\zeta) = \zeta^{2}-1-\frac{\zeta_n^{d-2}}{\zeta^{d-2}}\,(\zeta_n^{2}-1) \ , \ \zeta_n = \frac{1}{d n}\Bigl(1 + \sqrt{1+d(d-2)\,n^{2}}\Bigr)\ .
\end{equation}
The Euclidean time circle shrinks to zero size at $ \zeta = \zeta_n $ so that $\zeta > \zeta_n$. In terms of the coordinate $u$ defined via
\begin{equation}
    \zeta = \zeta_n + \frac{1}{2n}u^2 \ ,
    \label{eq:ucoordinate}
\end{equation}
the metric \eqref{eq:10gn0} behaves close to $u\rightarrow 0$ as
\begin{equation}
    ds^{2}_{10} = du^2 + \frac{u^2}{n^2}\,d\tau^2 + \zeta^{2}_n\,(dv^{2}+\sinh^{2}{v}\,d\Omega^{2}_{d-2}) + ds^{2}_{X_{q}}+\ldots\ ,\quad u\rightarrow 0\ ,
\end{equation}
and comparing with \eqref{eq:shrinkingcondition} we can identify $\sigma_n$ as a product of $X_q$ and a hyperboloid of curvature radius $\zeta_n$. At $n=1$, the hyperbolic black hole geometry is simply $\text{AdS}_{d+1}$ in Rindler coordinates
\begin{equation}
    ds^2_{\text{AdS}_{d+1}} = (\zeta^2-1)\,d\tau^{2}+\frac{d\zeta^{2}}{\zeta^2-1}+\zeta^{2}\,(dv^{2}+\sinh^{2}{v}\,d\Omega^{2}_{d-2}) \ .
\label{eq:rindlerads}
\end{equation}
The Rindler coordinates $(\tau,\zeta,v,\Omega_{d-2})$ are related to Poincar\'e coordinates $(r,\vec{x})$ by a coordinate transformation \eqref{eq:RindlertoPoincare} given in Appendix~\ref{subapp:bannondetails}. From this transformation we see that the RT surface \eqref{eq:rho0} in the non-backreacted background $g_{1,0}$ is then located at $\zeta = 1$.

Our branes extend along all directions of $ \text{BH}_{d+1} $ spanning the domain $\mathcal{W}_n$ in $(\tau,\zeta,v)$ coordinates and wrap a 3-cycle $ M_{3} \subset X_{q} $ in the internal space determined by a single angle $ \xi_{n,0}(\tau,\zeta,v) $. The regularized on-shell D$p$-brane action thus takes the form
\begin{equation}
I_{\text{D}p}^{\text{reg}}[\xi_{n,0},g_{n,0}] = \mathcal{N}_{\text{D}p}\int_{\mathcal{W}_n} d\tau d\zeta dv\,\mathcal{L}_{n}(\xi_{n,0},\partial_i\xi_{n,0}) \ ,
\label{eq:Dpactiontext2}
\end{equation}
where the Lagrangian $ \mathcal{L}_n $ depends explicitly on $ n $ through the blackening factor $ f_n(\zeta) $, we denote $i = \{\tau,\zeta, v\}$, and we have defined the constant
\begin{equation}
	\mathcal{N}_{\text{D}p} \equiv \frac{\vol{(S^{d-2})}\vol{(M_3)}}{(2\pi)^{p}g_{\text{s}}}\biggl(\frac{\ell}{\ell_{\text{s}}}\biggr)^{p+1} \ .\label{eq:NDp}
\end{equation}
Here the volume factor $ \vol{(S^{d-2})} $ comes from the integration over the $S^{d-2}$ in the $ \mathbb{H}_{d-1} $ part of the black hole metric and we used \eqref{eq:tension} along with $\phi = 0$ for our backgrounds. For our D7- and D6-branes, we get explicitly
\begin{equation}
    \mathcal{N}_{\text{D7}} = \frac{\pi}{4}\frac{\vol{(M_3)}}{[\vol{(X_5)}]^{2}}\,g_{\text{YM}}^{2}N^{2} \ , \quad \mathcal{N}_{\text{D6}} = \frac{1}{4}N\sqrt{2\lambda} \ .
    \label{eq:NDp_explicit}
\end{equation}
At $n = 1$, the brane fills the region $\mathcal{W}_1 = \{ (\tau,\zeta,v)\,\lvert\,r(\tau,\zeta,v) > r_\text{q}\}$ where $r$ is the Poincar\'e radial coordinate given in \eqref{eq:RindlertoPoincare}. The form of the region $\mathcal{W}_n$ when $n>1$ is more complicated and can be obtained in principle from the embedding $\xi_{n,0}$, but we will not need to do it explicitly.

The regularized action \eqref{eq:Dpactiontext2} is divergent in both $\Lambda_v\rightarrow \infty$ and $\Lambda_\zeta \rightarrow \infty$ limits. We will renormalize the divergence in $\Lambda_\zeta \rightarrow \infty$ by defining the renormalized action
\begin{equation}
    I_{\text{D}p}^{\text{ren}}[\xi_{n,0},g_{n,0}] = \lim_{\Lambda_\zeta\rightarrow \infty}\bigl(I_{\text{D}p}^{\text{reg}}[\xi_{n,0},g_{n,0}] + I_{\text{D}p}^{\text{ct}}[\xi_{n,0},\gamma_{n,0}] \bigr) \ ,
    \label{eq:renormalizedbraneaction}
\end{equation}
where the counterterms in the coordinates $(\tau,\zeta,v,\Omega_{d-2})$ take the form
\begin{equation}
I_{\text{D}p}^{\text{ct}}[\xi_{n,0},\gamma_{n,0}] =\mathcal{N}_{\text{D}p}\int_{0}^{2\pi} d\tau  \int_{0}^{\Lambda_{v}} dv\,\mathcal{L}_n^{\text{ct}}(\xi_{n,0},\partial_i\xi_{n,0})\biggr\lvert_{\zeta = \Lambda_\zeta} \ ,
\label{eq:Dpactioncttext}
\end{equation}
with $\gamma_{n,0}$ being the induced metric of the cut-off surface. The renormalized action is still divergent in $\Lambda_v\rightarrow \infty$ which will be related to the UV divergence of entanglement entropy $\epsilon \rightarrow 0$. The relation between the cut-offs we will use is $e^{\Lambda_v} = 2R\slash \epsilon$ as discussed in Appendix \ref{subapp:bannondetails}. We emphasize that this cut-off $\epsilon$ differs from the Poincar\'e coordinate cut-off $ r = 1\slash \epsilon$ used in Section \ref{sec:backreacted}, but this will not affect the form of universal regulator independent terms in the end result.

Now the leading flavor correction $S^{(1)}(R)$ to the entanglement entropy of the spherical subregion is obtained from the renormalized on-shell brane action \eqref{eq:renormalizedbraneaction} using the formula \eqref{eq:bannonentropy}. The derivative with respect to $n$ acts on the Lagrangian of the brane action \eqref{eq:Dpactiontext2} as $\frac{d}{dn}\mathcal{L}_n$, but also on the integration domain $\mathcal{W}_n$ which produces a boundary term localized on $\partial \mathcal{W}_n$ as dictated by the Reynolds transport theorem \cite{flanders1973differentiation,Reddiger:2019lqh} (see also Appendix \ref{app:boundaryterms}). Then in $\frac{d}{dn}\mathcal{L}_n$ the derivative acts both on the explicit $n$-dependence of $\mathcal{L}_n$ in the blackening factor $f_n(\zeta)$ and on the embedding $\xi_{n,0}$. The variation of the embedding produces total derivatives which integrate to additional boundary terms on top of the ones coming from the Reynolds transport theorem. Keeping only boundary terms that are non-zero the final result is given by (see Appendix \ref{subapp:bannondetails} for details and also \cite{Chalabi:2020tlw})
\begin{equation}
S^{(1)}(R)= \mathcal{I} + \mathcal{B}_{\text{RT}}^{(1)} + \mathcal{B}_{\text{RT}}^{(2)} + \mathcal{B}_{\zeta=\infty}^{(1)} + \mathcal{B}_{\zeta=\infty}^{(2)}\ ,
\label{eq:entropyfromprobes}
\end{equation}
with a $(d+1)$-dimensional bulk integral $\mathcal{I}$ and $d$-dimensional boundary terms whose expressions can be found in Appendix~\ref{subapp:bannondetails}. The boundary terms $\mathcal{B}_{\text{RT}}^{(1,2)}$ are localized at the RT surface $\zeta = 1$ and $\mathcal{B}_{\zeta=\infty}^{(1,2)}$ at the UV cut-off surface. The first term $\mathcal{B}_{\text{RT}}^{(1)}$ comes from the $n$-variation $\partial_n\zeta_n$ of the integration domain $\mathcal{W}_n$ while the first UV term $\mathcal{B}_{\zeta=\infty}^{(1)}$ arises from the $n$-variation of the UV counterterms. The rest of the boundary terms, on the other hand, arise from the total derivative term in the embedding variation $\partial_n\xi_{n,0}$ after integration by parts.

We will proceed to compute \eqref{eq:entropyfromprobes} in the two cases of a D7-brane on $\text{AdS}_5\times \text{S}^5$ and a D6-brane on $\text{AdS}_4\times \mathbb{C}\text{P}^3$ and compare with the backreacted calculations of Section~\ref{sec:backreacted}.\footnote{The probe D7-brane embedding and its on-shell action are more complicated in the $\text{AdS}_5\times \text{T}^{1,1}$ geometry so we leave the analogous computation of the entanglement entropy in that case to future work.} Details of the calculations are relegated to Appendices~\ref{subapp:probeD7} and \ref{subapp:probeD6} in the D7- and D6-brane cases, respectively. We find that the universal terms in the entanglement entropy match with the result of the backreacted method only when we omit the contribution of the IR boundary term $\mathcal{B}_{\text{RT}}^{(2)} $. More precisely, we find that the following formula
\begin{equation}
    \widetilde{S}^{(1)}(R) = S^{(1)}(R)-\mathcal{B}_{\text{RT}}^{(2)}  =\mathcal{I} + \mathcal{B}_{\text{RT}}^{(1)} + \mathcal{B}_{\zeta=\infty}^{(1)} + \mathcal{B}_{\zeta=\infty}^{(2)}\ ,
    \label{eq:Stilde}
\end{equation}
with $\mathcal{B}_{\text{RT}}^{(2)} $ subtracted reproduces universal terms in the results of Section~\ref{subsec:perturbativeRT}. Later in Section~\ref{subsec:backreactedCFfunctions}, we find that the same is true for the matching of A- and F-functions which capture only the universal part of the entanglement entropy.

The IR boundary term $\mathcal{B}_{\text{RT}}^{(2)}$ was originally argued to vanish in \cite{Karch:2014ufa}. However, in \cite{Chalabi:2020tlw} it was found to be non-zero in general and needed to match with the corresponding backreacted calculation of the entropy using the RT formula and the known backreacted geometry.\footnote{In contrast to our backreacted geometries which are sourced by smeared flavor branes, the backreacted solution in \cite{Chalabi:2020tlw} is sourced by fully localized branes.} In our case, we find, in agreement with \cite{Chalabi:2020tlw}, that $\mathcal{B}_{\text{RT}}^{(2)}$ is non-zero when conformal symmetry is broken (when the flavor is massive in our setup), but in contrast to \cite{Chalabi:2020tlw} we have to subtract it in order to match with the backreacted calculations. The main difference between the calculations is that our branes extend along three directions of the hyperbolic black hole geometry reaching the conformal boundary unlike the brane embeddings in \cite{Chalabi:2020tlw} which span two directions (the geometries considered in \cite{Chalabi:2020tlw} are domain wall geometries). Hence our brane on-shell actions also require holographic renormalization. Further discussion on the omission of the boundary term is given in Section \ref{sec:discussion}.


\subsection{Probe D7-brane on \texorpdfstring{$\text{AdS}_5\times \text{S}^5$}{}}\label{subsec:probeD7}

The non-backreacted background $g_{1,0}$ in this case is the $\text{AdS}_5\times \text{S}^5 $ geometry \eqref{eq:D3parameters}. Hence the geometry $g_{n,0}$ is given by
\begin{equation}
ds^{2} = \ell^{2}(ds^{2}_{\text{BH}_5} + ds^{2}_{\text{S}^5}) \ ,
\label{eq:D7background}
\end{equation} 
where $ds^{2}_{\text{BH}_5}$ is the metric of a hyperbolic black hole \eqref{eq:hypbh} for $ d=4 $ and the internal space is $ X_5 = \text{S}^{5} $ with the metric
\begin{equation}
ds^{2}_{\text{S}^{5}} = d\xi^{2} +\sin^{2}{\xi}\,d\theta^{2}+ \cos^{2}{\xi}\,d\Omega_{3}^{2} \ .
\end{equation}
The D7-brane dual to massive flavor wraps an $ M_3 = \text{S}^{3}\subset \text{S}^{5} $ inside the internal space and is parametrized by \cite{Karch:2005ms,Karch:2006bv}
\begin{equation}
\xi = \xi_{1,0}(r) = \arcsin{\frac{r_\text{q}}{r}}\ , \quad \theta = \text{constant} \ ,
\label{eq:D7n1}
\end{equation}
where $r$ is the Poincar\'e coordinate of AdS.  For $n> 1$, we assume (similarly to \cite{Chalabi:2020tlw}) that the brane still wraps the same internal space, but with a modified embedding $\xi_{n,0}$ that we parametrize in terms of the hyperbolic black hole coordinates as $\xi = \xi_{n,0}(\tau,\zeta,v)$. The resulting D7-brane DBI action \eqref{eq:braneactions} is of the form \eqref{eq:Dpactioncttext} with the Lagrangian
\begin{equation}
\mathcal{L}_n= \zeta^{3}\sinh^{2}{v}\,\cos^{3}{(\xi_{n,0})}\, \sqrt{1+\zeta^{-2}\,(\partial_v\xi_{n,0})^{2} + f_n(\zeta)\,(\partial_\zeta \xi_{n,0})^{2}+f_n(\zeta)^{-1}\,(\partial_{\tau}\xi_{n,0})^{2}}
\label{eq:D7lagrangiantext}
\end{equation}
and the constant \eqref{eq:NDp} being
\begin{equation}
\mathcal{N}_{\text{D7}} = \frac{1}{2\pi^{3}}\,g_{\text{YM}}^{2}N^{2},
\end{equation}
where we used $\vol{(X_5)} = \vol{(\text{S}^5)} = \pi^3 $ and $\vol{(M_3)} = \vol{(\text{S}^3)} = 2\pi^2 $. At $n=1$, the equations of motion for the embedding have the solution
\begin{equation}
 \xi_{1,0}(\tau,\zeta,v) = \arcsin{\biggl( \frac{\R }{\zeta\cosh{v} + \sqrt{\zeta^{2}-1}\,\cos{\tau}}\biggr)}\ ,
 \label{eq:D7xi10}
\end{equation}
which is the embedding \eqref{eq:D7n1} in non-backreacted $\text{AdS}_5\times \text{S}^5$ written in the Rindler coordinates \eqref{eq:RindlertoPoincare}. For $n > 1$, the embedding $\xi_{n,0}(\tau,\zeta,v)$ in the background \eqref{eq:D7background} can in principle be determined from the equations of motion, but we will not need to know its value for all values of $r$, and an asymptotic solution of the equation close to relevant values of the radial direction suffices.

\paragraph{Computation of the entanglement entropy.} Given the Lagrangian \eqref{eq:D7lagrangiantext}, we can then compute all terms in the entropy \eqref{eq:entropyfromprobes} which is done in detail in Appendix \ref{subapp:probeD7} where also the counterterms that renormalize the action \eqref{eq:D7lagrangiantext} are given. 

We will start with the boundary terms localized on the RT surface $\zeta = 1$. For the first term, we get
\begin{align}
    \mathcal{B}^{(1)}_{\text{RT},-} &=\frac{\pi}{3}\,\mathcal{N}_{\text{D7}}\,\biggl[\frac{R^2}{\epsilon^2}-(1+2\R^2)\log{\frac{2R}{\epsilon}}+2\R^2+\mathcal{O}(\epsilon^2)\biggr]\label{eq:BIR1smallmass}\\
    \mathcal{B}^{(1)}_{\text{RT},+} &=\frac{\pi}{3}\,\mathcal{N}_{\text{D7}}\,\biggl[\frac{R^{2}}{\epsilon^{2}}-(1+2\R^2)\log{\frac{2R}{\epsilon}}+2\R^2-3\R \sqrt{\R^2-1}\biggr.\nonumber\\
&\qquad\qquad\qquad\qquad\qquad\qquad\qquad\qquad\qquad\qquad\biggl.+(1+2\R^2)\arcosh{\R} + \mathcal{O}(\epsilon^{2})\biggr]\ ,
\label{eq:BIR1finalD7}
\end{align}
where $\mathcal{B}^{(1)}_{\text{RT},-}$ is the expression for $\R  < 1$ and $\mathcal{B}^{(1)}_{\text{RT},+}$ is for $\R  > 1$ (following notation conventions of Section \ref{subsec:perturbativeRT}). The second boundary term $\mathcal{B}_{\text{RT}}^{(2)}$ is given by
\begin{equation}
\mathcal{B}_{\text{RT}}^{(2)} = -\frac{\pi}{3}\,\mathcal{N}_{\text{D7}}\,
\begin{dcases}
\frac{1}{3}\R^2-\frac{2}{15}\R^4\ ,\quad &\R < 1 \\
\frac{1}{3}\R^2-\frac{2}{15}\R^4+ \frac{2}{15}\frac{\sqrt{\R^2-1}}{\R }\,\left(1-2\R^2+\R^4\right)\ ,\quad &\R \geq 1 
\end{dcases}
\label{eq:BIR2final} \ ,
\end{equation}
and it is non-zero for $\R  >0$ as mentioned above. For the UV boundary terms, we get
\begin{equation}
    \mathcal{B}_{\zeta=\infty}^{(1)} + \mathcal{B}_{\zeta=\infty}^{(2)}  = -\frac{\pi}{12}\,\mathcal{N}_{\text{D7}}\,\biggl(\frac{R^2}{\epsilon^2}-\log{\frac{2R}{\epsilon}}\biggr)\ ,
    \label{eq:BUVD7text}
\end{equation}
where the dependence on the $n$-derivative of the normalizable mode of $\xi_n$ (appearing in the $\zeta\rightarrow \infty$ expansion) is canceled between the two terms (see Appendix \ref{subapp:probeD7} for details). The computation of the bulk integral $\mathcal{I}$ is tedious and details are relegated to Appendix \ref{subapp:probeD7}. The results for $\R <1$ and $\R >1$ are given by
\begin{align}
    \mathcal{I}_- &= \frac{\pi}{3}\,\mathcal{N}_{\text{D7}}\, \R^2\biggl(\log\frac{2R}{\epsilon}-\frac{13}{6}+\frac{1}{10}\,\R^2\biggr)\\
    \mathcal{I}_+ &= 
    \frac{\pi}{3}\,\mathcal{N}_{\text{D7}}\,  \biggl[\R^2\log{\frac{2R}{\epsilon}}-\frac{4}{15\R }\sqrt{\R^2-1}\bigg(1-\frac{97}{16}\R^2+\frac{3}{8}\R^4\bigg)-\frac{13}{6}\R^2+\frac{1}{10}\R^4\nonumber \\
     &\quad -\frac{1}{4}(1+\R^2)\arcosh{\R}\biggr]\ .
     \label{eq:IbulkD7text}
\end{align}
We can now compute the flavor correction to the entanglement entropy. As explained above, we will not include $\mathcal{B}_{\text{RT}}^{(2)}$ to the entropy and use the formula $S^{(1)}(R) = \mathcal{I} + \mathcal{B}_{\text{RT}}^{(1)} + \mathcal{B}_{\zeta=\infty}^{(1)} + \mathcal{B}_{\zeta=\infty}^{(2)}$. The result is
\begin{align}
S_-^{(1)}(R) &= \frac{g_{\text{YM}}^{2}\Nc^{2}}{8\pi^{2}}\,\biggl[\frac{R^{2}}{\epsilon^{2}}-\biggl(1+\frac{4}{3}\,\R^2\biggr)\log{\frac{2R}{\epsilon}}-\frac{2}{9}\,\R^2+\frac{2}{15}\,\R^4\biggr]\label{eq:SD7S5minusprobe}\\
S_+^{(1)}(R) &= \frac{g_{\text{YM}}^{2}\Nc^{2}}{8\pi^{2}}\,\biggl[\frac{R^{2}}{\epsilon^{2}}-\biggl(1+\frac{4}{3}\,\R^2\biggr)\log{\frac{2R}{\epsilon}}+\frac{2}{9}\,\R^2+\frac{2}{15}\,\R^4\biggr.\label{eq:SD7S5plusprobe}\\
&\biggl.\qquad\qquad\quad-\frac{16}{45\R}\sqrt{\R^2-1}\,\biggl(1+\frac{83}{16}\,\R^2+\frac{3}{8}\,\R^4\biggr) + \biggl(1+\frac{4}{3}\,\R^2\biggr)\arcosh{\R}\biggr]\nonumber \ .
\end{align}
The difference between \eqref{eq:SD7S5minusprobe}--\eqref{eq:SD7S5plusprobe} (call it $S^{(1)}_{\text{probe},\pm}(R)$) obtained using the probe brane method and the result \eqref{eq:SD7S5minus}--\eqref{eq:SD7S5plus} of the backreacted calculation (call it $S^{(1)}_{\text{backreacted},\pm}(R)$) is given by
\begin{equation}
    S^{(1)}_{\text{probe},\pm}(R)-S^{(1)}_{\text{backreacted},\pm}(R)= \frac{g_{\text{YM}}^2N^2}{8\pi^2}\biggl(\frac{R^2}{2\epsilon^2}+\frac{1}{4}-\frac{16}{9}\,\R^2\biggr)\ .
    \label{eq:entropydifference}
\end{equation}
We see that the entanglement entropies do not directly match, but the difference is only by non-universal terms \cite{Solodukhin:2008dh,Grover:2011fa,Casini:2015woa} which are quadratic in the radius $R$. In particular, there exists a redefinition of the cut-off $\epsilon$ and the radius $R$ such that the redefined entanglement entropy matches. Denoting by $\epsilon,R$ the cut-off and radius in the results of the backreacted method, the redefinition of the probe cut-off and radius that matches the entanglement entropies is given by
\begin{align}
    \epsilon &\rightarrow  \sqrt{2}\,\epsilon\,\biggl(1-\frac{16}{9}\,r_{\text{q}}^2\epsilon^2+\mathcal{O}(\epsilon^3)\biggr) \label{eq:epsredef}\\
    R&\rightarrow  R\,\biggl\{1-\frac{\epsilon^2}{4R^2}\biggl[1+\biggl(2+\frac{8}{3}\,\mathcal{R}^2\biggr)\log{2}\biggr]+\mathcal{O}(\epsilon^3)\biggr\} \ .
    \label{eq:D7redefinition}
\end{align}
In Section~\ref{subsec:backreactedCFfunctions}, we define an A-function, which is only sensitive to the universal part of the entanglement entropy, and we show the difference \eqref{eq:entropydifference} cancels in the A-functions. Therefore the A-functions obtained from a probe computation and the application of RT prescription in the backreacted geometries match.

It is worth noting that the redefinition of the cut-off parameter $\epsilon$ in (\ref{eq:epsredef}) remains independent of $R$, as underscored in \cite{Liu:2012eea, Liu:2013una}. Conversely, the redefinition of $R$ in (\ref{eq:D7redefinition}) is permitted to be contingent on $\epsilon$, a point elaborated in \cite{Casini:2015woa}. This allowance is intuitively clear, especially in lattice formulations, where the UV cut-off $\epsilon$ is interpreted as the lattice spacing. Consequently, any length possesses an inherent uncertainty, stemming from the finite lattice spacing.

Lastly, we compare our results with previous calculations in the literature. The same flavor contribution to the entanglement entropy was computed already in \cite{Karch:2014ufa}, for small masses $\mathcal{R} < 1$, and in \cite{Jones:2015twa}, for all masses $\mathcal{R}>0$, by utilizing the graviton propagator method of \cite{Chang:2013mca}. We have checked that our result \eqref{eq:SD7S5minusprobe}--\eqref{eq:SD7S5plusprobe} with the boundary term omitted matches with both of these calculations up to non-universal terms (similarly to \eqref{eq:entropydifference}) implying that there also exists a redefinition of $(\epsilon,R)$ matching the results exactly. Thus these independent graviton propagator calculations support the omission of the boundary term $\mathcal{B}_{\text{RT}}^{(2)}$. In addition, \cite{Karch:2014ufa} performs the exact same calculation we do using the probe brane method for $\mathcal{R} < 1$, but they do not include $\mathcal{B}_{\text{RT}}^{(2)}$, however, their answer differs from both our result above and the result of the graviton propagator method by an $\mathcal{R}^4$-term (and by other non-universal terms). We have traced this difference to come from the calculation of the $\mathcal{I}$ integral: our evaluation \eqref{eq:IbulkD7text} of the integral $\mathcal{I}_-$ differs from the one in \cite{Karch:2014ufa} by the $\mathcal{R}^4$-term and a non-universal $\mathcal{R}^2$-term.

\subsection{Probe D6-brane on \texorpdfstring{$\text{AdS}_4\times \mathbb{C}\text{P}^3$}{}}\label{subsec:probeD6}

The non-backreacted background $g_{1,0}$ in this case is the ABJM $\text{AdS}_4\times \mathbb{C}\text{P}^3 $ geometry \eqref{eq:ABJMparameters}. Hence the geometry $g_{n,0}$ is given by
\begin{equation}
ds^{2} = \ell^{2}(ds^{2}_{\text{BH}_4} + ds^{2}_{X_6}) \ ,
\label{eq:D6branebackground}
\end{equation} 
where $ds^{2}_{\text{BH}_4}$ is the hyperbolic black hole metric \eqref{eq:hypbh} for $d = 3$ and the metric on the internal space is four times the Fubini--Study metric $ds^{2}_{X_6} = 4 ds^2_{\mathbb{C}\text{P}^3}$ (see Appendix \ref{app:ABJMdetails} for details). The background \eqref{eq:D6branebackground} is supported by the same matter fields as the solution in \eqref{eq:ABJMparameters}, but in $F_4$ the volume form of $\text{AdS}_4$ is replaced by the volume form of the hyperbolic black hole. The D6-brane action is given by \eqref{eq:braneactions} and is a sum of a DBI and a WZ part denoted by $I_{\text{DBI}}$ and $I_{\text{WZ}}$ respectively.

When $n = 1$ and the background metric $g_{1,0}$ is the ABJM solution, it is known that the D6-brane dual to massive flavor wraps the subspace $M_3 = \mathbb{R}\text{P}^3$ of the internal space given by \cite{Hikida:2009tp,Jokela:2021knd}
\begin{equation}
\xi = \xi_{1,0}(r) =\arccos{\frac{r_\text{q}}{r}}\ ,\quad \theta_1 = \theta_2\ , \quad \phi_1 = -\phi_2\ , 
\label{eq:braneembedding}
\end{equation}
where $r$ is the Poincar\'e coordinate of AdS. For $n> 1$, we assume (similarly to \cite{Chalabi:2020tlw}) that the brane still wraps the same internal space, but with a modified embedding $\xi_{n,0}$ that we parametrize in terms of the hyperbolic black hole coordinates as $\xi = \xi_{n,0}(\tau,\zeta,v)$. The resulting DBI action is
\begin{eqnarray}
I_{\text{DBI}} & = &\, \mathcal{N}_{\text{D6}}\int_{0}^{2\pi} d\tau \int_{\zeta_n}^{\Lambda_{\zeta}} d\zeta \int_{v_0}^{\Lambda_{v}} dv\,\zeta^{2}\, \sinh{v}\,\sin{(\xi_{n,0})} \nonumber\\
& & \times\sqrt{1+\zeta^{-2}\,(\partial_v\xi_{n,0})^{2} + f_n(\zeta)\,(\partial_\zeta \xi_{n,0})^{2}+f_n(\zeta)^{-1}\,(\partial_{\tau}\xi_{n,0})^{2}} \ ,
\label{eq:D6DBIapp}
\end{eqnarray}
where we have integrated over the internal space $ M_3 $ and the $ \varphi $-circle of the hyperbolic black hole. The coefficient $ \mathcal{N}_{\text{D6}} $ is defined in \eqref{eq:NDp} for which we have used that $ds^2_{M_3} = 4ds^2_{\mathbb{R}\text{P}^3} $ so that $\vol{(M_3)} = 2^3\vol{(\mathbb{R}\text{P}^3)} = 8\pi^2$.

The 8-form flux in the background \eqref{eq:D6branebackground} is given by \cite{Hikida:2009tp}
\begin{equation}
F_8 = *F_2 = k\ell^{6}g_{\text{s}}\ell_{\text{s}}\,J\wedge J \wedge \epsilon_{\text{BH}_4} \ ,
\label{eq:F8text}
\end{equation}
where $ \epsilon_{\text{BH}_4} $ is the volume form of the hyperbolic black hole \eqref{eq:hypbh} with unit curvature radius,\footnote{As we are working with the background \eqref{eq:D6branebackground}, the volume form of the hyperbolic black hole appears instead of the volume form of $ \text{AdS}_4 $ as in \cite{Hikida:2009tp}.} 
\begin{equation}
\epsilon_{\text{BH}_4} = \zeta^{2}\,d\zeta\wedge \epsilon_{S^{1}\times\mathbb{H}^{2}}, \quad \epsilon_{S^{1}\times\mathbb{H}^{2}} = \sinh{v}\,d\tau\wedge dv\wedge d\varphi \ .
\end{equation}
To compute the WZ action, we need to find a regular $ C_7 $ such that $ dC_7 = F_8 $ and compute its pullback to the worldvolume of the brane. The regularity condition is implemented at the tip $\zeta = \zeta_n$ of the Euclidean hyperbolic black hole cigar, which is the origin of the $(u,\tau)$ polar coordinate system with $u$ defined in \eqref{eq:ucoordinate}. We will follow the analysis of \cite{Jokela:2021knd} which derived a $C_7$ which is regular at the center of spherically sliced $\text{AdS}_4$.

Let us denote by $M_4$ the subspace spanned by $\{\tau, \zeta, v, \xi\}$ with the conditions $\theta_1 = \theta_2,\phi_1 = -\phi_2$, and let us denote by $M_3$ the subspace that the brane wraps, described by $\xi=\xi_{n,0}(\tau,\zeta,v)$. In the subspace $M_4$, the 8-form flux \eqref{eq:F8text} becomes
\begin{equation}
F_8 = \frac{1}{2}k\ell^{6}g_{\text{s}}\ell_{\text{s}}\,\sin{\xi}\,\cos{\xi}\,d\xi\wedge \epsilon_{M_3}\wedge \epsilon_{\text{BH}_4} \ ,
\end{equation}
where $\epsilon_{M_3}$ is the volume form of $M_3$. We will look for a $C_7$ of the form
\begin{equation}
C_7 = \frac{1}{2}k\ell^{6}g_{\text{s}}\ell_{\text{s}}\left[ A(\zeta,\xi)\,d\xi\wedge \epsilon_{M_3}\wedge \epsilon_{S^{1}\times\mathbb{H}^{2}} + B(\zeta,\xi)\, d\zeta\wedge \epsilon_{M_3}\wedge \epsilon_{S^{1}\times\mathbb{H}^{2}}\right] \ ,
\end{equation}
for which the condition $F_8  = dC_7$ amounts to
\begin{equation}
\partial_\zeta A-\partial_\xi B = \zeta^{2}\,\sin{\xi}\,\cos{\xi}\ .
\end{equation}
Regularity at $\zeta = \zeta_n$ then amounts to (see also \cite{Jokela:2021knd})
\begin{equation}
A(\zeta_n,\xi) = B(\zeta,0) = B(\zeta,\pi) = 0 \ .
\end{equation}
The general solution is
\begin{equation}
A(\zeta,\xi) = -\partial_\xi U(\zeta,\xi)\ , \quad B(\zeta,\xi) = -\frac{1}{2}\,\zeta^{2}\sin^{2}{\xi} - \partial_\zeta U(\zeta,\xi) \ ,
\end{equation}
where $ U $ is a regular gauge transformation which we set to zero. Hence in the subspace $M_4$ we have
\begin{equation}
C_7 = -\frac{1}{4}k\ell^{6}g_{\text{s}}\ell_{\text{s}}\,\zeta^{2}\,\sin^{2}{\xi}\,d\zeta\wedge \epsilon_{M_3}\wedge \epsilon_{S^{1}\times\mathbb{H}^{2}}\ ,
\end{equation}
so that its pullback to the brane $\xi = \xi_{n,0}(\tau,\zeta,v)$ gives
\begin{equation}
\widehat{C}_7 = -\frac{1}{4}k\ell^{6}g_{\text{s}}\ell_{\text{s}}\,\zeta^{2}\sin^{2}{(\xi_{n,0})}\,\sinh{v}\,d\zeta\wedge \epsilon_{M_3}\wedge d\tau\wedge dv\wedge d\varphi \ .
\end{equation}
Hence the WZ action becomes
\begin{equation}
I_{\text{WZ}} = -T_{\text{D6}}\int \widehat{C}_7 =\, \frac{1}{2}\,\mathcal{N}_{\text{D6}} \int_{0}^{2\pi} d\tau  \int_{\zeta_n}^{\Lambda_{\zeta}} d\zeta\int_{v_0}^{\Lambda_v} dv\,\zeta^{2}\,\sinh{v}\,\sin^{2}{(\xi_{n,0})} \ ,
\label{eq:D6WZapp}
\end{equation}
where we used \eqref{eq:ABJMparameters}, which implies
\begin{equation}
T_{\text{D6}}\,k\ell^{6}g_{\text{s}}\ell_{\text{s}}\vol{(M_3)}\vol{(S^{1})} = \frac{k}{(2\pi)^{5}}\biggl(\frac{\ell}{\ell_{\text{s}}}\biggr)^{6}=\frac{1}{2}N\sqrt{2\lambda} = 2\,\mathcal{N}_{\text{D6}} \ ,
\end{equation}
and we used $\vol{(M_3)} = 8\pi^2$. As a result, the D6-brane action takes the form \eqref{eq:Dpactiontext2} with the Lagrangian
\begin{equation}
\mathcal{L}_n = \zeta^{2}\sinh{v}\,\sin{(\xi_{n,0})}\,\biggl( \sqrt{1+\zeta^{-2}\,(\partial_v\xi_{n,0})^{2} + f_n(\zeta)\,(\partial_\zeta \xi_{n,0})^{2}+f_n(\zeta)^{-1}\,(\partial_{\tau}\xi_{n,0})^{2}} + \frac{1}{2}\sin{(\xi_{n,0})} \biggr) \ .
\label{eq:D6lagrangiantext}
\end{equation}
One can check that at $ n=1 $, the equations of motion for the embedding coming from this Lagrangian have the solution
\begin{equation}
\xi_{1,0}(\tau,\zeta,v) = \arccos{\biggl( \frac{\R }{\zeta\cosh{v} + \sqrt{\zeta^{2}-1}\,\cos{\tau}}\biggr)}\ ,
\label{eq:D6xi10}
\end{equation}
which is the known massive embedding \eqref{eq:braneembedding} written in the Rindler coordinates \eqref{eq:RindlertoPoincare}.

\paragraph{Computation of the entanglement entropy.}

Given the Lagrangian \eqref{eq:D6lagrangiantext}, we can then compute all the terms in the entropy \eqref{eq:entropyfromprobes}. This is done in detail in Appendix \ref{subapp:probeD6}, where also the counterterms that renormalize the action \eqref{eq:D6lagrangiantext} are given.

Let us start with the IR boundary terms. For the first RT term, we get
\begin{equation}
\mathcal{B}^{(1)}_{\text{RT}} =\frac{3\pi}{2}\,\mathcal{N}_{\text{D6}}\,\times
\begin{dcases}
\frac{R}{\epsilon}-1-\frac{1}{3}\R^2\ , \quad &\R  < 1 \\
\frac{R}{\epsilon}-\frac{4}{3}\R \ , \quad &\R  \geq 1 \ 
\end{dcases},
\label{eq:BIR1finalD6text}
\end{equation}
while the second RT term becomes
\begin{equation}
\mathcal{B}_{\text{RT}}^{(2)} = -\frac{\pi}{6}\,\mathcal{N}_{\text{D6}}\,
\begin{dcases}
\R^2 \ ,\quad &\R < 1 \\
\frac{1}{\R } \ ,\quad &\R \geq 1 
\end{dcases} \ ,
\label{eq:BIR2finalD6text}
\end{equation}
which is non-zero for massive flavor as mentioned earlier. For the UV boundary terms, we get
\begin{equation}
\mathcal{B}_{\zeta=\infty}^{(1)}+\mathcal{B}_{\zeta=\infty}^{(2)} = -\frac{\pi}{2}\,\mathcal{N}_{\text{D6}}\,\biggl(\frac{R}{\epsilon} - 1\biggr) \ ,
\label{eq:BUVtext}
\end{equation}
where, similarly to the D7-brane case, the dependence on the $n$-derivative of the normalizable mode of $\xi_n$ is canceled between the two terms. The computation of $\mathcal{I}$ is again involved and the details of the computation can be found in Appendix \ref{subapp:probeD6}. The result is
\begin{equation}
\mathcal{I} = \pi\,\mathcal{N}_{\text{D6}}\,\times
\begin{dcases}
\frac{1}{6}\,\R^2\ , \quad &\R  < 1 \\
\R -\frac{1}{3\R}-\frac{1}{2}\ , \quad &\R  \geq 1 
\end{dcases}\ .
\label{eq:IfinalD6text}
\end{equation}
We can now compute the flavor correction to the entanglement entropy. As explained above, we will not include $\mathcal{B}_{\text{RT}}^{(2)}$ to $S^{(1)}(R)$, which gives the result
\begin{equation}
S^{(1)}(R) = \mathcal{I} + \mathcal{B}_{\text{RT}}^{(1)} + \mathcal{B}_{\zeta=\infty}^{(1)} + \mathcal{B}_{\zeta=\infty}^{(2)} = \frac{\pi}{4}N\sqrt{2\lambda}\times
\begin{dcases}
\frac{R}{\epsilon}-1-\frac{1}{3}\,\R^2\ , \quad &\R  < 1 \\
\frac{R}{\epsilon}-\R -\frac{1}{3\R}\ , \quad &\R  \geq 1 
\end{dcases} \ ,
\label{eq:S1D6probe}
\end{equation}
where finite terms match exactly with the finite terms of the entropy \eqref{eq:S1backreactedABJM} obtained from the backreacted calculation. The coefficient of the UV divergence in \eqref{eq:S1D6probe} differs from the coefficient in \eqref{eq:S1backreactedABJM} by a factor of $5\slash 4$. This mismatch is not a problem since the coefficient is regularization dependent and non-universal. The redefinition
\begin{equation}
    \epsilon \rightarrow \frac{4}{5}\,\epsilon\ , \quad R\rightarrow R\,,
\end{equation}
in the entanglement entropy obtained using the probe brane method gives the backreacted result. In Section~\ref{subsec:backreactedCFfunctions}, we define an F-function which is only sensitive to the finite universal part of the entanglement entropy so it matches in the two calculations.

\section{Renormalization group monotones for flavor}\label{subsec:backreactedCFfunctions}

In this section, we center our attention to the construction of renormalization group monotones from the holographic entanglement entropy. We will focus on the Liu--Mezei functions \cite{Liu:2012eea,Liu:2013una} which are obtained by appropriately differentiating entanglement entropy of a spherical subregion with respect to its radius. We will first review the construction of the Liu--Mezei functions and then study their monotonicity in our holographic field theories at leading order in the number of massive flavor degrees of freedom.

\subsection{Liu--Mezei functions}

Consider a regularized quantum field theory with a UV cut-off length $ \epsilon $. In the continuum limit $ \epsilon \rightarrow 0 $ there is an infinite number of degrees of freedom that are entangled across the entangling surface so that the entanglement entropy of any subregion is divergent in this limit. Assuming entanglement comes mostly from EPR pairs that are locally entangled across a smooth entangling surface, the divergence structure can be classified by local diffeomorphism invariant functionals supported on the surface \cite{Grover:2011fa,Liu:2013una,Liu:2012eea}. For a spherical subregion of radius $ R $, the resulting divergence structure is given by
\begin{equation}
	S(R) = 
	\begin{dcases}
		s_1 \frac{R}{\epsilon} + s_2+ \ldots\ , \quad &d = 3\\
		s_1 \frac{R^{2}}{\epsilon^{2}}  + s_2\log{\frac{2R}{\epsilon}} + s_3 + \ldots\ , \quad &d = 4
	\end{dcases} \ ,
	\label{SRexpansions}
\end{equation}
where ellipsis denote terms that vanish in the continuum limit $ \epsilon \rightarrow 0 $ and $s_1,s_2,s_3$ are unknown coefficients \cite{Grover:2011fa,Liu:2012eea,Liu:2013una}. For our theories with massive flavor, the theory has an additional length scale $ 1\slash r_\text{q} $ determined by the flavor mass $ m \propto r_\text{q} $. Hence in addition to $R\slash \epsilon$, there are two possible dimensionless parameters, the dimensionless radius $ \R = r_\text{q}R $ and $ r_\text{q}\epsilon $, that can appear in the coefficients of $S(R)$. However, $ r_\text{q}\epsilon $ can only appear in the subleading non-divergent terms (ellipsis) in the expansion \eqref{SRexpansions} and does not contribute in the continuum $\epsilon\rightarrow 0$ limit. Thus the parameters $ s_2,s_3 $ in \eqref{SRexpansions} are functions of $ \R $ only while $s_1$ is independent of $\R$.

In our context, a renormalization group monotone is a function of the dimensionless radius $\R $ \eqref{eq:dimensionlessradius} of the disk that monotonically decreases from the central charge of the UV fixed point CFT to the central charge in the IR. The role of such function is to presumably capture the variation of the number of degrees of freedom with the energy scale. In four dimensions (SYM), the central charge refers to the coefficient of the Euler characteristic in the Weyl anomaly (type A anomaly coefficient), while in three dimensions (ABJM), the role of the central charge is played by the free energy on a sphere. Hence we will call a RG monotone interpolating between the two central charges as F- and A-functions in three and four dimensions, respectively.

The first obstacle in constructing a RG monotone from $S(R)$ is the fact that it is UV divergent. However, since the structure of the UV divergences is fixed and depends only on the dimension $d$, they can be removed systematically. In the Liu--Mezei functions, the UV divergent pieces are annihilated by the action of a differential operator $\mathcal{D}_R$ which in three and four dimensions is given by \cite{Liu:2012eea}
\begin{equation}
    \mathcal{D}_R =
    \begin{dcases}
        R\,\partial_R-1\ ,\quad &d = 3\\
        \frac{1}{2}\,R\,\partial_R\,(R\,\partial_R-2) \ ,\quad &d = 4 
    \end{dcases} \ .
\end{equation}
Hence the three-dimensional F-function is given by
\begin{equation}
    \mathcal{F}(\R ) = \lim_{\epsilon\rightarrow 0}(R\,\partial_R-1)\,S(R)\ ,\quad d = 3 \ ,
    \label{Ffunction}
\end{equation}
and the four-dimensional A-function is
\begin{equation}
    \mathcal{A}(\R ) = \lim_{\epsilon\rightarrow 0}\frac{1}{2}\,R\,\partial_R\,(R\,\partial_R-2)\,S(R) \ ,\quad d = 4 \ .
    \label{Afunction}
\end{equation}
The differential operator removes all UV divergent $R\slash \epsilon$ terms in $\mathcal{F}$ so that it is a function of the dimensionless radius \eqref{eq:dimensionlessradius} only. The same is true for the A-function $\mathcal{A}$ if and only if the coefficient of the logarithmic divergence is quadratic in the dimensionless radius
\begin{equation}
    s_2(\R) = d_1 + d_2\mathcal{R}^2\,,
    \label{eq:s2}
\end{equation}
which is true in all our examples. The resulting expressions are
\begin{equation}
    \mathcal{F}(\R ) = (\mathcal{R}\,\partial_{\mathcal{R}}-1)\,s_2(\mathcal{R})\ ,\quad \mathcal{A}(\R ) =(\mathcal{R}\,\partial_{\mathcal{R}}-1)\,s_2(\mathcal{R})+\frac{1}{2}\,\mathcal{R}\,\partial_{\mathcal{R}}(\mathcal{R}\,\partial_{\mathcal{R}}-2)\,s_3(\mathcal{R})\,,
\end{equation}
where $s_2$ is assumed to be quadratic in $\R$ in $\mathcal{A}$. In our case, the theory with massless flavor $\R  = 0$ is a CFT (the UV fixed point) while the theory with infinitely massive flavor $\R \rightarrow \infty $ is the same CFT (the IR fixed point), but without flavor degrees of freedom included. 

We emphasize that in the D3-D7 case it makes sense to call the massless UV theory a fixed point only when working at linear order in an expansion in $\Nf\slash N $, because for finite $\Nf$ the theory contains a Landau pole at high energies. Holographically, as shown at the end of Section \ref{subsec:backreactedsols}, the backreacted D3-D7 geometries are indeed AdS everywhere (when $r_{\text{q}} = 0$) at linear order in $\Nf$.

In three dimensions we denote the sphere free energies of these CFTs by $F_{\text{UV}}$ and $F_{\text{IR}}$, respectively, and the coefficient of the Euler character by $A_{\text{UV}}$ and $A_{\text{IR}}$. The fact that they reduce to central charges at UV fixed points amounts to the requirement
\begin{equation}
    \mathcal{F}(\R ) =
    \begin{cases}
        F_{\text{UV}}\ ,\quad &\R \rightarrow 0\\
        F_{\text{IR}}\ ,\quad &\R \rightarrow \infty
    \end{cases}\ ,\quad
    \mathcal{A}(\R ) =
    \begin{cases}
        A_{\text{UV}}\ ,\quad &\R \rightarrow 0\\
        A_{\text{IR}}\ ,\quad &\R \rightarrow \infty
    \end{cases}\ ,
    \label{eq:UVIRlimits}
\end{equation}
and monotonicity of the F- and A-functions amounts to
\begin{equation}
    \partial_{\R }\mathcal{F}(\R ) < 0\ ,\quad \partial_{\R }\mathcal{A}(\R ) < 0\ ,\quad 0 < \R  < \infty\ .
    \label{eq:AFmonotonicity}
\end{equation}
We are interested in the limit $\Nf\rightarrow 0$. The UV central charges have an expansion in flavor
\begin{equation}
    F_{\text{UV}} = F_{\text{UV}}^{(0)}+ \Nf\,F_{\text{UV}}^{(1)}+\mathcal{O}(\Nf^2)\ ,
\end{equation}
where the zeroth order term coincides with the central charge of the IR theory $ F_{\text{UV}}^{(0)} = F_{\text{IR}} $ and $ A_{\text{UV}}^{(0)} = A_{\text{IR}} $. In the limit $\Nf \rightarrow 0$, the functions have the expansions
\begin{equation}
    \mathcal{F}(\R ) = \mathcal{F}^{(0)}(\R ) + \Nf\,\mathcal{F}^{(1)}(\R ) +\mathcal{O}(\Nf^2) \ ,\quad \mathcal{A}(\R ) =\mathcal{A}^{(0)}(\R ) + \Nf\,\mathcal{A}^{(1)}(\R ) +\mathcal{O}(\Nf^2)\ ,
\end{equation}
where the zeroth order terms coincide with the central charges of the unflavored CFTs
\begin{equation}
    \mathcal{F}^{(0)}(\R ) \equiv F_{\text{IR}}\ ,\quad \mathcal{A}^{(0)}(\R )\equiv A_{\text{IR}}\ ,
\end{equation}
and they are independent of $\R $, because all mass is carried by flavor degrees of freedom. In the infinite mass limit it follows by \eqref{eq:UVIRlimits} that
\begin{equation}
    \lim_{\R \rightarrow \infty}\mathcal{F}^{(1)}(\R ) = 0\ ,\quad \lim_{\R \rightarrow \infty}\mathcal{A}^{(1)}(\R ) = 0\ .
\end{equation}
In the no mass limit, we find
\begin{equation}
    F_{\text{UV}} = F_{\text{IR}}+ \Nf\,F_{\text{UV}}^{(1)}+\mathcal{O}(\Nf^2)\ .
\end{equation}
Thus at leading order in flavor, monotonicity \eqref{eq:AFmonotonicity} amounts to
\begin{equation}
    \partial_{\R }\mathcal{F}^{(1)}(\R ) < 0\ ,\quad \partial_{\R }\mathcal{A}^{(1)}(\R ) < 0\ ,\quad 0 < \R  < \infty\ .
    \label{eq:AFmonotonicityleading}
\end{equation}
In the following we will specialize to the different field theory examples and give explicit expressions for their Liu--Mezei functions.

\paragraph{Invariance under redefinitions.} In certain cases, the A-function can be invariant under specific redefinitions of $(\epsilon,R)$ that preserve the divergence structure \eqref{SRexpansions} of the entanglement entropy. One such redefinition in a $d = 4$ theory with a single mass scale $r_{\text{q}}$ is given by
\begin{equation}
    \epsilon \rightarrow a_0\,\epsilon\,\bigl(1 + a_2\,r_{\text{q}}^2\epsilon^2 + \mathcal{O}(\epsilon^3)\bigr),\quad R \rightarrow b_0R\,\biggl(1 +b_2(\mathcal{R})\,\frac{\epsilon^2}{R^2}+\mathcal{O}(\epsilon^3)\biggr)
    \label{eq:Aredefinition}
\end{equation}
where the dimensionless coefficients $a_{0},a_2,b_0$ are independent of $R$ and $b_{2}(\mathcal{R})$ is an arbitrary function of $\mathcal{R}$. Under this substitution, the divergence structure \eqref{SRexpansions} remains the same and the A-function \eqref{Afunction} is invariant $\mathcal{A}(\mathcal{R}) \rightarrow \mathcal{A}(\mathcal{R}) $ if and only if $b_2$ is quadratic in the dimensionless radius
\begin{equation}
    b_2(\mathcal{R}) = c_1+c_2\mathcal{R}^2\,.
    \label{eq:allowedb2}
\end{equation}
It follows that two different entropies can have the same A-function if they differ by a redefinition of this type as is the case for the entropies obtained from the probe and backreacted calculations related by \eqref{eq:epsredef}--\eqref{eq:D7redefinition}.

\begin{figure}[t]
	\begin{subfigure}{0.45\textwidth}
		\centering
		\includegraphics[scale=0.65]{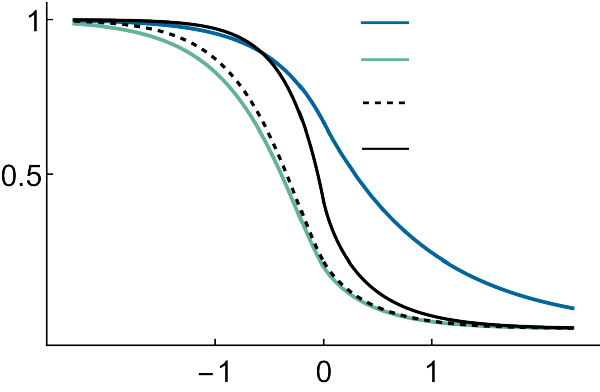}
    \put(-57,112){$\mathcal{F}_\text{D6($\mathbb{C}\text{P}^{3}$)}$}
    \put(-57,100){$\mathcal{A}_\text{D7($\text{S}^{5}$)}$}
    \put(-57,87){$\mathcal{A}_\text{D7($\text{T}^{1,1}$), chiral}$}
    \put(-57,75){$\mathcal{A_\text{D7($\text{T}^{1,1}$), non-chiral}}$}
	\end{subfigure}
 \hspace{0.2cm}
	\begin{subfigure}{0.45\textwidth}
		\centering
		\includegraphics[scale=0.65]{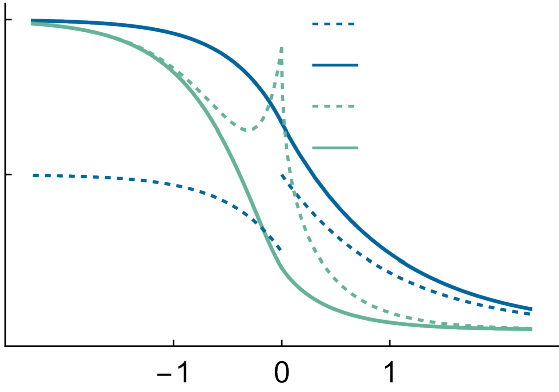}    
         \put(-16,0){$\log{(\R )}$}
        \put(-61,112){$\mathcal{F}_\text{D6($\mathbb{C}\text{P}^{3}$), probe}$}
    \put(-61,100){$\mathcal{F}_\text{D6($\mathbb{C}\text{P}^{3}$), backreacted}$}
    \put(-61,87){$\mathcal{A}_\text{D7($\text{S}^{5}$), probe}$}
    \put(-61,75){$\mathcal{A}_\text{D7($\text{S}^{5}$), backreacted}$}
	\end{subfigure}
	\caption{\textbf{(Left)}: Entanglement A- and F-functions extracted from backreacted geometry as a function of $\log{(\R )}$, normalized and plotted in a  logarithmic scale on the $\R  $-axis. Notice that the concavity changes at around the location of the cavity  \cite{Balasubramanian:2018qqx}. \textbf{(Right)}: Entanglement A- and F-functions extracted from backreacted computation compared with those extracted from probe computation including the $\mathcal{B}_{\text{RT}}^{(2)}$ terms. Notice that if the $\mathcal{B}_{\text{RT}}^{(2)}$ contribution is included then functions are not monotonic nor continuous, while a vanishing value of this term yields full matching between the A- and F-functions of probe and backreacted computations. This further demonstrates why we require the omission of the $\mathcal{B}_{\text{RT}}^{(2)}$ term.}
 \label{fig:a-f-functions}
\end{figure}
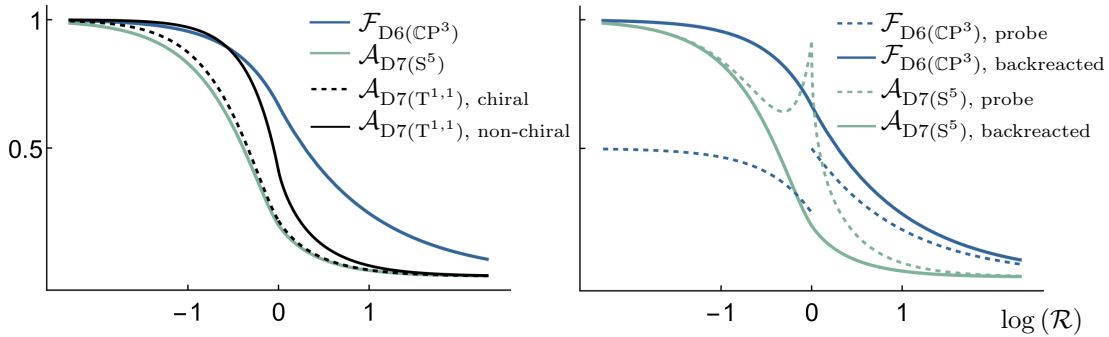

In Fig.~\ref{fig:a-f-functions} we depict the entanglement A- and F-functions (or flavor-correction thereof) for all the cases at once. They possess the expected behavior, that is being continuous, stationary at fixed points, and monotonically decreasing from UV to IR. On the contrary, did we include the boundary term $\mathcal{B}_{\text{RT}}^{(2)}$, the functions cease to be monotonic (even continuous) and moreover their derivatives are predominantly non-vanishing at the fixed points; see Fig.~\ref{fig:a-f-functions} (right).

\subsection{Smeared D7-branes on the conifold}
First, let us quote the results in the backreacted D3-D7 case in the conifold.
From \eqref{eq:S0conifold}, we get for the leading term
\begin{equation}
\mathcal{A}^{(0)}(\R ) = \frac{27}{16}\Nc^{2} \ ,
\end{equation}
which is the central charge of the unflavored KW theory.

\paragraph*{Chiral profile.}

The subleading corrections to the A-function obtained from \eqref{eq:S1conifoldchiral1} and \eqref{eq:S1conifoldchiral2} are given by
\begin{align}
\mathcal{A}^{(1)}_-(\R ) &= \frac{81g_{\text{YM}}^{2}\Nc^{2}}{256\pi^{2}}\,\biggl[1+\frac{9\pi}{64}\,\R^3\,\biggl(1+4\log{\frac{\R }{2}}\biggr)\biggr]\\
\mathcal{A}^{(1)}_+(\R ) &=\frac{81g_{\text{YM}}^{2}\Nc^{2}}{256\pi^{2}}\,\biggl[1-\frac{9}{14\R}\sqrt{\R^2-1}\,\biggl(1-\frac{3}{2}\,\R^2\biggr)+\frac{1}{7}\,{}_3F_2\left(\begin{matrix} -\frac{1}{2}, \frac{3}{2}, \frac{3}{2}\\\frac{5}{2}, \frac{5}{2}\end{matrix};\frac{1}{\R^2}\right)\biggr.\nonumber\\
&\biggl.\qquad\qquad\qquad\;\;-\frac{9}{7}\,\R^2\,{}_3F_2\left(\begin{matrix} \frac{1}{2}, \frac{1}{2}, \frac{1}{2}\\\frac{3}{2}, \frac{3}{2}\end{matrix};\frac{1}{\R^2}\right)+\frac{9}{28}\,\R^3\arcsin{\R^{-1}}\biggr]\ .
\end{align}
where the subscript minus (plus) refers to $\R  < 1$ ($\R  > 1$). We get the large-$\mathcal{R}$ expansion
\begin{equation}
    \mathcal{A}^{(1)}(\R ) = \frac{81g_{\text{YM}}^2\Nc^2}{256\pi^2}+\frac{729g_{\text{YM}}^2\Nc^2}{12800\pi^2}\frac{1}{\mathcal{R}^2}+\mathcal{O}(\mathcal{R}^{-4})\label{eq:limR0chiral}
\end{equation}
where the leading term is the contribution of flavor to the IR central charge.

\paragraph*{Non-chiral profile.}

The subleading corrections to the A-function obtained from \eqref{eq:S1conifoldnonchiral1} and \eqref{eq:S1conifoldnonchiral2} are given by
\begin{align}
\mathcal{A}^{(1)}_-(\R ) &= \frac{81g_{\text{YM}}^{2}\Nc^{2}}{256\pi^{2}}\,\biggl(1-\frac{3\pi}{16}\,\R^3\biggr)\\
\mathcal{A}^{(1)}_+(\R ) &=\frac{81g_{\text{YM}}^{2}\Nc^{2}}{256\pi^{2}}\,\biggl[1-\frac{3}{4\R}\sqrt{\R^2-1}\,\biggl(1-\frac{1}{2}\,\R^2\biggr)-\frac{3}{4}\,\R^3\arctan{\left(\R -\sqrt{\R^2-1}\right)}\biggr] \ .
\end{align}
The limit of $\mathcal{A}$ when $\mathcal{R} \to 0$ is taken from the expression outside the cavity ($-$) and matches with \eqref{eq:limR0chiral}:
\begin{equation}
   \mathcal{A}^{(1)}(\R ) = \frac{81g_{\text{YM}}^2\Nc^2}{256\pi^2}+\frac{243g_{\text{YM}}^2\Nc^2}{2560\pi^2}\frac{1}{\mathcal{R}^2} + \mathcal{O}(\mathcal{R}^{-4}) 
  \ . \label{eq:limR0nonchiral}
\end{equation}

\subsection{Smeared D7-branes on the five-sphere}
When the field theory is in flat space we find similar results as for the conifold case at the deep UV modulo an overall constant that signals the different counting for the flavor degrees of freedom at the fixed point.

The A-function of unflavored SYM theory is
\begin{equation}
    \mathcal{A}^{(0)}(\R ) = \Nc^{2} \ .
\end{equation}
which matches with the result in \cite{Ryu:2006bv}. When the energy is lowered, the A-function differs from the conifold case also qualitatively. To this end, the leading flavor correction is
\begin{align}
\mathcal{A}^{(1)}_-(\R ) &=\frac{g_{\text{YM}}^{2}\Nc^{2}}{8\pi^{2}}\biggl(1-\frac{4}{3}\,\R^2+\frac{8}{15}\,\R^4\biggr)\label{eq:CD7S5minus}\\
\mathcal{A}^{(1)}_+(\R ) &= \frac{g_{\text{YM}}^{2}\Nc^{2}}{8\pi^{2}}\biggl[1-\frac{4}{3}\,\R^2+\frac{8}{15}\,\R^4-\frac{8}{15\R}\sqrt{\R^2-1}\,\biggl(1-2\,\R^2+\R^4 \biggr)\biggr] \ .
\label{eq:CD7S5plus}
\end{align}
The correction due to massless flavor is explicitly
\begin{equation}
    A_{\text{UV}}^{(1)} = \frac{g_{\text{YM}}^{2}\Nc^{2}}{8\pi^{2}} \ .
\end{equation}

\subsection{Smeared D6-branes on the complex projective three-space} 
Let us finally present the results in three dimensions for flows between ABJM fixed points. The zeroth order F-function is simply
\begin{equation}
    \mathcal{F}^{(0)}(R) = F_{\text{UV}}^{(0)} = \frac{\pi\sqrt{2}}{3}\,N^{3\slash 2}k^{1\slash 2}\ ,
\end{equation}
which is the free energy of the ABJM CFT without flavor as computed before in the literature \cite{Santamaria:2010dm,Conde:2011sw}. The flavor correction is
\begin{equation}
\mathcal{F}^{(1)}(R) = \frac{\pi}{4}N \sqrt{2\lambda}\times
\begin{dcases}
1-\frac{1}{3}\,\R^2 \ , \quad &\R  <1\\
\frac{2}{3\R} \ , \quad &\R  >1 
\end{dcases} \ .
\label{eq:FABJMbackreacted}
\end{equation}
We see that the flavor correction to the free energy of the UV fixed point is
\begin{equation}
    F_{\text{UV}}^{(1)} =\frac{\pi}{4}N \sqrt{2\lambda} \ ,
\end{equation}
which matches with \cite{Santamaria:2010dm,Conde:2011sw,Jokela:2021knd}.

\section{Discussion}\label{sec:discussion}

In this paper, we used entanglement entropy to study proposals for renormalization group monotones in quantum field theories with massive flavor degrees of freedom. We considered three examples of strongly coupled quantum field theories, namely, flavor deformed $\mathcal{N}=1$ Klebanov--Witten theory and $ \mathcal{N} = 4$ SYM in four dimensions, and ABJM theory in three dimensions. Both of these theories admit holographic dual descriptions in type II supergravity where the flavor degrees of freedom were introduced by flavor D7- and D6-branes, respectively. To account for the backreaction of flavors, we considered the Veneziano limit, $N,\Nf\to \infty$ with $\Nf\slash N $ finite. In this regime, the geometries of the gravity duals were constructed by smearing the flavor sources. To gain better analytic control, we specialized on a few flavor limit and compared the findings with those obtainable from the 't Hooft limit, $N\to\infty$.

We computed the entanglement entropy of a spherical subregion using two complementary methods: one is based on the RT prescription on the smeared backreacted geometry, the other on a probe computation which does not necessitate the backreaction of fundamentals. A study of universal terms in the HEE, captured by the renormalization group monotones, is performed in the limit of small $\Nf$, in which both the probe and backreacted computations match regardless of the flavor mass. We found that, to leading order in $\Nf$, the renormalization group monotones proposed in \cite{Liu:2012eea} are monotonically decreasing in all the cases addressed in our work.

In addition, we computed analytically the leading flavor correction to the entanglement entropy in the Klebanov--Witten theory, which, to best of our knowledge, has not been obtained elsewhere. Here there are two distinct options to incorporate flavor and we find that the corresponding A-functions behave monotonically as expected. The probe brane calculation in the KW theory is left for future work.


We obtained an agreement between the probe and backreacted calculations of the universal parts of flavor entanglement entropy only if we omit a non-zero boundary term $\mathcal{B}_{\text{RT}}^{(2)}$ in the probe calculation. This term was found to be necessary for a similar matching of entanglement entropies in a rather different setup in \cite{Chalabi:2020tlw} which begs the question why in all of our cases it has to be omitted. We have not been able to conclusively answer the question, but we believe the reason is connected to how we compute the bulk integral term $\mathcal{I}$ in the formula \eqref{eq:entropyfromprobes}: we compute it using Poincar\'e coordinates by performing a coordinate transformation while the formula \eqref{eq:entropyfromprobes}, including $\mathcal{B}_{\text{RT}}^{(2)}$, is derived strictly working in hyperbolic coordinates. Preliminary checks suggests that there is a subtlety involving a pole of the integrand of $\mathcal{I}$ which is treated differently in Poincar\'e/hyperbolic coordinates and the difference appears to be $\mathcal{B}_{\text{RT}}^{(2)}$. Therefore if we were able to compute $\mathcal{I}$ directly in hyperbolic coordinates, we surmise that $\mathcal{B}_{\text{RT}}^{(2)}$ should be included in the calculation as in \cite{Chalabi:2020tlw}. Similar discrepancy in computations of the term analogous to $\mathcal{I}$ using Poincar\'e and hyperbolic coordinates was observed by the authors of \cite{Chalabi:2020tlw}.\footnote{Private communication.} We emphasize that this question appears to be of purely technical nature and we hope to address it in future work.

Numerous promising avenues beckon. One trajectory involves exploring brane intersections that give rise to massive flavor deformations within defect conformal field theories (dCFT). Noteworthy candidates for such dCFTs involve the interplay of flavor D3- or D5-branes with (color) D3-branes, wherein partial results for the backreacted geometries have already surfaced (see \cite{Conde:2016hbg,Jokela:2019tsb,Jokela:2021evo,Garbayo:2022pqp}). This avenue holds particular interest because, despite commendable efforts \cite{Chu:2019uoh,Jeong:2022zea,Caceres:2022hei,de-la-Cruz-Moreno:2023mew}, there remains a conspicuous absence of compelling evidence for the existence of a defect c-function \cite{Hoyos:2021vhl}, let alone an entanglement monotone \cite{Hoyos:2020zeg}, in scenarios where Lorentz invariance is broken. Of significance, holographic entanglement entropy has been successfully derived for scenarios involving massless flavor, as demonstrated in \cite{Estes:2014hka,Rodgers:2018mvq}. Exploring analogous calculations in the context of massive deformations within dCFTs promises to unveil novel insights into the interplay between holography and broken symmetries.

Additional fruitful extensions invite for exploration, such as delving into the analysis of entanglement monotones within backgrounds featuring backreacted flavor branes with varying masses \cite{Filev:2014nza}. Furthermore, there is potential for valuable insights by scrutinizing flavored quantum field theories on curved spacetimes. A prototypical example is de Sitter space where the quantity of interest is entanglement entropy across the equator of the spatial $(d-1)$-dimensional sphere. In three dimensions, this entropy has been related to a certain F-function defined using a Euclidean QFT living on the Wick rotated de Sitter space, a $d$-sphere \cite{Ghosh:2018qtg}. Extending backreacted computations of the present paper to flavored theories on de Sitter space would require finding completely new backreacted geometries. The probe brane method is more amenable to a generalization in this direction as has been initiated in \cite{Vaganov:2015vpq,Jokela:2021knd} and we hope to return to this in future work.

\vspace{1.0cm}
\begin{acknowledgments}
We thank Adam Chalabi, Dimitrios Giataganas, Tony Liimatainen, Karapet Mkrtchyan, Carlos N\'u\~nez, Ronnie Rodgers, Miika Sarkkinen, and Javier Subils for useful discussions. N.~J. has been supported in part by Research Council of Finland grants no.~345070 and 354533. J.~K. is supported by the Deutsche Forschungsgemeinschaft (DFG, German Research Foundation) under Germany’s Excellence Strategy through the W\"{u}rzburg-Dresden
Cluster of Excellence on Complexity and Topology in Quantum Matter - ct.qmat (EXC 2147, project-id 390858490), as well as through the German-Israeli Project Cooperation (DIP) grant ‘Holography and the Swampland’. J.~K. thanks the Osk. Huttunen foundation and the Magnus Ehrnrooth foundation for support during earlier stages of this work. H.~R. is supported in part by the Finnish Cultural Foundation. 
\end{acknowledgments}

\appendix

\section{Details of the geometries}\label{app:geometries}
In this section we present the backreacted flavored geometries of relevance to this work expanded to leading order in the $\Nf \rightarrow 0$ limit. We give the details of the metric components and provide a discussion about their regularity by computing curvature invariants.

\subsection{D3-D7 geometries}\label{app:T11S5details}

The D3-D7 backreacted geometries used in this work can be written in the form:
\begin{eqnarray}
ds^2=h^{-1/2}dx^2_{1,3}+h^{1/2}e^{2f} d\varrho^2+h^{1/2}[e^{2g}ds^2_{KE}+e^{2f}(d\tau+\mathcal{A})^2]\ .
\end{eqnarray}
We center our attention in the compactifications in the $\text{T}^{1,1}$ and $\text{S}^5$ manifolds. After backreaction, the corresponding Sasaki-Einstein manifolds are deformed, and the flavor deformation is encoded in a squashing between the K\"ahler-Einstein base $ds^{2}_{KE}$ and the fiber $(d\tau+\mathcal{A})$. This squashing appears when $g\neq f$ in the metric above. These geometries have also a non-constant dilaton $\phi(\varrho)$, where $\varrho$ is the holographic coordinate, which has the UV at $\varrho \rightarrow +\infty$ and the IR at $\varrho \rightarrow -\infty$, and non vanishing fluxes $F_5$ and $F_1$, whose expressions in terms of the metric functions $h, f, g$, and $\phi$ can be found in \cite{Bigazzi:2008zt}. The metric functions are obtained from the solution of the following system of first order BPS equations
\begin{align}
g'(\varrho) &=  e^{2f-2g}\\
 f'(\varrho)  &=  3-2e^{2f-2g}-\frac{Q_\text{f}}{2}p(\varrho)e^\phi\\
\phi'(\varrho)  &=  Q_\text{f} \ p(\varrho)e^\phi\\
h'(\varrho)  &=  -Q_\text{c}e^{-4g} \ ,
\end{align}
where $p(\varrho)$ denotes the flavor profile, which has a non-zero value only outside a cavity with radius $\varrho_\text{q}$. The value of $Q_\text{f}$ is proportional to the number of flavors $\Nf$ and is determined by the quantization condition of the D3-brane charge. Denoting by $\vol{(X_5)}$ the volume of the submanifold ($M_3 \subset X_5$) wrapped by any D7-brane, we have
\begin{eqnarray}
Q_\text{f}=\frac{1}{4}g_{\text{s}} \Nf \frac{\vol{(M_3)}}{\vol{(X_5)}}\quad , \quad Q_\text{c} = \frac{(2\pi)^4 g_{\text{s}} \ell_s^4\Nc}{\vol{(X_5)}}\equiv 4\ell^4 \ .
\end{eqnarray}
For the sphere and conifold: 
$\vol{(M_3(\text{S}^5))}=2\pi^2, \vol{(M_3(\text{T}^{1,1}))}=\frac{16}{9}\pi^2, \vol{(X_5(\text{S}^5))}=\pi^3$ and $\vol{(X_5(\text{T}^{1,1}))}=\frac{16}{27}\pi^3$.
The solution for the $\text{S}^5$ has been obtained elsewhere, and we will use its expanded version for small $\Nf$. We will construct the perturbative solution in $\Nf$ for the $\text{T}^{1,1}$, for two different types of flavor brane embeddings. Accordingly, we will write:
\begin{eqnarray}
f=f_0+\Nf f_1\ , \ \ \ g=g_0+\Nf g_1\ , \ \ \ \phi=\phi_0+\Nf \phi_1\ , \ \ \ h=h_0+\Nf h_1 \ .  
\end{eqnarray}
The solution at order ${\cal{O}}(\Nf^0)$ is insensitive to the flavor, thus it is identical for both geometries, and it is the same inside and outside the cavity, and given by:
\begin{eqnarray}
&f_0=g_0=\log (\ell_s e^\varrho)\ , \ \ \phi_0=0\ , \ \ h_0=h_{0a}+\dfrac{4e^{-4\varrho}g_{\text{\tiny{s}}} \Nc \pi^4}{\vol{(X_5)}} \ .
\end{eqnarray}
The integration constant $h_{0a}$ affects the UV and corresponds to turning on a source term for an irrelevant operator. We will set it to zero ($h_{0a}=0$).

\subsubsection{D3-D7 on the \texorpdfstring{$\text{T}^{1,1}$}{} (chiral profile)}
The $\text{T}^{1,1}$ is described by the metric
\begin{eqnarray}
&ds^2_{KE}=\frac{1}{6}\sum_{i=1}^2[\sin^2\theta_i (d\phi_i)^2+(d\theta_i)^2]\quad, \quad d\tau+\mathcal{A}=\frac{1}{3}[d\psi+\sum_{i=1}^2\cos \theta_i d\phi_i] \ ,
\end{eqnarray}
where $0<\theta_1, \theta_2<\pi, \ \  0<\phi_1, \phi_2<2\pi,\ \  0< \psi< 4\pi$. The geometries inside the cavity (where $\varrho<\varrho_q$) will be denoted with the superscript ($+$). The geometries outside the cavity (where $\varrho>\varrho_q$) will be denoted by ($-$). After integrating the BPS system to order $\Nf$ we get
\begin{equation}
    p^+ =  0\quad, \quad f^+_1  =  g^+_1=0\quad, \quad
    \phi^+  =  0\quad , \quad
    h^+_1  =  \frac{9e^{-4\rho_\text{q}}g_s^2N_c\pi^4\vol{(M_3)}}{490\vol{(X_5)}^2} \ ,
\end{equation}
and
\begin{align}
p^-  =&  \ 1-e^{3(\varrho_\text{\tiny{q}}-\varrho)}[1+3(\varrho-\varrho_\text{q})]  \\
 f^-_1  =& \ \frac{g_{\text{s}} \vol{(M_3)}}{72\vol{(X_5)}}\big[e^{6(\varrho_\text{\tiny{q}}-\varrho)}+1+3(\varrho_\text{q}-\varrho)+e^{3(\varrho_\text{\tiny{q}}-\varrho)}(-2+3(\varrho-\varrho_\text{q}))\big]  \\
 g^-_1  =& \ - \frac{g_{\text{s}} \vol{(M_3)}}{144\vol{(X_5)}}\big[e^{6(\varrho_\text{\tiny{q}}-\varrho)}-5+6(\varrho-\varrho_\text{q})+4e^{3(\varrho_\text{\tiny{q}}-\varrho)}(1+3(\varrho-\varrho_\text{q})) \big]  \\
 \phi^-_1  =& \ \frac{g_{\text{s}} \vol{(M_3)}}{12\vol{(X_5)}}[-2+3(\varrho-\varrho_\text{q})+e^{3(\varrho_\text{\tiny{q}}-\varrho)}(2+3(\varrho-\varrho_\text{q}))]  \\
 h^-_1  =& \frac{2g_{\text{s}}^2\Nc\pi^4\vol{(M_3)}e^{-4\varrho}}{45\vol{(X_5)}^2}\Big[e^{6(\varrho_\text{\tiny{q}}-\varrho)}+\frac{40}{49}e^{3(\varrho_\text{q}-\varrho)}(10 +21(\varrho-\varrho_\text{q})) -\frac{5}{4} (7-12(\varrho-\varrho_\text{q}))\Big]\ .
\end{align}
We outline the procedure used to construct the solution above. The geometry is obtained to ${\cal{O}}(\Nf)$ by integrating the BPS system in both the inside and outside region. In the process we obtain 4 $\times$ 2 integration constants (the factor 2 comes from integrating in both regions): 4 coming from the integration of $g_1$, 2 from the integration of $\phi_1$, and 2 coming from the integration of $h_1$ ($g_1$ and $f_1$ satisfy decoupled second order differential equations, whereas $\phi_1, h_1$ satisfy first order equations. $f_1$ can be solved in terms of $g_1$ or vice versa). By imposing continuity of the metric functions at the cavity (and, automatically, their derivatives), the number of independent constants is halved to 4. Imposing regularity of the internal geometry in the IR forces one of the constants coming from the integration of $g_1$ to vanish. We will impose an extra (arbitrary) normalization requirement given by $e^{g}(\varrho \rightarrow 0)=1$, which to order $\Nf$ reads $e^{g_0}(1+\Nf g_1)|_{\varrho=0}=0$ and fixes the remaining constant coming from the integration of $g_1$ to be zero. We are left with the integration constant coming from the dilaton equation and the one coming from the $h$ equation. The latter can be fixed in the same way we did for the order zero, that is, setting $h(\varrho \rightarrow +\infty)=0$.

The $\varrho$ coordinate is not suitable to determine the entanglement entropy, since it is not the natural holographic coordinate to be associated to the energy scale. For this purpose, we introduce the $r$ coordinate, whose most prominent feature is to render $h=\frac{1}{\ell^4r^4}$ both inside ($+$) and outside ($-$) the cavity. The relation between both coordinates is given by
\begin{align}
r^+  =& \ \frac{\ell_s}{\ell^2}e^\varrho\bigg[1-\Nf
\frac{9g_{\text{s}}e^{4(\varrho-\varrho_\text{\tiny{q}})} \vol{(M_3)}}{7840\vol{(X_5)}} \bigg]\\
r^-  =& \ \frac{\ell_s}{\ell^2}e^\varrho\bigg[1-\Nf
\frac{g_{\text{s}} \vol{(M_3)}}{360\vol{(X_5)}}\bigg(e^{6(\varrho_\text{\tiny{q}}-\varrho)}+\frac{40}{49}e^{3(\varrho_\text{\tiny{q}}-\varrho)}(10+21(\varrho-\varrho_\text{q}))-\frac{5}{4}(7-12(\varrho-\varrho_\text{q}))  \bigg) \bigg] \ .
\end{align}
In terms of the new coordinate, we have, to order $\mathcal{O}(\Nf^0)$
\begin{eqnarray}
f_0=g_0=\log \big(r\ell^2 \big)\quad , \quad \phi_0=0\quad, \quad h_0=\frac{1}{\ell^4 r^4} \ .
\label{eq:geomr}
\end{eqnarray}
Let us denote by $r_\text{q}$ the position of the cavity in this new coordinate. The flavor correction to the geometry then reads, inside the cavity
\begin{equation}
 p^+  =  0\quad, \quad
f_1^+  =  g_1^+=\frac{9g_{\text{s}} \vol{(M_3)}}{7840\vol{(X_5)}}\frac{r^4}{r_\text{q}^4}\quad, \quad 
 \phi_1^+  =  0 \quad, \quad
h_1^+  =  0 \ .
\end{equation}
and outside the cavity
\begin{align}
p^-  &=  1-\frac{r_\text{q}^3}{r^3}\bigg(1+3\log\frac{r}{r_\text{q}}\bigg)
 \\
 f_1^-  &=  g_{\text{s}} \frac{\vol{(M_3)}}{\vol{(X_5)}}\bigg[-\frac{1}{96}-\frac{1}{196}\frac{r_\text{q}^3}{r^3}+\frac{1}{60}\frac{r_\text{q}^6}{r^6}+\frac{5}{56}\frac{r_\text{q}^3}{r^3}\log\frac{r}{r_\text{q}}\bigg]  \\
 g_1^-  &=  g_{\text{s}} \frac{\vol{(M_3)}}{\vol{(X_5)}}\bigg[\frac{1}{96}-\frac{1}{196}\frac{r_\text{q}^3}{r^3}-\frac{1}{240}\frac{r_\text{q}^6}{r^6}-\frac{1}{28}\frac{r_\text{q}^3}{r^3}\log\frac{r}{r_\text{q}}\bigg]  \\
 \phi_1^-  &= -\frac{g_{\text{s}} \vol{(M_3)}}{12\vol{(X_5)}}\bigg[2-2\frac{r_\text{q}^3}{r^3}-3\bigg(1+\frac{r_\text{q}^3}{r^3}\bigg)\log \frac{r}{r_\text{q}}  \bigg] \\
 h_1^-  &=  0\ .
\end{align}
The line element in the $r$ coordinate reads
\begin{equation}
ds^2_{10}=h^{-1/2}dx^2_{1,3}+h^{1/2}\Sigma^2dr^2+h^{1/2}[e^{2g} ds^2_{KE}+e^{2f}(d\tau+\mathcal{A})^2] \ ,
\label{eq:linelementr}
\end{equation}
where $\Sigma$ is defined in \eqref{eq:Sigma} and has the expansion $\Sigma(r) = \Sigma_0\,[1 + \Nf\,\Sigma_1 + \mathcal{O}(\Nf^2)]$, where
\begin{equation}
    \Sigma_0(r) = \ell^2\ , \quad \Sigma_1(r) = 
    \begin{dcases}
        \frac{9g_{\text{s}}\vol{(X_3)}}{1568\vol{(X_5)}}\frac{r^4}{r_\text{q}^4}\ ,\quad &r < r_\text{q}\\
        \frac{g_{\text{s}}\vol{(X_3)}}{32\vol{(X_5)}}  \bigg(1-\frac{40}{49}\frac{r_\text{q}^3}{r^3}-\frac{12}{7}\frac{r_\text{q}^3}{r^3}\log{\frac{r}{r_\text{q}}}\bigg)\ ,\quad &r \geq r_\text{q}
        \end{dcases}\ .
\label{eq:sigmaT11chiral}
\end{equation}
\paragraph{Discussion about the regularity.}
The geometry we have constructed is finite and does not have any curvature singularities. We can check it explicitly by constructing several curvature invariants. The Ricci scalar is
\begin{equation}
g^{\mu\nu}R_{\mu\nu}^+=0 \quad , \quad
g^{\mu\nu}R_{\mu\nu}^-=\frac{2\Nf \sqrt{g_{\text{s}}}}{\sqrt{3\Nc \pi^3}\ell_s^2}\bigg(1-\frac{r_\text{q}^3}{r^3}-\frac{3r_\text{q}^3}{4r^3}\log\frac{r}{r_\text{q}} \bigg) \ ,
\end{equation}
with $R$ continuous across the cavity. It is finite in the UV region ($r\rightarrow +\infty$) 
and vanishing in the IR ($r \rightarrow 0$). In addition, we compute $R_{\mu \nu}R^{\mu \nu}$:
\begin{align}
R_{\mu\nu}R^{\mu\nu+}&=\frac{640}{27g_{\text{s}}\Nc\pi \ell_s^4}\bigg(1-\frac{27g_{\text{s}} \Nf}{392\pi}\frac{r^4}{r_\text{q}^4}\bigg)\\
R_{\mu\nu}R^{\mu\nu-}&=\frac{640}{27g_{\text{s}}\Nc\pi \ell_s^4}\bigg[1-\frac{9g_{\text{s}} \Nf }{40\pi}\bigg(1-\frac{34}{49}\frac{r_\text{q}^3}{r^3}+\frac{7}{6}\frac{r_\text{q}^3}{r^3}\log\frac{r}{r_\text{q}}\bigg)\bigg]\ ,
\end{align}
This product is also continuous across the cavity, and finite and non-vanishing in both the UV and IR region. In particular, $(R_{\mu \nu}R^{\mu \nu})_{IR}=\frac{160}{\ell^4}$. (The value of $\ell$ can be obtained from \eqref{eq:geomr} and the definition of the $r$ coordinate, $\ell=\frac{4g_s N\pi^4\ell_s^4}{\vol{(X_5)}}$.) The Kretschmann scalar $K=R_{\mu \nu \rho \sigma}R^{\mu \nu \rho \sigma}$ reads
\begin{align}
K^+  &=  \frac{704}{27g_{\text{s}}\Nc\pi \ell_s^4}\bigg(1-\frac{837g_{\text{s}} \Nf r^4}{21560\pi r_\text{q}^4}\bigg)\\
K^-  &=  \frac{704}{27g_{\text{s}}\Nc\pi \ell_s^4}\bigg[1-\frac{15g_{\text{s}} \Nf}{88\pi}\bigg(1-\frac{30}{49}\frac{r_\text{q}^3}{r^3}-\frac{4}{25}\frac{r_\text{q}^6}{r^6}-\frac{66}{35}\frac{r_\text{q}^3}{r^3}\log\frac{r}{r_\text{q}}\bigg)\bigg] \ ,
\end{align}
and it is continuous across the cavity and finite and non-vanishing in the UV and IR regions with value $K_{IR}=\frac{176}{\ell^4}$.

\subsubsection{D3-D7 on the \texorpdfstring{$\text{T}^{1,1}$}{} (non-chiral profile)}
The $\text{T}^{1,1}$ is described by the metric:
\begin{eqnarray}
&ds^2_{KE}=\frac{1}{6}\sum_{i=1}^2[\sin^2\theta_i (d\phi_i)^2+(d\theta_i)^2]\quad, \quad d\tau+\mathcal{A}=\frac{1}{3}[d\psi+\sum_{i=1}^2\cos \theta_i d\phi_i] \ , 
\end{eqnarray}
where $0<\theta_1, \theta_2<\pi, \quad  0<\phi_1, \phi_2<2\pi$ and $  0< \psi< 4\pi$. The geometries inside (outside) the cavity, denoted with the superscript $+$ ($-$), are given by
\begin{align}
    p^+ =  0\quad, \quad f^+_1  =  g^+_1=0\quad, \quad
    \phi^+  =  0\quad , \quad
    h^+_1  =  \frac{3g_{\text{s}}^2\Nc\pi^4\vol{(M_3)}e^{-4\varrho_\text{\tiny{q}}}}{70\vol{(X_5)}^2}  \ ,
\end{align}
and
\begin{align}
p^-=& \ 1-e^{-3(\varrho-\varrho_\text{\tiny{q}})} \\
f^-_1=& \ -\frac{g_{\text{s}} \vol{(M_3)}}{72\vol{(X_5)}}\big[3(\varrho-\varrho_\text{q})+e^{-6(\varrho-\varrho_\text{\tiny{q}})}-e^{-3(\varrho-\varrho_\text{q})}\big] \\
g^-_1= & \ - \frac{g_{\text{s}} \vol{(M_3)}}{144\vol{(X_5)}}\big[6(\varrho-\varrho_\text{q})-3-e^{-6(\varrho-\varrho_\text{\tiny{q}})}+4e^{-3(\varrho-\varrho_\text{\tiny{q}})} \big]  \\
\phi^-_1 =& \ \frac{g_{\text{s}} \vol{(M_3)}}{12\vol{(X_5)}}[e^{-3(\varrho-\varrho_\text{\tiny{q}})}+3(\varrho-\varrho_\text{q})-1] \\ 
h^-_1 =& \ \frac{2g_{\text{s}}^2\Nc \pi^4\vol{(M_3)}e^{-4\varrho}}{45\vol{(X_5)}^2}\bigg[15(\varrho-\varrho_\text{q})-\frac{15}{4} -e^{-6(\varrho-\varrho_\text{\tiny{q}})}+\frac{40}{7}e^{-3(\varrho-\varrho_\text{\tiny{q}})} \bigg] \nonumber \ .
\end{align}
The procedure above we use to construct the geometry is identical to the one followed for the chiral profiles, thus we will not repeat it here. Following the previous case, we introduce the coordinate $r$ that satisfies $h=\frac{1}{\ell^4r^4}$ both inside ($+$) and outside ($-$) the cavity. The change of coordinates is given by
\begin{align}
r^+=\frac{\ell_s}{\ell^2}e^\varrho\bigg[1-\Nf &
\frac{3g_{\text{s}}e^{4(\varrho-\varrho_\text{\tiny{q}})} \vol{(M_3)}}{1120\vol{(X_5)}} \bigg]\\
r^-=\frac{\ell_s}{\ell^2}e^\varrho\bigg[1-\Nf&
\frac{g_{\text{s}} \vol{(M_3)}}{360\vol{(X_5)}}\bigg(15(\varrho-\varrho_\text{q}) -\frac{15}{4}+\frac{40}{7}e^{3(\varrho-\varrho_\text{\tiny{q}})}-e^{-6(\varrho-\varrho_\text{\tiny{q}})}\bigg)  \bigg] \ ,
\end{align}
which gives the following geometry:
\begin{eqnarray}
&f_0=g_0=\log \left(r \ell^2 \right)\quad , \quad  \phi_0=0\quad ,  \quad h_0=\frac{1}{\ell^4r^4} \ .
\end{eqnarray}
Once again, we denote the position of the cavity with $r_\text{q}$ in this new coordinate. Using this, the functions inside ($+$) and outside ($-$) cavity read
\begin{equation}
p^+=0 \quad, \quad
f_1^+=g_1^+=\frac{3g_{\text{s}} \vol{(M_3)}}{1120\vol{(X_5)}}\frac{r^4}{r_\text{q}^4}\quad , \quad
\phi_1^+=0\quad , \quad
h_1^+=0 \ ,
\end{equation}
and
\begin{align}
p^-&=1-\frac{r_\text{q}^3}{r^3} \\
f_1^-&=-g_{\text{s}} \frac{\vol{(M_3)}}{\vol{(X_5)}}\bigg[\frac{1}{96}-\frac{5r_\text{q}^3}{168r^3}+\frac{r_\text{q}^6}{60r^6}\bigg]  \\
g_1^-&=g_{\text{s}} \frac{\vol{(M_3)}}{\vol{(X_5)}}\bigg[\frac{1}{96}-\frac{r_\text{q}^3}{84r^3}+\frac{r_\text{q}^6}{240r^6}\bigg] \\
\phi_1^-&=-\frac{g_{\text{s}} \vol{(M_3)}}{12\vol{(X_5)}}\bigg[1-\frac{r_\text{q}^3}{r^3}-3\log \frac{r}{r_\text{q}}  \bigg] \\
h_1^-&=0\nonumber \ ,
\end{align}
respectively. The line element has the same form as in \eqref{eq:linelementr},
with $\Sigma(r) = \Sigma_0\,[1 + \Nf\,\Sigma_1 + \mathcal{O}(\Nf^2)]$ where
\begin{equation}
    \Sigma_0(r) = \ell^2\ , \quad \Sigma_1(r) = 
    \begin{dcases}
        \frac{3g_{\text{s}}\vol{(X_3)}}{224\vol{(X_5)}}\frac{r^4}{r_\text{q}^4}\ ,\quad &r < r_\text{q}\\
        \frac{g_{\text{s}}\vol{(X_3)}}{32\vol{(X_5)}}  \biggl(1-\frac{4}{7}\frac{r_\text{q}^3}{r^3}\biggr)\ ,\quad &r \geq r_\text{q}
        \end{dcases}\ .
\label{eq:sigmaT11nonchiral}
\end{equation}

\paragraph{Discussion about the regularity.}

We have computed the following curvature invariants for this geometry. The Ricci scalar reads
\begin{equation}
g^{\mu\nu}R_{\mu\nu}^+=0\quad , \quad g^{\mu\nu}R_{\mu\nu}^-=\frac{2\Nf \sqrt{g_{\text{s}}}}{\sqrt{3\Nc \pi^3}\ell_s^2}\bigg(1-\frac{r_\text{q}^3}{4r^3} \bigg) \ ,
\end{equation}
where the product is continuous across the cavity, finite in the UV region and vanishing in the IR. $R_{\mu \nu}R^{\mu \nu}$ is given by:
\begin{align}
R_{\mu\nu}R^{\mu\nu+}&=\frac{640}{27g_{\text{s}}\Nc\pi \ell_s^4}\bigg(1-\frac{9g_{\text{s}} \Nf}{56\pi}\frac{r^4}{r_\text{q}^4}\bigg)\\
R_{\mu\nu}R^{\mu\nu-}&=\frac{640}{27g_{\text{s}}\Nc\pi \ell_s^4}\bigg[1-\frac{9g_{\text{s}} \Nf }{40\pi}\bigg(1-\frac{2r_\text{q}^3}{7r^3}\bigg)\bigg] \ .
\end{align}
This product is continuous across the cavity, and finite and non-vanishing in both the UV and IR region with value $(R_{\mu \nu}R^{\mu \nu})_{IR}=\frac{160}{\ell^4}$. Furthermore, the Kretschmann scalar in this case has the form
\begin{align}
K^+&=\frac{704}{27g_{\text{s}}\Nc\pi \ell_s^4}\bigg(1-\frac{279g_{\text{s}} \Nf r^4}{3080\pi r_\text{q}^4}\bigg)\\
 K^-&=\frac{704}{27g_{\text{s}}\Nc\pi \ell_s^4}\bigg[1-\frac{15g_{\text{s}} \Nf}{88\pi}\bigg(1-\frac{110}{175}\frac{r_\text{q}^3}{r^3}+\frac{28}{175}\frac{r_\text{q}^6}{r^6}\bigg)\bigg]\ ,
\end{align}
and it is as well continuous across the cavity and finite and non-vanishing in the UV and IR regions with value $K_{IR}=\frac{176}{\ell^4}$.

\subsubsection{D3-D7 on the \texorpdfstring{$\text{S}^5$}{}}
The $\text{S}^{5}$ can be described by the metric:
\begin{eqnarray}
&ds^2_{KE}=\frac{1}{4}\cos^2\frac{\chi}{2}(\omega_1^2+\omega_2^2)+\frac{1}{4}\cos^2\frac{\chi}{2}\sin^2\frac{\chi}{2}\omega_3^2+\frac{1}{4}d\chi^2\ ,
\end{eqnarray}
with
\begin{equation}
    d\tau+\mathcal{A}=d\tau+\frac{1}{2}\cos^2\frac{\chi}{2}\omega_3\ ,
\end{equation}
where $\omega_i, \ i=\{1, 2, 3\}$, are left-invariant one-forms satisfying $d\omega_i=\frac{1}{2}\epsilon_{ijk}\omega^j\wedge \omega^k$:
\begin{align}
    \omega_1&=\cos \psi_i d\theta_i+\sin \psi_i \sin \theta_i d\phi_i \\
\omega_2&=\sin \psi_i d\theta_i-\cos \psi_i \sin \theta_i d\phi_i  \\
\omega_3&=d\psi_i+\cos \theta_i d\phi_i \ ,
\label{eq:S5detailsomegas}
\end{align}
where $0<\chi, \theta_i < \pi, \quad  0<\tau, \phi_i <2\pi,\quad  0< \psi_i< 4\pi$. The $\text{S}^5$ geometry can be solved from the BPS system perturbatively in $\Nf$. However, this geometry has been already obtained in \cite{Bigazzi:2009bk} for any value of $\Nf$. To order ${\cal{O}}(\Nf)$ the geometry inside the cavity reads:
\begin{align}
p^+&=0 \\
e^{f_+}&=e^{g_+}=\ell_s e^\varrho \bigg[1+\Nf \frac{g_{\text{s}} \vol{(M_3)}}{24\vol{(X_5)}}\bigg(-\frac{3}{4}-\frac{1}{4}e^{4(\varrho_\text{\tiny{q}}-\varrho_*)}+e^{2(\varrho_\text{\tiny{q}}-\varrho_*)}-\varrho_\text{q}+\varrho_* \bigg)\bigg] \\
\phi^+&=-\Nf \frac{g_{\text{s}}\vol{(M_3)}}{4\vol{(X_5)}}\bigg[-\frac{3}{4}-\frac{1}{4}e^{4(\varrho_\text{\tiny{q}}-\varrho_*)}+e^{2(\varrho_\text{\tiny{q}}-\varrho_*)}-\varrho_\text{q}+\varrho_* \bigg] \ ,
\end{align}
whereas outside we have
\begin{align}
p^{-}=& \ (1-e^{2(\varrho_\text{\tiny{q}}-\varrho)})^2 \\
e^{f_-}=& \ \ell_s e^\varrho \bigg[1-\Nf \frac{g_{\text{s}} \vol{(M_3)}}{24\vol{(X_5)}}\bigg(\frac{1}{3}+\varrho-\varrho_*-e^{2(\varrho_\text{\tiny{q}}-\varrho_*)}+\frac{1}{4}e^{4(\varrho_\text{\tiny{q}}-\varrho_*)}\\ & \qquad \qquad \qquad \qquad \qquad -\frac{1}{3}e^{6(\varrho_\text{\tiny{q}}-\varrho)}+\frac{3}{4}e^{4(\varrho_\text{\tiny{q}}-\varrho)}\bigg) \bigg]\nonumber \\
e^{g_-}= & \ \ell_s e^\varrho \bigg[1+\Nf \frac{g_{\text{s}} \vol{(M_3)}}{24\vol{(X_5)}}\bigg(\frac{1}{6}-\varrho+\varrho_*-\frac{3}{2}e^{2(\varrho_\text{\tiny{q}}-\varrho)}+\frac{3}{4}e^{4(\varrho_\text{\tiny{q}}-\varrho)}-\frac{1}{6}e^{6(\varrho_\text{\tiny{q}}-\varrho)}\\ & \qquad \qquad \qquad \qquad \qquad -\frac{1}{4}e^{4(\varrho_\text{\tiny{q}}-\varrho_*)}+e^{2(\varrho_\text{\tiny{q}}-\varrho_*)}\bigg)\bigg] \nonumber \\
\phi_-=& \ -\Nf\frac{g_{\text{s}} \vol{(M_3)}}{4\vol{(X_5)}}\bigg[ \varrho_*-\varrho -e^{2(\varrho_\text{\tiny{q}}-\varrho)}+\frac{1}{4}e^{4(\varrho_\text{\tiny{q}}-\varrho)}+e^{2(\varrho_\text{\tiny{q}}-\varrho_*)}-\frac{1}{4}e^{4(\varrho_\text{\tiny{q}}-\varrho_*)}\bigg] \ .
\end{align}
The previous geometry depends on $\varrho_*$, which is an arbitrary value of reference of the holographic coordinate introduced in \cite{Bigazzi:2009bk}, useful to determine the range of validity of the solution. Our entropy computations will be done in another coordinate $r$, in which the only place where the $\varrho_*$ plays a role is the dilaton, and it will be hidden in the entropy computation and absorbed into the string coupling. As in \cite{Kontoudi:2013rla}, we introduce the $r$ coordinate, which can be identified with the energy scale, as
\begin{align}
r^+=& \ \frac{e^\varrho}{\ell^2}\bigg[1+\Nf \frac{\vol{(M_3)}g_{\text{s}}}{96\vol{(X_5)}^3}\bigg(-\frac{e^{4\varrho }\vol{(X_5)}^2}{160\Nc^2\pi^8r_\text{q}^4\ell_s^4 g_{\text{s}}^2}+16g_{\text{s}}\Nc e^{-2\varrho_*}\pi^4r_\text{q}^2\vol{(X_5)}\ell_s^2 \nonumber \\ & \ -16e^{-4\varrho_*}g_{\text{s}}^2\Nc^2\pi^8r_\text{q}^4\ell_s^4-4\vol{(X_5)}^2\log \bigg(r_\text{q} \frac{\ell^2}{\ell_s}\bigg)+\vol{(X_5)}^2(4\varrho_*-3)\bigg) \bigg] \\
r^-=& \ \frac{e^\varrho}{\ell^2}\bigg[1-\Nf \frac{g_{\text{s}} \vol{(M_3)}}{288\vol{(X_5)}^4}\bigg( 48(e^{-2\varrho}-e^{2\varrho_*})\Nc g_{\text{s}} \pi^4r_\text{q}^2\vol{(X_5)}^2\ell_s^2 \nonumber\\ & \ +24(-3e^{-4\varrho} +2e^{-4\varrho_*})g_{\text{s}}^2\Nc^2\pi^8r_\text{q}^4 \vol{(X_5)}\ell_s^4+\frac{256}{5}e^{-6\varrho}g_{\text{s}}^3\Nc^3\pi^{12}r_\text{q}^6\ell_s^6\nonumber \\ & \ +\vol{(X_5)}^3(1+12(\varrho-\varrho_*)) \bigg) \bigg] \ ,  
\end{align}
with which the geometry reads, to order ${\cal{O}}(\Nf^0)$,
\begin{eqnarray}
f_0=g_0=\log \big(r\ell^2 \big)\quad, \quad \phi_0=0\quad, \quad h_0=\frac{1}{r^4\ell^4} \ .
\end{eqnarray}
For the flavor corrections inside ($+$) and outside ($-$) the cavity, we have
\begin{align}
p^+=& \ h_1^+ = 0  \\
f_1^+= & \ g_1^+=\frac{g_{\text{s}} \vol{(M_3)}r^4}{960 \vol{(X_5)}r_\text{q}^4}  \\
\phi_1^+=& \  \frac{g_{\text{s}} \vol{(M_3)}}{\vol{(X_5)}^3} \bigg[-e^{-2\rho_*}g_{\text{s}} \Nc \pi^4r_\text{q}^2\vol{(X_5)}\ell_s^2 +e^{-4\rho_*}g_{\text{s}}^2\Nc^2\pi^8r_\text{q}^4\ell_s^4+\frac{1}{16}\vol{(X_5)}^2 (3-4\rho_*)\nonumber \\ & \qquad \qquad \qquad \qquad \qquad +\frac{1}{4}\vol{(X_5)}^2\log \big(r_\text{q}\frac{\ell^2}{\ell_s}\big)\bigg] \ ,
\end{align}
and
\begin{align}
h_1^-=& \ 0 \\
p^-=& \ \big(1-\frac{r_\text{q}^2}{r^2}\big)^2 \\ 
f_1^-=& \ \frac{g_{\text{s}} \vol{(M_3)}}{96\vol{(X_5)}}\bigg(-1+4\frac{r_\text{q}^2}{r^2}-\frac{9}{2}\frac{r_\text{q}^4}{r^4} +\frac{8}{5}\frac{r_\text{q}^6}{r^6}\bigg)  \\
g_1^-=& \ \frac{g_{\text{s}} \vol{(M_3)}}{96\vol{(X_5)}}\bigg(1-2\frac{r_\text{q}^2}{r^2}+\frac{3}{2}\frac{r_\text{q}^4}{r^4}-\frac{2}{5}\frac{r_\text{q}^6}{r^6}\bigg)  \\
\phi_1^-=& \ \frac{g_{\text{s}} \vol{(M_3)}}{\mathrm{16Vol(X_5)}^3}\bigg[\vol{(X_5)}^2\bigg(\frac{4r_\text{q}^2}{r^2}-\frac{r_\text{q}^4}{r^4} \bigg)-16e^{-2\rho_*}g_{\text{s}} \Nc \pi^4r_\text{q}^2\vol{(X_5)}\ell_s^2+16e^{-4\rho_*}g_{\text{s}}^2\Nc^2\pi^8r_\text{q}^4\ell_s^4\nonumber \\ &\qquad \qquad \qquad \qquad \qquad -4\rho_* \vol{(X_5)}^2+4\vol{(X_5)}^2\log(r\frac{\ell^2}{\ell_s}) \bigg] \ ,
\end{align}
respectively. As before, $r_\text{q}$ denotes the position of cavity in this new coordinate. The line element in this coordinate has identical form to \eqref{eq:linelementr}, where the Sasaki-Einstein manifold is now the one corresponding to the $\text{S}^5$ and using the similar flavor expansion for $\Sigma(r)$ as in the previous geometries gives
\begin{equation}
    \Sigma_0(r) = \ell^2\ , \quad \Sigma_1(r) = 
    \begin{dcases}
        \frac{g_{\text{s}}\vol{(X_3)}}{192\vol{(X_5)}}\frac{r^4}{r_\text{q}^4}\ ,\quad &r < r_\text{q}\\
        \frac{g_{\text{s}}\vol{(X_3)}}{32\vol{(X_5)}}  \bigg(1-\frac{4}{3}\frac{r_\text{q}^2}{r^2}+\frac{1}{2}\frac{r_\text{q}^4}{r^4}\bigg)\ ,\quad &r \geq r_\text{q}
        \end{dcases}\ .
\label{eq:sigmaS5}
\end{equation}

\paragraph{Discussion about the regularity.}

The geometry we have constructed is finite and does not have any curvature singularities. Using the same notation as in the previous cases, we have, for the curvature invariants
\begin{equation}
g^{\mu\nu}R_{\mu\nu}^+=0\quad , \quad 
g^{\mu\nu}R_{\mu\nu}^-=\frac{e^{\Phi_{*}}\sqrt{g_{\text{s}}}\Nf}{\sqrt{\Nc}\pi^\frac{3}{2}\ell_s^2}\bigg(1-\frac{r_\text{q}^2}{r^2}\bigg)  \ ,
\end{equation}
which is continuous across the cavity. It is finite in the UV region and vanishing in the IR. We also have
\begin{align}
R_{\mu\nu}R^{\mu\nu+}&=\frac{40}{g_{\text{s}} \Nc\pi \ell_s^4}\bigg(1-\frac{g_{\text{s}} \Nf}{24\pi }\frac{r^4}{r_\text{q}^4} \bigg) \\
R_{\mu\nu}R^{\mu\nu-}&=\frac{40}{g_{\text{s}} \Nc\pi \ell_s^4}\bigg[1-\frac{3g_{\text{s}} \Nf}{20\pi}\bigg(1-\frac{8}{9}\frac{r_\text{q}^2}{r^2}+\frac{1}{6}\frac{r_\text{q}^4}{r^4}\bigg) \bigg]\ .
\end{align}
This product is continuous across the cavity, and finite and non-vanishing in both the UV and IR region with value $(R_{\mu \nu}R^{\mu \nu})_{IR}=\frac{160}{\ell^4}$. For the Kretschmann scalar, we get
\begin{align}
K^+&=\frac{20}{g_{\text{s}}\Nc\pi \ell_s^4}\bigg(1-\frac{g_{\text{s}} \Nf}{24\pi}\frac{r^4}{r_\text{q}^4} \bigg)\\
K^-&=\frac{20}{g_{\text{s}} \Nc\pi \ell_s^4}\bigg[1-\frac{3g_{\text{s}} \Nf}{20\pi}\bigg(1-\frac{8}{9}\frac{r_\text{q}^2}{r^2}+\frac{1}{6}\frac{r_\text{q}^4}{r^4}\bigg) \bigg] \ ,
\end{align}
which, as well, is continuous across the cavity and finite and non-vanishing in the UV and IR regions with value $K_{IR}=\frac{80}{\ell^4}$.

\subsection{ABJM geometry}\label{app:ABJMdetails}

In this appendix, we present the details of the flavored ABJM geometry.

\subsubsection{ABJM solution from dimensional reduction}

Consider the action of 11-dimensional supergravity
\begin{equation}
    I_{\text{11D}} =-\frac{1}{(2\pi)^8\ell_{\text{s}}^9}\int d^{11}x\sqrt{g}\,\biggl[R_{11}-\frac{1}{2\cdot 4!}\,\widetilde{F}_4^2\biggr]+\frac{1}{(2\pi)^8\ell_{\text{s}}^9}\int C_3\wedge \widetilde{F}_4\wedge \widetilde{F}_4\ ,
\end{equation}
and an 11-dimensional metric of the form
\begin{equation}
ds^{2}_{11} = e^{-\frac{2\Phi}{3}}\,g_{ab}\,dx^adx^b+e^{\frac{4\Phi}{3}}\,(d\varphi+A_1)^{2}\ ,
\label{eq:generalreduction}
\end{equation}
where $\varphi$ parametrizes a circle $ \varphi \sim \varphi + 2\pi\ell_{\text{s}} $. Under dimensional reduction on the circle, the action reduces to the action of type IIA supergravity \cite{Polchinski:1998rr}
\begin{equation}
	I_{\text{IIA}} = -\frac{1}{(2\pi)^7\ell_{\text{s}}^8}\int d^{10}x\sqrt{g}\,\biggl[e^{-2\Phi}\,(R_{10}+4\,g^{ab}\,\partial_a\Phi\,\partial_b\Phi) - \dfrac{1}{2\cdot 4!}\,\widetilde{F}_4^{2} - \dfrac{1}{2\cdot 2!}\,\widetilde{F}_2^{2}\biggr]
\end{equation}
with the ten-dimensional fields $ds^{2}_{10} = g_{ab}\,dx^adx^b$, $\widetilde{F}_2 = dA_1$ and the same $\widetilde{F}_4$. We will also define 
\begin{equation}
    e^{\Phi} = g_{\text{s}}\,e^\phi\ ,\quad F_2 = g_{\text{s}}\widetilde{F}_2\ ,\quad F_4 = g_{\text{s}}\widetilde{F}_4\ ,
    \label{eq:framechange}
\end{equation}
to write the action in the form \eqref{eq:typeAaction} with the 10-dimensional Newton's constant defined as $2\kappa_{10}^2 = (2\pi)^7g_{\text{s}}^2\ell_{\text{s}}^8 $.

The ABJM solution is obtained in this way by starting with the 11-dimensional solution \cite{Aharony:2008ug}
\begin{gather}
ds^{2}_{11} = L^{2}(ds^{2}_{\text{AdS}_4} + 4ds^{2}_{\mathbb{C}\text{P}^3})+\biggl(\frac{2L}{k\ell_{\text{s}}}\biggr)^{2}\,(d\varphi+k\ell_{\text{s}}\,\omega)^{2}\\
\widetilde{F}_4 = 3L^3\,\epsilon_{\text{AdS}_4}\ ,\quad L^{6} = \frac{\pi^{2} N k}{2}\,\ell_{\text{s}}^{6}\ 
\label{eq:11Dabjm}
\end{gather}
where $ \varphi \sim \varphi + 2\pi\ell_{\text{s}} $, $\epsilon_{\text{AdS}_4}$ is the volume element of $\text{AdS}_{4}$, and $\omega$ is related to the Kähler form as explained below. We can write it in the form \eqref{eq:generalreduction} by defining $ \ell $ and $ \Phi $ via
\begin{equation}
L^{2} = e^{-\frac{2\Phi}{3}}\,\ell^{2}\ ,\quad \frac{2L}{k\ell_{\text{s}}} =  e^{\frac{2\Phi}{3}}\ ,
\end{equation}
which gives
\begin{equation}
\ell^{2} = \frac{2L^{3}}{k\ell_{\text{s}}} = \sqrt{\frac{2\pi^{2}N}{k}}\,\ell_{\text{s}}^2\ ,\quad e^{\Phi} = \frac{2\ell}{k\ell_{\text{s}}} = \biggl(\frac{2^{11}\pi^{5}}{3\vol{(X_{6})}}\dfrac{N}{k^{5}}\biggr)^{1\slash 4}\ .
\label{eq:10Dparameters}
\end{equation}
where we used \eqref{eq:11Dabjm} and defined $\vol{(X_6)} = 2^6\vol{(\mathbb{C}\text{P}^3)} = \frac{64}{3}\,\pi^3 $. The ten-dimensional solution becomes
\begin{gather}
ds^{2}_{10} = \ell^{2}(ds^{2}_{\text{AdS}_4} + 4ds^{2}_{\mathbb{C}\text{P}^3})\\
e^{\Phi} = \frac{2\ell}{k\ell_{\text{s}}} \, \quad \widetilde{F}_2 = k\ell_{\text{s}}\,J \ ,\quad  \widetilde{F}_4 = \frac{3}{2}\,k\ell^2\ell_{\text{s}}\,\epsilon_{\text{AdS}_4}\ .
\label{eq:10Dabjm}
\end{gather}
where $J = d\omega$ and we used \eqref{eq:10Dparameters} to write $\widetilde{F}_4$ in terms of $\ell$. This is the form of the ABJM solution often used in the literature (with string length also set to unity). In terms of \eqref{eq:framechange} the solution becomes
\begin{equation}
    g_{\text{s}} = \frac{2\ell}{k\ell_{\text{s}}}\ , \quad \phi = 0\ ,\quad F_2 = 2\ell\,J\ ,\quad  F_4 =3\ell^3\,\epsilon_{\text{AdS}_4}\ ,
    \label{eq:ABJMsolutionourconventions}
\end{equation}
which is presented in the main text and which has similar form as the D3-brane solution of type IIB supergravity. The explicit expression for the metric on the internal space is given by \cite{Hikida:2009tp}
\begin{align}
    ds^2_{X_6} \equiv 4ds^{2}_{\mathbb{C}\text{P}^3} = d\xi^{2} +\sin^{2}{\xi}\,&\left( d\psi + \frac{\cos{\theta_{1}}}{2}\,d\phi_1 - \frac{\cos{\theta_{2}}}{2}\,d\phi_2 \right)^{2} + \cos^{2}{\Bigl(\frac{\xi}{2}\Bigr)}\,(d\theta_{1}^{2} + \sin^{2}{\theta_{1}}\,d\phi_{1}^{2})\nonumber\\
&\qquad\qquad\qquad\qquad\qquad+ \sin^{2}{\Bigl( \frac{\xi}{2}\Bigr) }\,(d\theta_{2}^{2} + \sin^{2}{\theta_{2}}\,d\phi_{2}^{2}) \ ,
\label{eq:CP3app}
\end{align}
while the K\"ahler form is \cite{Hikida:2009tp,Zafrir:2012yg}
\begin{align}
J = -\frac{1}{4}\sin{\xi}\, d\xi \wedge (2d\psi + \cos{\theta_1}\,d\phi_1 - \cos{\theta_2}\,d\phi_2)& - \frac{1}{2}\cos^{2}{\Bigl( \frac{\xi}{2}\Bigr) }\,\sin{\theta_1}\,d\theta_1 \wedge d\phi_1\nonumber\\
&- \frac{1}{2}\sin^{2}{\Bigl( \frac{\xi}{2}\Bigr) }\,\sin{\theta_2}\,d\theta_2 \wedge d\phi_2 \ .
\label{eq:kahlerapp}
\end{align}
The coordinate ranges are\footnote{Notice that in \cite{Hikida:2009tp} the $\xi$-coordinate is one-half of our $\xi$-coordinate. We follow the conventions of \cite{Jokela:2021knd}.}
\begin{equation}
0 \leq \xi \leq \pi\ , \quad 0\leq \psi < 2\pi\ , \quad 0\leq \theta_{1,2} < \pi\ , \quad 0 \leq \phi_{1,2} < 2\pi \ .
\end{equation}
We have assumed that the coefficient of the 11D action and the circumference of the compactification circle are both given by the string length $\ell_{\text{s}}$. We used this convention, because it is used in \cite{Aharony:2008ug}. Another common convention is to take the coefficient of the 11D action to be $1\slash ((2\pi)^8\ell_{\text{p}11}^9)$ and the circumference to be $2\pi \ell_{\varphi}$ where the 11-dimensional Planck length $\ell_{\text{p}11}$ and $ \ell_{\varphi}$ are two independent length scales. These two scales determine the string length and the 10D string coupling by the standard formulae $\ell_{\text{p}11} = g_{\text{s}}^{1\slash 3}\ell_{\text{s}}$ and $ \ell_{\varphi} = g_{\text{s}}\ell_{\text{s}}$. The relation to the above convention is given by a scaling of the fields: the 11-dimensional metric is scaled by a factor of $g_{\text{s}}^{-2\slash 3}$ while the forms are scaled by $g_{\text{s}}^{-1}$.

\subsubsection{Backreaction sourced by smeared D6-branes}

The starting point is to write the Fubini--Study metric \eqref{eq:CP3app} as a fibration over $\text{S}^4$ as
\begin{equation}
    ds^2_{X_6} = ds^{2}_{\text{S}^{4}}+\sum_{i=1}^2(E^{i})^{2}\ ,
\end{equation}
where the one-forms are
\begin{align}
E_1&=d\theta +\frac{Z^2}{1+Z^2}\,(\sin{\varphi}\, \omega_1-\cos{\varphi} \,\omega_2) \\
E_2&=\sin{\theta}\, \biggl(d\varphi-\frac{Z^2}{1+Z^2}\,\omega_3\biggr) +\frac{Z^2}{1+Z^2}\,(\cos{\varphi}\,\omega_1+\sin{\varphi} \,\omega_2) \ .
\end{align}
and $\omega_i, (i=\{1,2,3\})$ are the $SU(2)$ invariant forms introduced in \eqref{eq:S5detailsomegas}. This basis parametrizes $\mathbb{C}\rm{P}^{3}$. The angles cover the ranges: $0 < Z < \infty, 0 < \varphi, \phi_i < 2\pi, 0 < \theta, \theta_i < \pi, 0 < \psi_i < 4\pi$, where $(\phi_i, \theta_i, \psi_i)$ are the angles parametrizing the left-invariant forms. The metric on the four sphere can be written as $ds^2_{\text{S}^4}=\sum_{i=1}^4\mathcal{S}_i^2$ where \cite{Conde:2011sw}
\begin{align}
\mathcal{S}_1&=\frac{Z}{1+Z^2}\,(\sin{\varphi}\,\omega_1-\cos{\varphi}\,\omega_2) \\
\mathcal{S}_2&=\frac{Z}{1+Z^2}\,(\sin{\theta}\,\omega_3 -\cos{\theta}\, (\cos{\varphi}\,\omega_1+\sin{\varphi}\,\omega_2)) \\
\mathcal{S}_3&=\frac{Z}{1+Z^2}\, (-\cos{\theta}\,\omega_3 -\sin{\theta}\, (\cos{\varphi}\, \omega_1+\sin{\varphi}\, \omega_2)) \\
\mathcal{S}_4&=\frac{2}{1+Z^2}\,dZ \ .
\end{align}
In this notation, the Kähler form \eqref{eq:kahlerapp} can be written as \cite{Conde:2011sw}
\begin{equation}
    J = \frac{1}{4}\Bigl(E^1\wedge E^2-\mathcal{S}^Z\wedge \mathcal{S}^3-\mathcal{S}^1\wedge \mathcal{S}^2\Bigr).
\end{equation}
After including backreaction of the smeared D6-branes, the ABJM solution \eqref{eq:10Dabjm} becomes \cite{Conde:2011sw,Bea:2013jxa,Balasubramanian:2018qqx}
\begin{gather}
ds^{2}_{10} = h^{-1\slash 2}ds^{2}_{1,2} +h^{1\slash 2} e^{2f} d\varrho^2 + h^{1\slash 2}\Bigl[ e^{2g}ds^{2}_{\text{S}^{4}}+e^{2f}\sum_{i=1}^2(E^{i})^{2}\Bigr]\label{eq:backreactedabjmapp}\\
\phi = \phi(\varrho)\ , \quad F_2 = \frac{\ell}{2}\Bigl[E^1\wedge E^2-\eta\,\bigl(\mathcal{S}^Z\wedge \mathcal{S}^3+\mathcal{S}^1\wedge \mathcal{S}^2\bigr)\Bigr],\quad F_4 = 3\ell^5\,h^{-2}\,e^{-4g-f}\,d\varrho\wedge \epsilon_{\mathbb{R}^3}\nonumber
\end{gather}
where $h = h(\varrho)$, $f = f(\varrho)$, and $g = g(\varrho)$ are functions of the radial coordinate and $\eta = \eta(\varrho)$ is the profile function of the massive brane distribution (given below). Following \cite{Conde:2011sw,Balasubramanian:2018qqx}, instead of the coordinate $\varrho$ we write all functions in terms of the radial coordinate
\begin{equation}
    x = e^{\varrho-\varrho_{\text{\tiny{q}}}} \  
    \label{eq:abjmxcoordinate}
\end{equation}
for which the metric \eqref{eq:backreactedabjmapp} contains $d\varrho^2 = dx^2\slash x^2$. Then the functions can be determined by solving a master equation for a master function $W(x)$,
\begin{equation}
W''+4\eta' + (W'+4\eta)\left(\frac{W'+10\eta}{3W}-\frac{W'+4\eta+6}{x(W'+4\eta)} \right) = 0\ ,
\label{eq:master}
\end{equation}
where the profile function $\eta(x)$ has the form
\begin{equation}
\eta(x) = 1+\frac{3}{4}\frac{\Nf}{k}\left(1-\frac{1}{x^{2}} \right)\Theta(x-1)\ ,
\label{eq:profilefunc}
\end{equation} 
where $\Theta(x)$ is a step function and the relation to the profile function \eqref{eq:ABJMp} introduced in the main text is given by $\eta(x)-1 = \frac{3}{4}\frac{\Nf}{k}\,p(\varrho)$, where a change of coordinates (\ref{eq:abjmxcoordinate}) is implied. This describes a radial distribution of flavor branes terminating at $x = 1$ (corresponding to $\varrho = \varrho_{\text{q}}$). In the cavity region $ x < 1 $, we have \cite{Balasubramanian:2018qqx}:\footnote{Compared to \cite{Balasubramanian:2018qqx}, we have multiplied $e^\phi$ by a factor of $\frac{k}{2\ell}$ so that $\phi$ vanishes when $\Nf = 0$ in agreement with our conventions \eqref{eq:ABJMsolutionourconventions}.}
\begin{eqnarray}
e^{f_+} &  = & \ \ell^{2} r_\text{q}\,\frac{\sqrt{1+4\gamma}+1}{2}\frac{x}{\sqrt{1+4\gamma x}}\\
e^{g_+} & = & \ \ell^{2}r_\text{q}\,\frac{\sqrt{1+4\gamma}+1}{\sqrt{2}}\frac{x}{\left[ \sqrt{1+4\gamma x}\,(1+\sqrt{1+4\gamma x})\right]^{1\slash 2}}\\
e^{\phi_+} & = & \ \frac{\ell r_\text{q}}{2k}\frac{\sqrt{1+4\gamma}+1}{2}\frac{xh(x)^{1\slash 4}}{1+4\gamma x} \\
h^+ & = & \ 4\ell^{4}\left(\frac{\sqrt{1+4\gamma}-1}{2r_\text{q}\ell^{2}} \right)^{4}\left(1 + \frac{1}{4\gamma x} \right)\Bigg[\left(\frac{1}{2}+6\gamma x + \frac{1+(1-6 \gamma x)\sqrt{1+4 \gamma x}}{4\gamma x} \right) \frac{\sqrt{1+4 \gamma x}+1}{\gamma^{2}x^{2}}\nonumber \\
& & + 24\log{\left(\frac{\sqrt{4 \gamma x}}{\sqrt{1 + 4 \gamma x}+1} \right) }+\alpha \Bigg] \ ,
\end{eqnarray}
where $\gamma$ and $\alpha$ are constants \cite{Bea:2013jxa} with closed form expressions in the few flavor limit \cite{Bea:2016ekp,Balasubramanian:2018qqx}; see below. In the region outside cavity $x\geq 1$, the functions are given by:
\begin{align}
e^{f_-} &= \ell^{2} r_\text{q}\left(\frac{(\sqrt{1+4\gamma}+1)^{2}}{2\sqrt{1+4\gamma}} \right)^{1\slash 3}\frac{x}{W(x)^{1\slash 3}}\exp{\left(\frac{2}{3}\int_{1}^{x} dx'\frac{\eta(x')}{W(x')} \right) }\\
e^{g_-} &=\ell^{2} r_\text{q}\left(\frac{(\sqrt{1+4\gamma}+1)^{2}}{2\sqrt{1+4\gamma}} \right)^{1\slash 3}\sqrt{\frac{3x}{W'(x)+4\eta(x)}}W(x)^{1\slash 6}\exp{\left(\frac{2}{3}\int_{1}^{x} dx'\,\frac{\eta(x')}{W(x')} \right) }\\
e^{\phi_-} &= \ell r_\text{q}\left(\frac{(\sqrt{1+4\gamma}+1)^{2}}{2\sqrt{1+4\gamma}} \right)^{1\slash 3}\frac{6x\,h(x)^{1\slash 4}}{W(x)^{1\slash 3}(W'(x)+4\eta(x))}\exp{\left(\frac{2}{3}\int_{1}^{x} dx'\,\frac{\eta(x')}{W(x')} \right) }\\
h^- &= 2\ell^{4}\,e^{-g(x)}(W'(x)+4\eta(x))\left(\int_{x}^{\infty} dx'\,\frac{x'\,e^{-3g(x')}}{W(x')^{2}}\right) \ . 
\end{align}
Here $r_\text{q}$ is the location of the cavity in an $r = r(x)$ coordinate to be defined below.

In the probe limit $  \Nf \rightarrow 0 $, the master equation \eqref{eq:master} can be solved perturbatively. Expanding the equation \eqref{eq:master} in powers of $\Nf$ by using
\begin{equation}
    W(x) = W_0(x) + \Nf\,W_1(x) + \mathcal{O}(\Nf^2) \ ,
\end{equation}
we find the solution
\begin{equation}
    W_0(x) = 2x\ ,\quad W_1(x) = 
    \begin{dcases}
        0\ ,\quad &x < 1\\
        \frac{3}{40}\frac{35x^{2}-10x^{2}-1}{x^{3}}\frac{1}{k}\ ,\quad &x\geq 1
    \end{dcases} \ .
\label{eq:Wads}
\end{equation}
The constant $ \gamma $ is fixed by requiring that the solution for $ x\geq 1 $ is asymptotically AdS$ _4 $ and $ \alpha $ is fixed by the continuity $ \lim_{x\rightarrow 1^{-}}h(x) = \lim_{x\rightarrow 1^{+}}h(x) $. The result is 
\begin{equation}
\gamma = \frac{3}{10}\,\frac{\Nf}{k}+ \mathcal{O}(\Nf^2)\quad , \quad \alpha = \frac{200}{63}\frac{k^2}{\Nf^2} + \mathcal{O}(1\slash \Nf) \ .
\end{equation}
At order ${\cal O}(\Nf^0)$, we have
\begin{eqnarray}
f_0=\log{(r_\text{q}\ell^2\, x )}\quad , \quad g_0=\log{(r_\text{q}\ell^2\, x )}\quad , \quad \phi_0=0\quad , \quad h_0=\frac{1}{r_\text{q}^4\ell^4x^4} \ .
\label{eq:flavorlessABJMapp}
\end{eqnarray}
The expansions for cavity region $ x < 1 $ are obtained from \eqref{eq:Wads} and are given by
\begin{align}
f_1^+ &=-\frac{1}{k}\frac{3(2x-1)}{10} \\
g_1^+&=-\frac{1}{k}\frac{3(3x-2)}{20} \\
\phi_1^+&=\frac{1}{k}\frac{3x(x^2-42)}{140} \\
h_1^+&=\frac{3}{35k r_\text{q}^4x^4\ell^4}(x^3+14x-14) \ .
\end{align}
The expansions for outside cavity $ x\geq 1 $ are given by
\begin{align}
f_1^- &=\frac{3}{320kx^4}(1+20x^2-53x^4-20x^4\log x) \\
g_1^-&=-\frac{3}{320kx^4}(3-20x^2+33x^4+20x^4\log x) \\
\phi_1^-&=-\frac{3}{560kx^4}(3-84x^2+245x^4) \\
h_1^-&=\frac{3}{560kr_\text{q}^4x^8\ell^4}(140x^4\log x+91x^4-84x^2+9) \ .
\end{align}
As in the D3-D7 cases, we will introduce the coordinate $r = r(x)$ such that the warp factor has the form $h=\frac{1}{\ell^4r^4}$ for all $\Nf$. The relation between the coordinates is
\begin{align}
x^+&=\frac{r}{r_\text{q}}-\frac{9r}{28kr_\text{q}}\Nf\bigg(1-\frac{r^3}{15r_\text{q}^3}-\frac{14r}{15r_\text{q}} \bigg)  \\
x^-&=\frac{r}{r_\text{q}}+\Nf \frac{45}{448k}\frac{r}{r_\text{q}}\bigg(1-\frac{28}{25}\frac{r_\text{q}^2}{r^2}+\frac{3}{25}\frac{r_\text{q}^4}{r^4}+\frac{28}{15}\log\frac{r}{r_\text{q}}\bigg)
\end{align}
so that the cavity $x = 1$ is now located at $ r = r_\text{q}$. With this change, the unflavored geometry \eqref{eq:flavorlessABJMapp} becomes
\begin{eqnarray}
f_0=\log{(\ell^2 r )}\quad , \quad g_0=\log{(\ell^2 r )}\quad , \quad \phi_0=0\quad, \quad h_0=\frac{1}{\ell^4r^4} \ . 
\end{eqnarray}
The first flavor correction inside the cavity is given by
\begin{align}
p^{+}&=1 \\
f^{+}_1&= \frac{3r}{10kr_\text{q}}\left(-1+\frac{r^2}{14r_\text{q}^2} \right)  \\
g^{+}_1&= -\frac{3r}{20kr_\text{q}}\left(1-\frac{r^2}{7r_\text{q}^2} \right)  \\
\phi^{+}_1&=-\frac{9}{10k}\left(\frac{r}{r_\text{q}}-\frac{r^3}{42r_\text{q}^3} \right)    \\
h^{+}_1&=0 \ ,
\end{align}
and outside cavity as
\begin{align}
p^{-}&=1+\frac{3\Nf}{4k}\left(1-\frac{r_\text{q}^2}{r^2}\right)\nonumber \\
f^{-}_1&=-\frac{3}{8k}\bigg(1-\frac{1}{5}\frac{r_\text{q}^2}{r^2}-\frac{2}{35}\frac{r_\text{q}^4}{r^4} \bigg)  \nonumber \\
 g^{-}_1&=-\frac{3}{16k} \bigg(1-\frac{2}{5}\frac{r_\text{q}^2}{r^2}+\frac{3}{35}\frac{r_\text{q}^4}{r^4} \bigg) \nonumber \\
 \phi^{-}_1&=-\frac{21}{16k} \bigg(1-\frac{12}{35}\frac{r_\text{q}^2}{r^2}+\frac{3}{245}\frac{r_\text{q}^4}{r^4} \bigg) \nonumber \\
 h^{-}_1&=0 \ .
\end{align}
In the coordinate $r$, the metric and 4-form of the solution \eqref{eq:backreactedabjmapp} become
\begin{gather}
ds^{2}_{10} = \ell^{2}\biggl[ r^{2}ds^{2}_{1,2} + \Sigma(r)^{2}\,dr^{2} + \frac{1}{\ell^{4}r^{2}} \biggl(e^{2g}ds^{2}_{\text{S}^{4}}+e^{2f}\sum_{i=1}^2(E^{i})^{2}\biggr)\biggr]\nonumber\\
F_4 = 3\ell^5\,h^{-2}\,e^{-4g-2f}\,\Sigma\,dr\wedge \epsilon_{\mathbb{R}^3}\ ,
\label{eq:backreactedABJMr}
\end{gather}
where $\Sigma(r)$ is defined in \eqref{eq:Sigma}. In terms of the $x$-coordinate \eqref{eq:abjmxcoordinate}, we get
\begin{equation}
    \Sigma(r) =e^{f(r)}\,\frac{x'(r)}{x(r)} \ ,
\end{equation}
which is perturbatively expanded in flavor as for the D3-D7 cases
\begin{equation}
    \Sigma_0(r) = \ell^2 \ , \quad \Sigma_1(r) = \begin{dcases}
        \frac{3}{35}\frac{r^3}{r_\text{q}^3}\frac{1}{k}\ ,\quad &r < r_\text{q}\\
        -\frac{3}{16}\biggl(1-\frac{8}{5}\frac{r_\text{q}^2}{r^2}+\frac{1}{7}\frac{r_\text{q}^4}{r^4} \biggr)\frac{1}{k}\ ,\quad &r \geq r_\text{q}
    \end{dcases} \ . 
\end{equation}
Hence we see that the no flavor limit of $F_4$ in \eqref{eq:backreactedABJMr} reduces to \eqref{eq:ABJMsolutionourconventions} since $\epsilon_{\text{AdS}_4} = r^2\,dr\wedge \epsilon_{\mathbb{R}^3}$.

Let us finish this appendix with a comment on regularity. In contrast to D3-brane backgrounds, the Type IIA description of the ABJMf geometry is completely regular for any masses \cite{Conde:2011sw,Bea:2013jxa}. In particular, there is no associated Landau pole, a fact that can be verified directly in the field theory \cite{Bianchi:2009ja,Bianchi:2009rf}.

\section{Details on entanglement entropy from backreacted geometries}\label{app:RTdetails}
In this appendix we present the detailed computation of the entanglement entropies in our backreacted backgrounds using Ryu-Takayanagi prescription. We will only present the results of the computation for the ABJM and D3-D7 geometries in the $\text{T}^{1,1}$ with chiral and non-chiral profiles, since for the D3-D7 geometry in the $\text{S}^5$, the computation has been done in \cite{Kontoudi:2013rla}.

\subsection{Entanglement entropy for the backreacted flavored D3-D7 for \texorpdfstring{chiral $\text{T}^{1,1}$}{}}\label{app:EET11}
The entanglement entropy functional obtained from \eqref{eq:entexp} reads:
\begin{equation}
\mathcal{S}=\mathcal{N}_S\,(\mathcal{S}^{(0)}+\Nf \mathcal{S}^{(1)}) \ ,
\label{eq:EEfunctional}
\end{equation}
where we identify $\mathcal{S}^{(0)}$ and $\mathcal{S}^{(1)}$, appearing in \eqref{eq:S0andS1functionals}, as
\begin{align}
\mathcal{S}^{(0)}&=
r\rho^2\sqrt{1+r^4\rho'^2}\label{eq:S0T11} \\
\mathcal{S}^{(1)}&=
r\rho^2\sqrt{1+r^4\rho'^2}\bigg(4g_1+f_1+\frac{1}{1+r^4\rho'^2}\Sigma_1+\frac{2g_{\text{s}} \Nc\pi^4\ell_s^4 r^4}{\vol{(X_5)}}\frac{2+r^4\rho'^2}{1+r^4\rho'^2}h_1\bigg)\ .
\label{eq:S1T11}
\end{align}
We perturbatively expand the embedding in flavor to order ${\mathcal{O}}(\Nf)$ as
\begin{eqnarray}
\rho=\rho_0(r)+\Nf \rho_1(r)=\sqrt{R^2-\frac{1}{r^2}}+\Nf \rho_1(r)\ .
\label{eq:embeddingexpansion}
\end{eqnarray}
The results for the contribution to the entropy to ${\mathcal{O}}(\Nf^0)$ are equal to the ones for the case of the $\text{S}^5$, since the effect of the internal space is only seen from the corrections in flavor. On-shell, we have
\begin{eqnarray}
\mathcal{S}^{(0)}=
R\sqrt{r^2R^2-1}\ .
\label{order0S}
\end{eqnarray}
Integrating this Lagrangian over the embedding, from the tip value $1/R$ to the UV cut-off $\epsilon$, yields
\begin{eqnarray}
\frac{S^{(0)}}{\mathcal{N}_S}=\int_{1/R}^{1/\epsilon}dr\, \mathcal{S}_0=
\frac{1}{2}\bigg[\frac{R^2}{\epsilon^2}-\log\frac{2R}{\epsilon}-\frac{1}{2}\bigg]\ .
\label{eq:order0Sintegrated}
\end{eqnarray}
Note that this expression is exact in the cut-off parameter or lattice spacing $\epsilon$, but subsequently it will appear in conjuction with subleading terms when the AdS isometries are broken with flavor corrections. Then we aim at keeping only the low order terms relevant for regulating the continuum limit.

We will now compute the boundary term $B_1$ defined in \eqref{eq:RTboundaryterms}, 
for which we must determine the correction to the embedding
\begin{equation}
B_1=\frac{\partial\mathcal{S}^{(0)}\left(\rho_0,\rho_0',r\right)}{\partial \rho_0'(r)} \rho_1(r)\bigg\lvert_{r=\frac{1}{\epsilon}}= \bigg(\frac{R^2}{\epsilon^2}-1\bigg)\rho_1\left(\frac{1}{\epsilon}\right)
\label{eq:B1T11}
\end{equation}

\paragraph{RT surface lying outside the cavity.}

We will determine $\rho_1$ by minimizing the entanglement entropy functional. The equation for the correction of the embedding outside the cavity is given by
\begin{align}
\rho_1^{-}{}''(r)+&\frac{5rR^2}{r^2R^2-1}\rho_1^{-}{}'(r)+\frac{2R^2}{(r^2R^2-1)^2}\rho_1^-(r)\nonumber -\frac{g_{\text{s}} \vol{(M_3)}}{16\vol{(X_5)}}\frac{1}{r^3(r^2R^2-1)^{3/2}}\nonumber\\
&\times\bigg(1-\frac{58}{49}\frac{\R^3}{r^3R^3}+2r^2R^2-\frac{80}{49}\frac{\R^3}{rR}+\frac{30}{7}\frac{\R^3}{r^3R^3}(1+\frac{4}{5}r^2R^2)\log \frac{\R}{rR} \bigg)=0 \ .
\end{align}
The solution to this equation contains two constants, one of which is fixed such that $\rho_1(r \rightarrow \infty)=0$. This requirement means that to have a meaningful comparison of flavor corrections we demand to compare the subsystems with the same radii $R$, {\emph{i.e.}}, boundary condition at the asymptotic boundary is the same but the embedding of the RT surface in the bulk alters. In particular, the tip value of the embedding is sensitive to the backreaction of flavor branes on to the geometry, the flavor degrees of freedom dictate how low in the scale of energy can the entanglement entropy probe the system. The other integration constant is fixed by the regularity of the quantity
\begin{equation}
    \rho_0(r)+\Nf \rho_1(r), \quad \text{when} \quad r \rightarrow 1/R+\Nf \ r_{*,1} \ ,
\end{equation}
{\emph{i.e.}}, the RT embedding smoothly caps off. After a straightforward albeit tedious computation the solution reads
\begin{align}
\rho_1^-(r)r&=\frac{g_{\text{s}}\vol{(M_3)}}{9408\vol{(X_5)}}\bigg[\frac{1323 \pi rR\R^3\log\frac{\R }{2}}{2\left(r^2R^2-1\right)^\frac{3}{2}}(-2r R+2r^3R^3-(2r^2R^2-1)\sqrt{r^2R^2-1}) \\
&\qquad+\frac{3(2r^2R^2-1)}{\sqrt{r^2R^2-1}}\bigg(196\bigg(1-\frac{29\R^3}{98r^3R^3}-r^2R^2-\frac{299\R^3}{196rR}\bigg)-210\frac{\R^3}{r^3R^3}\bigg(1+\frac{11}{10}r^2R^2\bigg)\nonumber \\
&\qquad\times\log\frac{rR}{\R}+\frac{21rR\R^3}{\sqrt{r^2R^2-1}}\bigg(9\arctan \sqrt{r^2R^2-1}-\mathrm{arccsc} (rR)\left(8+21\log\frac{rR}{\R}\right) \bigg)\bigg)\nonumber \\
&\qquad+\frac{1701}{8}\pi r R\R^3\frac{2r^2R^2-1}{R^2r^2-1}+\frac{294}{\sqrt{r^2R^2-1}}\bigg[1-\frac{40\R^3}{49r^3R^3}-6r^2R^2+\frac{80\R^3}{49rR}+4r^4R^4\nonumber\\
&\qquad+\frac{992}{49}rR\R^3-\frac{27}{8}\pi r^2R^2\R^3\bigg]-\frac{504\R^3}{\left(r^2R^2-1\right)^\frac{3}{2}r^3R^3}\left(1-3r^2R^2-6r^4R^4+8r^6R^6\right)\log \frac{\R}{rR}\nonumber\\
&\qquad -\frac{56\R^3\left(2r^2R^2-1\right)}{r^2R^2\left(r^2R^2-1\right)}\bigg(5 \  {}_3 F_2\left(\begin{matrix}-\frac{1}{2}, \frac{3}{2}, \frac{3}{2}\\\frac{5}{2}, \frac{5}{2}\end{matrix};\frac{1}{r^2R^2}\right)+18r^2R^2 {}_3 F_2\left(\begin{matrix} \frac{1}{2}, \frac{1}{2}, \frac{1}{2}\\\frac{3}{2}, \frac{3}{2}\end{matrix};\frac{1}{r^2R^2}\right) \bigg)\bigg]\ ,\nonumber
\end{align}
with
\begin{eqnarray}\label{eq:r1starT11}
r_{*,1} R=\frac{g_{\text{s}} \vol{(M_3)}}{32\vol{(X_5)}}\bigg[1+\frac{3}{392}\R^3\left(-2752+147\pi\left(3+4\log 2\right)\right)+\frac{9}{14}\left(24-7\pi\right)\R^3\log\left(\R\right) \bigg]\ .    
\end{eqnarray}
By substituting \eqref{eq:r1starT11} into the expression of $B_1$ \eqref{eq:B1T11} we get:
\begin{equation}
B_1=-\frac{g_{\text{s}} \vol{(M_3)}}{32\vol{(X_5)}} \  .
\end{equation}
Furthermore, $\mathcal{S}^{(1)}$ \eqref{eq:S1T11} reads:
\begin{equation} 
\mathcal{S}^{(1)}=\frac{g_{\text{s}}\vol{(M_3)}}{1568r^5R\vol{(X_5)}}\left(49r^3-40\frac{\R^3}{R^3}+84\frac{\R^3}{R^3}\log\frac{\R}{rR}\right)\sqrt{r^2R^2-1}\left(2r^2R^2-1\right)\ . 
\end{equation}
This expression can be integrated analytically
\begin{align}
\int_{1/R}^{1/\epsilon} dr \mathcal{S}^{(1)}&=\frac{27g_{\text{s}}\vol{(M_3)}}{32 \vol{(X_5)}}\bigg[ -\frac{4\R^3}{567r^3R^3}{}_3 F_2\left(\begin{matrix} -\frac{1}{2}, \frac{3}{2}, \frac{3}{2}\\\frac{5}{2}, \frac{5}{2}\end{matrix} ;\frac{1}{r^2R^2}\right)
+\frac{4\R^3}{63rR}F_2\left(\begin{matrix} \frac{1}{2}, \frac{1}{2}, \frac{1}{2}\\\frac{3}{2}, \frac{3}{2}\end{matrix};\frac{1}{r^2 R^2}\right)\nonumber\\
&\qquad\qquad- \frac{2}{27}\log\left(r R+\sqrt{r^2R^2-1}\right)+\sqrt{r^2R^2-1}\bigg(\frac{1}{27rR}-\frac{10\R^3}{1323r^4R^4}+\frac{rR}{27}\nonumber\\
&\qquad\qquad+\frac{29\R^3}{441r^2R^2}+\bigg(\frac{\R^3}{14r^2R^2}-\frac{\R^3}{63r^4R^4}\bigg)\log\frac{rR}{\R}\bigg)+\frac{\R^3}{378}\bigg[12\mathrm{arccsc}(rR)\nonumber\\
&\qquad\qquad+\mathrm{arccot}\sqrt{r^2R^2-1}\left(10-21\log\frac{\R}{rR}\right)\bigg] \bigg]\Bigg|_{1/R}^{1/\epsilon} \ .
\label{eq:S1integralT11}
\end{align}
Evaluating the expression above in the given integration limits, expanding for small $\epsilon$ and adding the remaining contributions $S^{(0)}$ \eqref{eq:order0Sintegrated}, $B_1,$ \eqref{eq:B1T11}, and $B_2=0$ we arrive at
\begin{equation}
\frac{S}{\mathcal{N}_S}=\bigg[\frac{R^2}{2\epsilon^2}+\frac{1}{2}\log\frac{\epsilon}{2R}-\frac{1}{4}+\Nf\frac{g_{\text{s}} \vol{(M_3)}}{4\vol{(X_5)}}\bigg( \frac{R^2}{8\epsilon^2}+\frac{1}{4}\log \frac{\epsilon}{2R}+\frac{3}{32}\pi \R^3\log \frac{\R  }{2}-\frac{13}{128}\pi \R^3-\frac{1}{16}\bigg) \bigg]\ .
\end{equation}

\paragraph{RT surface partially penetrating the cavity.}

To obtain the entropy in this configuration, we must first determine the correction to the embedding. We now have two distinct regions, the interior and the exterior of the  cavity. Therefore, we must solve the differential equation in both regions. In the process we obtain four constants of integration. One of them is fixed so that $\rho_1(r \rightarrow \infty)=0$. Two of the remaining ones are fixed by requiring that $\rho^+(r_\text{q})=\rho^-(r_\text{q})$ and $\rho^+{}'(r_\text{q})=\rho^-{}'(r_\text{q})$, that is, the embedding is continuous and smooth (without kinks) across the cavity. The leftover constant is determined together with $r_{*,1}$ so that $\rho_{0}+\Nf\rho_1$ is finite when $r\rightarrow 1/R+\Nf \ r_{*,1}$. The equation to solve inside the cavity ($+$) is
\begin{eqnarray}
\rho_1^+{}''(r)+\frac{5rR^2}{r^2R^2-1}\rho_1^+{}'(r)+\frac{2R^2}{(r^2R^2-1)^2}\rho_1^+(r)-\frac{9g_{\text{s}}  \vol{(M_3)}r(2r^2R^2-1)}{784r_\text{q}^4\vol{(X_5)}(r^2R^2-1)^{\frac{3}{2}}}=0 \ .
\end{eqnarray}
The solution to the equation inside the cavity gives
\begin{eqnarray}
\rho_1^+(r)& = & \frac{g_{\text{s}} \vol{(M_3)}rR^2}{\vol{(X_5)}\R^2}\bigg[-\frac{5\R^2}{84} {}_3 F_2\left(\begin{matrix}-\frac{1}{2}, \frac{3}{2}, \frac{3}{2}\\\frac{5}{2}, \frac{5}{2}\end{matrix};\frac{1}{\R^2}\right)-\frac{3}{14}\R^4 {}_3 F_2\left(\begin{matrix}\frac{1}{2}, \frac{1}{2}, \frac{1}{2}\\\frac{3}{2}, \frac{3}{2}\end{matrix};\frac{1}{\R^2}\right)\nonumber\\
& & +\frac{3r^2R^2}{1568\R^2}-\frac{\R^2}{8}+\frac{3}{4}\R^4-\frac{51}{448}\pi \R^5-\frac{3}{224}\R (-6+23\R^2)\sqrt{\R^2-1}\nonumber\\
& & +\frac{3}{224}\R^5[8\arcsec (\R)+9\arctan \sqrt{\R^2-1}] \bigg] \ .
\end{eqnarray}
The equation determining the outer part of the embedding ($-$) coincides with the one derived for the non-penetrating embeddings, although the solution differs by the choice of boundary conditions, and it is given by
\begin{align}
    \rho_1^-(r)&=\frac{g_{\text{s}} \vol{(M_3)}R}{\vol{(X_5)}\R}\bigg[\frac{3\sqrt{\R^2-1}(23\R^2-6)[2rR(rR-\sqrt{r^2R^2-1})-1]}{448(r^2R^2-1)}-\frac{21\R^4(2r^2R^2-1)}{3136(r^2R^2-1)}\nonumber\\
    &\qquad\times\bigg(9\arctan\sqrt{\R^2-1}-8\arcsin\frac{1}{\R  }-9\arctan\sqrt{r^2R^2-1}+\arccsc (rR)(8+21\log\frac{rR}{\R})\bigg)\nonumber\\
    &\qquad-\frac{\R  }{168}\bigg(\frac{2rR}{\sqrt{r^2R^2-1}}-\frac{2r^2R^2-1}{r^2R^2-1}\bigg)\bigg(5 {}_3 F_2\left(\begin{matrix}-\frac{1}{2}, \frac{3}{2}, \frac{3}{2}\\\frac{5}{2}, \frac{5}{2}\end{matrix};\frac{1}{\R^2}\right)+18 \R^2 {}_3 F_2\left(\begin{matrix}\frac{1}{2}, \frac{1}{2}, \frac{1}{2}\\\frac{3}{2}, \frac{3}{2}\end{matrix};\frac{1}{\R^2}\right) \bigg)\nonumber\\
    &\qquad-\frac{\R^4(2r^2R^2-1)}{168r^3R^3(r^2R^2-1)}\bigg(5{}_3 F_2\left(\begin{matrix}-\frac{1}{2}, \frac{3}{2}, \frac{3}{2}\\\frac{5}{2}, \frac{5}{2}\end{matrix};\frac{1}{r^2R^2}\right)+ 18r^2R^2{}_3 F_2\left(\begin{matrix}\frac{1}{2}, \frac{1}{2}, \frac{1}{2}\\\frac{3}{2}, \frac{3}{2}\end{matrix};\frac{1}{r^2R^2}\right) \bigg)\nonumber\\
    &\qquad+\frac{1}{3136\sqrt{r^2R^2-1}}\bigg(\frac{21\R^4}{r^4R^4}\bigg(2r^5R^5\left(9\arctan\sqrt{\R^2-1}-8\arccsc(\R)\right)\nonumber\\
    &\qquad+\log\frac{rR}{\R}(2+7r^2R^2+42r^4R^4)\bigg)-\frac{22\R^4}{r^4R^4}+\frac{343\R^4}{r^2R^2}-\frac{98\R}{rR}\bigg)\bigg] 
\end{align}
with
\begin{align}
r_{*,1}R &= \frac{5g_{\text{s}} \vol{(M_3)}}{84\vol{(X_5)}}\bigg[{}_3 F_2\left(\begin{matrix}-\frac{1}{2}, \frac{3}{2}, \frac{3}{2}\\\frac{5}{2}, \frac{5}{2}\end{matrix};\frac{1}{\R^2}\right)+\frac{18}{5}\R^2 {}_3 F_2\left(\begin{matrix}\frac{1}{2}, \frac{1}{2}, \frac{1}{2}\\\frac{3}{2}, \frac{3}{2}\end{matrix};\frac{1}{\R^2}\right)\nonumber\\
&\qquad\qquad+\frac{9}{40\R}\sqrt{\R^2-1}(23\R^2-6)-\frac{9}{280\R^4}+\frac{21}{10}-\frac{63\R^2}{5}\nonumber\\
&\qquad\qquad+\frac{9}{80}\pi \R^3[16\arccsc(\R  )+9\pi-18\arctan \sqrt{\R^2-1}] \bigg] \ . 
\end{align}

The contribution ${\cal{O}}(\Nf^0)$ to the entropy is the same as before. To get the contribution $\int dr \mathcal{S}^{(1)}$ we have to split the integral in two regions: the exterior ($-$) and the interior ($+$) of the cavity. $B_1$ \eqref{eq:B1T11} is computed as in the case before, since it involves the geometry close to $r=1/\epsilon$. We have already integrated $\mathcal{S}^{(1)}$ outside the cavity. To get the needed contribution from the interior part of the cavity, we must only evaluate \eqref{eq:S1integralT11} with the limits of integration $(r_\text{q},1/\epsilon)$. Namely, our integrand is
\begin{equation}
\mathcal{S}_+^{(1)}=\frac{9g_{\text{s}}r^2R^3\vol{(M_3)}}{1568\R^4\vol{(X_5)}}(r^2R^2-1)^\frac{1}{2}(2r^2R^2-1) \ ,
\end{equation}
which integrates to:
\begin{equation}
\int_{1/R}^{r_\text{\tiny{q}}} dr \mathcal{S}_+^{(1)}=\frac{3g_{\text{s}}  \vol{(M_3)}r^3}{1568\R  \vol{(X_5)}}(r^2R^2-1)^\frac{3}{2}\Bigg\lvert_{1/R}^{r_\text{\tiny{q}}}\ .
\end{equation}
To get the final result, we must evaluate the already solved integrals in the given limits of integration, add all the contributions and expand for small $\epsilon$, which gives
\bea
S & = & \mathcal{N}_S \bigg[\frac{R^2}{2\epsilon^2}-\frac{1}{2}\log\frac{2R}{\epsilon}-\frac{1}{4}+\frac{\Nf g_{\text{s}} \vol{(M_3)}}{4\vol{(X_5)}}\bigg(\frac{R^2}{8\epsilon^2}+\frac{1}{4}\log \frac{\epsilon (\R  +\sqrt{\R^2-1})}{2R}-\frac{1}{16}\nonumber \\
& & \qquad -\sqrt{\R^2-1}\big(\frac{3}{28\R  }+\frac{19\R  }{56}\big)+\frac{1}{42}{}_3 F_2\left(\begin{matrix} -\frac{1}{2}, \frac{3}{2}, \frac{3}{2}\\\frac{5}{2}, \frac{5}{2}\end{matrix} ;\frac{1}{\R^2}\right)\nonumber \\
& & \qquad -\frac{3}{14}\R^2F_2\left(\begin{matrix} \frac{1}{2}, \frac{1}{2}, \frac{1}{2}\\\frac{3}{2}, \frac{3}{2}\end{matrix};\frac{1}{\R^2}\right)-\frac{11}{56}\R^3 \mathrm{arccsc}\R\bigg) \bigg]\ . 
\eea

\subsection{Entanglement entropy for the backreacted flavored D3-D7 for non-chiral \texorpdfstring{$\text{T}^{1,1}$}{}}\label{app:EET11NC}

The expressions \eqref{eq:EEfunctional}, \eqref{eq:S0T11}, and \eqref{eq:S1T11} also hold for the non-chiral profile. We can analogously expand the embedding as in \eqref{eq:embeddingexpansion}. The contribution to the entropy to order zero in flavor is also given by \eqref{order0S}, which after integration and expanding for small values of the cut-off gives \eqref{eq:order0Sintegrated}. The boundary term is given by \eqref{eq:B1T11}, for which we need the correction to the embedding. Let us determine such corrections.

\paragraph{RT surface lying outside the cavity.}\label{paragraphapp:RToutsideT11}

We will determine $\rho_1$ by minimizing the entanglement entropy functional. The equation to solve is:
\begin{align}
&\rho_1^{-}{}''(r)+\frac{5rR^2}{r^2R^2-1}\rho_1^{-}{}'(r)+\frac{2R^2}{(r^2R^2-1)^2}\rho_1^-(r)\nonumber \\ &\qquad\qquad-\frac{g_{\text{s}} \vol{(M_3)}}{16\vol{(X_5)}}\frac{1}{r^3(r^2R^2-1)^\frac{3}{2}}\bigg(1-\frac{10\R^3}{7r^3R^3}+2r^2R^2\bigg(1-\frac{4\R^3}{7r^3R^3}\bigg) \bigg)=0  \ .
\label{outer}
\end{align}
Imposing $\rho_1(r \rightarrow \infty)=0$ and regularity of $\rho_0(r)+\Nf \rho_1(r)$ at $r \rightarrow 1/R+\Nf \ r_{*,1}$ we can fix the two constants of integration of the equation and also determine the value of the correction $r_{*,1}$,
\begin{align}
\rho_1^-(r) &=\frac{g_{\text{s}} \vol{(M_3)}\R^3 R}{32(r^2R^2-1)^\frac{3}{2}\vol{(X_5)}}\bigg[-\frac{1}{7r^4R^4}-\frac{5}{14r^2R^2}+\frac{1}{rR\R^3}-\frac{5}{2}+rR\left(\frac{3\pi}{2}-\frac{1}{\R^3}\right)+3r^2R^2\nonumber\\
&\qquad\qquad -3(2r^2R^2-1)\sqrt{r^2R^2-1}\arctan (rR-\sqrt{r^2R^2-1})-\frac{3}{2}\pi r^3R^3\nonumber\\
&\qquad\qquad+\frac{3}{4}\pi\sqrt{r^2R^2-1}(2r^2R^2-1)\bigg] 
\end{align}
with
\begin{eqnarray}
r_{*,1}=\frac{g_{\text{s}} \vol{(M_3)}}{32\vol{(X_5)}R}\bigg[1+\frac{3}{14}(7\pi-24)\R^3 \bigg] \ .  
\end{eqnarray}
We find \eqref{eq:B1T11} becomes
\begin{equation}
B_1=-\frac{g_{\text{s}}\vol{(M_3)}}{32\vol{(X_5)}} \ .
\end{equation}
This produces the following integrand for the entanglement entropy correction $\mathcal{S}^{(1)}$
\begin{eqnarray}
\mathcal{S}^{(1)}=\frac{g_{\text{s}}\vol{(M_3)}}{224\vol{(X_5)}r^5R}\left(7r^3-4\frac{\R^3}{R^3}\right)\sqrt{r^2R^2-1}(2r^2R^2-1) \ .
\end{eqnarray}
A straightforward integration gives
\begin{align}\label{eq:S1integralT11-v2}
\int_{1/R}^{1/\epsilon} dr \mathcal{S}^{(1)}&=\frac{g_{\text{s}}\vol{(M_3)}r}{448\vol{(X_5)}}\bigg[ \sqrt{r^2R^2-1}\left(14-2\frac{\R^3}{r^3R^3}+14r^2R^2+9\frac{\R^3}{rR}\right) \\& \qquad+14rR\R^3\arctan (rR-\sqrt{r^2R^2-1})+28rR\log(rR-\sqrt{r^2R^2-1}) \bigg]\Bigg\lvert_{1/R}^{1/\epsilon}\ .\nonumber
\end{align}
Evaluating the expression above in the given integration limits, expanding for small $\epsilon$ and adding the remaining contributions $S^{(0)}$ and $B_1$ we finally find
\begin{equation}
S=\mathcal{N}_S\bigg[\frac{R^2}{2\epsilon^2}+\frac{1}{2}\log\frac{\epsilon}{2R}-\frac{1}{4}+\Nf\frac{g_{\text{s}} \vol{(M_3)}}{4\vol{(X_5)}}\bigg( \frac{R^2}{8\epsilon^2}+\frac{1}{4}\log \frac{\epsilon}{2R}-\frac{\pi}{32}\R^3-\frac{1}{16}\bigg) \bigg]\ . 
\end{equation}

\paragraph{RT surface partially penetrating the cavity.}

We will proceed as in the chiral-profile case, solving the embedding in the inner and outer region and fixing the constants using $\rho_1(r \rightarrow \infty)=0$, continuity of the embedding and its derivative across the cavity, and jointly fixing the remaining constant together with $r_{*,1}$ so that $\rho_{0}+\Nf\rho_1$ is finite when $r\rightarrow \frac{1}{R}+\Nf \ r_{*,1}$. Inside the cavity, we have to solve the following equation
\begin{equation}
\rho_1^+{}''(r)+\frac{5rR^2}{r^2R^2-1}\rho_1^+{}'(r)+\frac{2R^2}{(r^2R^2-1)^2}\rho_1^+(r)-\frac{3g_{\text{s}}  \vol{(M_3)}rR^4(2r^2R^2-1)}{112\R^4\vol{(X_5)}(r^2R^2-1)^{\frac{3}{2}}}=0 \ .
\end{equation}
Solving also \eqref{outer} for the outer embeddings ($-$) with the appropriate boundary conditions, we get
\begin{align}
\rho_1^+(r) &=\frac{g_{\text{s}} \vol{(M_3)}}{224\vol{(X_5)}r\sqrt{r^2R^2-1}}\bigg[\frac{r^4R^4}{\R^4}+\frac{7r^2R^2}{\R}(2-5\R^2)\sqrt{\R^2-1}+28r^2 R^2(2\R^2-1) \nonumber \\
&\qquad\qquad\qquad\qquad-42r^2R^2\R^3\arctan(\R -\sqrt{\R^2-1})\bigg]\\ 
\rho_1^-(r)&=\frac{g_{\text{s}} \vol{(M_3)}R(\R^2-1)}{448\vol{(X_5)}\R[\R(\R  -\sqrt{\R^2-1})-1]}\bigg[\frac{14rR(-2+5\R^2)}{\sqrt{r^2R^2-1}\sqrt{\R^2-1}}+\frac{1}{\sqrt{r^2R^2-1}}\nonumber\\
&\qquad\times\bigg(1-\frac{\R  }{\sqrt{\R^2-1}}\bigg)\bigg(-\frac{14\R}{rR}+2\frac{\R^4}{r^4R^4}-28rR\R+\frac{7\R^4}{r^2R^2}+70r R \R^3+42\R^4\bigg)\nonumber\\
&\qquad +\frac{7\R(2r^2R^2-1)(5\R^2-2)}{r^2R^2-1}-\frac{42\R^4(2r^2R^2-1)}{r^2R^2-1}\arctan(rR-\sqrt{r^2R^2-1})\nonumber\\
&\qquad+42\R^4\bigg(\frac{2r^2R^2-1}{r^2R^2-1}-\frac{2rR}{\sqrt{r^2R^2-1}}\bigg)\arctan(\R  -\sqrt{\R^2-1})\bigg(1-\frac{\R  }{\sqrt{\R^2-1}}\bigg)\nonumber\\
&\qquad +\frac{7(2r^2R^2-1)}{\sqrt{\R^2-1}}\frac{2-7\R^2+5\R^4+6\R^5\arctan(rR-\sqrt{r^2R^2-1})}{r^2R^2-1} \bigg] \ ,
\end{align}
respectively. We find
\begin{align}
r_{*,1}R &= \frac{5g_{\text{s}} \vol{(M_3)}}{224\vol{(X_5)}\R^4(\R^2-\R  \sqrt{\R^2-1}-1)}\bigg[1-\R^2-42\R^4+133\R^6-91\R^8\nonumber\\ 
&\qquad+42\R^7(-1+\R^2-\R  \sqrt{\R^2-1})\arctan(\R  -\sqrt{\R^2-1})\nonumber\\
&\qquad+\R  \sqrt{\R^2-1}(1+14\R^2-77\R^4+91\R^6) \bigg] \ .
\end{align}
As it happened in the chiral case, the contribution ${\mathcal{O}}(\Nf^0)$ is the same as before. Splitting of the $\int dr \mathcal{S}^{(1)}$ is needed, proceeding in an analogous way to the previous case (\ref{paragraphapp:RToutsideT11}), so we will omit  the intermediate details and just present the results. Inside the cavity we find
\begin{equation}
\mathcal{S}_+^{(1)}=\frac{3g_{\text{s}}r^2R^3\vol{(M_3)}}{224\R^4\vol{(X_5)}}(r^2R^2-1)^\frac{1}{2}(2r^2R^2-1)\ , 
\end{equation}
and integrating this result yields
\begin{equation}
\int_{1/R}^{r_\text{\tiny{q}}} dr\,  \mathcal{S}_+^{(1)}=\frac{g_{\text{s}}  \vol{(M_3)}r^3}{224\R  \vol{(X_5)}}(r^2R^2-1)^\frac{3}{2}\Bigg\lvert_{1/R}^{r_\text{\tiny{q}}}\ . 
\end{equation}
Substituting the endpoints of integration, adding all the contributions and expanding for small $\epsilon$, we find
\begin{align}
S&=\mathcal{N}_S \bigg[\frac{R^2}{2\epsilon^2}-\frac{1}{2}\log\frac{2R}{\epsilon}-\frac{1}{4}+\frac{\Nf g_{\text{s}} \vol{(M_3)}}{4\vol{(X_5)}}\bigg(\frac{R^2}{8\epsilon^2}+\frac{1}{4}\log \frac{\epsilon (\R  +\sqrt{\R^2-1})}{2R}-\frac{1}{16}\nonumber \\
&\qquad\qquad-\frac{1}{16\R  }\sqrt{\R^2-1}(2+3\R^2)-\frac{\R^3}{8}\arctan(\R  -\sqrt{\R^2-1})\bigg) \bigg]\ . 
\end{align}

\subsection{Entanglement entropy for the backreacted flavored ABJM} \label{app:EEABJM}

Doing the same decomposition as in \eqref{eq:EEfunctional}, we obtain:
\begin{align}
\mathcal{S}^{(0)}&=\frac{1}{2}\rho\sqrt{1+r^4\rho'^2} \ , \label{eq:S0ABJM} \\
\mathcal{S}^{(1)}&=\frac{1}{2}\rho\sqrt{1+r^4\rho'^2}\bigg(4g_1+2f_1-2\phi_1+\frac{1}{1+r^4\rho'^2}\Sigma_1+\frac{r^4\ell^4(3+2r^4\rho'^2)}{2(1+r^4\rho'^2)}h_1\bigg)\ .\label{eq:S1ABJM}
\end{align}
The expansion of the embedding is also given by \eqref{eq:embeddingexpansion}. The results for the contribution to the entropy to order zero in flavor are equal to the ones for the case of the $AdS_4\times\mathbb{C}\text{P}^3$,  since the effect of the internal space is only seen from the corrections in flavor. To order $\mathcal{O}(\Nf^0)$ we have, on-shell 
\be
\mathcal{S}^{(0)}=\frac{R}{2}\ .
\ee
Integrating this result leads to
\begin{equation}
\frac{S^{(0)}}{\mathcal{N}_S}=\int_{1/R}^{1/\epsilon}dr\, \mathcal{S}^{(0)}=\frac{1}{2}\bigg(\frac{R}{\epsilon}-1\bigg) \ . 
\end{equation}
The boundary term $B_1$ defined in \eqref{eq:RTboundaryterms} is
\begin{equation}
B_1=\frac{1}{2R}\sqrt{\frac{R^2}{\epsilon^2}-1}\,\rho_1 \bigg(\frac{1}{\epsilon}\bigg)\label{eq:B1ABJM}
\end{equation}

\paragraph{RT surface lying outside the cavity.} Extremization of the entanglement entropy functional leads to the equation:
\bea
& & \rho_1^{-}{}''(r)+\frac{4rR^2}{r^2R^2-1}\rho_1^{-}{}'(r)+\frac{R^2}{(r^2R^2-1)^2}\rho_1^-(r) \\
&& \qquad+\frac{1}{kr^3(r^2R^2-1)^{3/2}}\bigg(\frac{3}{8}(r^2R^2+1)+\frac{3}{10}(-4+3r^2R^2)\frac{\R^2}{r^2R^2}+\frac{9}{280}(5-3r^2R^2)\frac{\R^4}{r^4R^4}\bigg)=0 \ . \nonumber
\eea
Imposing analogous boundary and regularity conditions as in the D3-D7 geometries, we obtain
\begin{equation}
\rho_1^-(r)r=\frac{3}{16k\sqrt{r^2R^2-1}}\bigg[1+\frac{4}{3}\R^2-\frac{8\R^2}{15r^2R^2}-\frac{8}{3}rR\R^2+\frac{\R^4}{35r^4R^4}-\frac{8}{3}r^2R^2\R^2\log\frac{r R}{1+rR}\bigg] \ ,
\end{equation}
with
\begin{equation}\label{eq:r1starABJM}
r_{*,1} R=\frac{3}{16k}\bigg[-1+\frac{28}{15}\R^2\bigg(1-\frac{10}{7}\log 2\bigg) -\frac{\R^4}{35}\bigg] \ .
\end{equation}
Using this solution in the expression \eqref{eq:B1ABJM} we find, in this case, $B_1 = 0$. Further, the expression $\mathcal{S}^{(1)}$ \eqref{eq:S1ABJM} reads
\begin{equation}
\mathcal{S}^{(1)}=\frac{3}{1120k R r^2}\bigg[35(1+5r^2R^2)-28\frac{\R^2}{r^2R^2}(2+r^2R^2)+(5-3r^2R^2)\frac{\R^4}{r^4R^4}\bigg]\ .
\end{equation}
After the integration, we are left with
\begin{equation}
\int_{1/R}^{1/\epsilon} dr \mathcal{S}^{(1)}=\frac{1}{1120krR}\bigg[525r^2R^2-3\frac{\R^4}{r^4R^4}+\frac{\R^2}{r^2R^2}(56+3\R^2)-21(5-4\R^2) \bigg]\Bigg|_{1/R}^{1/\epsilon} \ .
\label{eq:S1integralABJM}
\end{equation}
By explicitly evaluating this expression, imposing the cut-off, and adding the remaining contributions $S^{(0)}$ and $ B_1$, we obtain
\begin{equation}
\frac{S}{\mathcal{N}_S}=\frac{1}{2}\bigg[\frac{R}{\epsilon}-1+\frac{\Nf}{k}\bigg( \frac{15R}{16\epsilon}-\frac{3}{4}-\frac{1}{4}\R^2\bigg) \bigg] \ .
\end{equation}

\paragraph{RT surface partially penetrating the cavity.} 

We will proceed as in the D3-D7 case, solving for the flavor contribution that affects the inner and outer embeddings. We impose  $\rho_1(r \rightarrow \infty)=0$, the same smoothness condition across the cavity, and the regularity of the embedding at $r\rightarrow \frac{1}{R}+\Nf \ r_{*,1}$. Inside the cavity, we must solve
\begin{equation}
\rho_1^+{}''(r)+\frac{4rR^2}{r^2R^2-1}\rho_1^+{}'(r)+\frac{R^2}{(r^2R^2-1)^2}\rho_1^+(r)+\frac{(3+21\R^2-6r^2R^2)R^3}{35k\R^3(r^2R^2-1)^{3/2}}=0 \ .
\end{equation}
The wanted solutions are
\begin{align}
\rho_1^+(r) =\frac{3R}{10k\R\sqrt{r^2R^2-1}}\bigg[1+\frac{r^2R^2}{7\R^2}-\frac{5}{3}rR\R-&\frac{5}{6}rR\R^3\log\frac{\R^2}{\R^2-1}+\frac{5}{4} rR\R^2\log\frac{\R  +1}{\R  -1}\nonumber \\&-\frac{5}{12}rR\log\frac{\R  +1}{\R  -1}\bigg] \ , 
\end{align}
and
\begin{align}\label{eq:embedding-inside-ABJM}
\rho_1^-(r) &=\frac{R}{4k\R\sqrt{r^2R^2-1}}\bigg[1+\frac{3\R}{4rR}-3\R^2-\frac{2\R^3}{5r^3R^3}+\frac{\R^3}{rR}+\frac{3\R^5}{140r^5R^5}-\frac{rR}{2}\log\frac{rR+1}{rR-1}\nonumber \\
&\qquad\qquad \qquad+\frac{3rR\R^2}{2}\log\frac{rR+1}{rR-1}-rR\R^3\log\frac{r^2R^2}{r^2R^2-1}\bigg] \ .
\end{align}
Thus,
\begin{align}\label{eq:r1starABJMinside}
r_{*,1}R =\frac{3}{70k \R^3}\bigg[-1-7\R^2+\frac{35}{3}\R^3+&\bigg(\frac{35}{12}\R^2-\frac{35}{4}\R^4\bigg)\log\frac{\R  +1}{\R  -1}+\frac{35}{6}\R^5\log\frac{\R^2}{\R^2-1} \bigg] \ .
\end{align}
The contribution of order $\mathcal{O}(\Nf^0)$ to the entropy is the same as in the non-penetrating configurations. $\int dr\, \mathcal{S}^{(1)}$ must be split into the interior and exterior regimes of the cavity. The boundary term $B_1$ \eqref{eq:B1ABJM} is computed as in the case before, since it involves the geometry close to $r=1/\epsilon$ and it vanishes. Evaluating \eqref{eq:S1integralABJM} with the limits of integration corresponding to this embedding $(r_\text{q},1/\epsilon)$, we get the contribution $\mathcal{S}^{(1)}_-$ outside the cavity. Inside the cavity we find
\begin{equation}
 \mathcal{S}^{(1)}_+=\frac{3rR^2}{70k\R^3}[-1+7\R^2+2r^2R^2]
\end{equation}
which produces
\begin{equation}
\int_{1/R}^{r_\text{\tiny{q}}} dr\,  \mathcal{S}^{(1)}_+=\frac{3r^2R^2}{140k\R^3}[-1+7\R^2+r^2R^2]\bigg|_{1/R}^{r_\text{\tiny{q}}} \ .
\end{equation}
Evaluating the integrals in the given endpoints of integration, imposing the cut-off and gathering all the contributions, we finally obtain
\begin{equation}
S=\frac{\mathcal{N}_S}{2} \bigg[\frac{R}{\epsilon}-1+\frac{\Nf}{k}\bigg(\frac{15R}{16\epsilon}-\frac{1}{4\R  }-\frac{3}{4}\R  \bigg) \bigg]\ .
\end{equation}

\section{Computation of sphere free energies}\label{app:spherefreenenergy}

In this appendix, we compute sphere free energies of the flavored CFTs using the gravity dual. By conformal symmetry, the sphere free energy is equal to the vacuum entanglement entropy of a spherical subregion in flat space.

\subsection{SYM free energy}\label{app:spherefreenenergySYM}

To compute the bulk on-shell action, we will use the five-dimensional effective action of type IIB supergravity on the Euclidean AdS$ _5 $ background. The reason for this is that on-shell the ten-dimensional supergravity action vanishes and the prescription for obtaining the correct value is still an open problem (see~\cite{Kurlyand:2022vzv,Mkrtchyan:2022xrm,Chakrabarti:2022jcb} for recent advances).

The effective action in five dimensions is obtained by dimensionally reducing ten-dimensional equations of motion and finding the five-dimensional action that produces them. For the D3-brane solution, the five-dimensional Euclidean on-shell action is simply Einstein gravity with a five-dimensional cosmological constant:
\begin{equation}
F_{\text{S}^{4}}^{(0)} = -\frac{1}{2\kappa_{5}^{2}}\int d^{5}x\sqrt{g_5}\,\biggl(R_5 + \frac{4\cdot 3}{\ell^{2}}\biggr)\ ,
\end{equation}
where the five-dimensional Newton's constant is related to the ten-dimensional one as
\begin{equation}
\frac{1}{2\kappa_{5}^{2}} = \frac{\ell^{5}}{2\kappa_{10}^{2}}\vol{(X_5)}\ .
\end{equation}
For the AdS$ _5 $, we get
\begin{equation}
F_{\text{S}^{4}}^{(0)} = \frac{1}{16\pi^{7}g_{\text{s}}^{2}}\,\biggl(\frac{\ell}{\ell_{\text{s}}}\biggr)^{8}\vol{(X_5)}\vol{(\text{AdS}_5)} \ .
\end{equation}
Substituting to the free energy gives
\begin{equation}
F_{\text{S}^{4}}^{(0)} = \frac{\pi N^{2}}{\vol{(X_5)}}\vol{(\text{AdS}_{5})} \ .
\end{equation}
The correction to the free energy on the sphere is computed by the on-shell D7-brane DBI action for the massless embedding that extends along all AdS directions and wraps an $ M_3 = \text{S}^3\subset \text{S}^5 $:
\begin{equation}
F_{\text{S}^{4}}^{(1)} = T_{\text{D7}}\,\ell^{8}\vol{(M_{3})}\vol{(\text{AdS}_{5})} = \frac{1}{(2\pi)^{7}g_{\text{s}}}\biggl(\frac{\ell}{\ell_{\text{s}}}\biggr)^{8} \vol{(M_{3})}\vol{(\text{AdS}_{5})} \ .
\end{equation}
Using \eqref{eq:D3parameters} gives
\begin{equation}
F_{\text{S}^{4}}^{(1)} = \frac{g_{\text{YM}}^{2}N^{2}}{16}\frac{\vol{(M_{3})}}{[\vol({X_5})]^{2}}\vol{(\text{AdS}_{5})} \ .
\end{equation}

\subsection{ABJM free energy}\label{app:spherefreenenergyABJM}

We will now compute the bulk Euclidean on-shell action of the ABJM solution by using the four-dimensional effective action of M-theory. First we use the fact that the non-backreacted ABJM solution \eqref{eq:nobackreaction} with \eqref{eq:ABJMparameters} has an uplift to 11-dimensions where it becomes the M2-brane solution on an orbifold singularity (see \cite{Aharony:2008ug} and Appendix \ref{app:ABJMdetails})
\begin{equation}
\label{eq:Mtheorysol}
ds^{2}_{11} = L^{2}\bigl(ds^{2}_{\text{AdS}_{4}} + 4ds^{2}_{\text{S}^{7}\slash\, \mathbb{Z}_k}\bigr) \ , \quad \ 
F_4 = 3 L^{3}\epsilon_{\text{AdS}_4} \ .
\end{equation}
where $L$ is given in \eqref{eq:11Dabjm} and it is related to the 10-dimensional curvature radius as $ \ell^3 = g_{\text{s}} L^3 $. We can reduce this over $\text{S}^{7}\slash \mathbb{Z}_k$ all the way to four dimensions. The free energy becomes \cite{Drukker:2010nc}
\begin{equation}
F_{\text{S}^{3}}^{(0)} = -\frac{1}{2\kappa_{4}^{2}}\int d^{4}x\sqrt{g_4}\,\biggl(R_4 + \frac{3\cdot 2}{L^{2}}\biggr) \ .
\end{equation}
The four-dimensional Newton's constant is given by
\begin{equation}
\frac{1}{2\kappa_{4}^{2}} = \frac{L^{7}}{2\kappa_{11}^{2}}\vol{(\text{S}^{7}\slash\,\mathbb{Z}_k)} \ ,
\end{equation}
where the 11-dimensional Newton's constant in these conventions is given by $ 2\kappa_{11}^{2} = (2\pi)^{8}\ell_{\text{s}}^{9} $ (see Appendix \ref{app:ABJMdetails}). The free energy becomes
\begin{equation}
F_{\text{S}^{3}}^{(0)} = \frac{3}{\pi^{8}}\,\biggl(\frac{L}{\ell_{\text{s}}}\biggr)^{9}\vol{(\text{S}^{7}\slash\,\mathbb{Z}_k)}\vol{(\text{AdS}_4)} \ .
\label{eq:Mtheoryint}
\end{equation}
Let us then consider the massless flavor contribution to the sphere free energy. The massless D6-brane extends along all AdS$_4$ directions and wraps an $M_3 = \mathbb{R}\text{P}^3 \subset \mathbb{C}\text{P}^3 $. Using \eqref{eq:NDp} for the coefficient $\mathcal{N}_{\text{D}p}$, setting $\xi = \pi\slash 2$ in the on-shell regularized D6-brane action \eqref{eq:Dpactiontext2} with the Lagrangian \eqref{eq:D6lagrangiantext} gives simply the regularized volume of AdS$_4$ up to an overall factor. Renormalizing amounts to replacing the regularized volume with the renormalized one. The resulting free energy is
\begin{equation}
F_{\text{S}^{3}}^{(1)} =  \frac{3\slash 2}{(2\pi)^{6}g_{\text{s}}}\biggl(\frac{\ell}{\ell_{\text{s}}}\biggr)^{7} \vol{(M_{3})}\vol{(\text{AdS}_{4})} \ ,
\end{equation}
where in $ 3\slash 2 = 1 + 1\slash 2 $ the $ 1\slash 2 $ comes from the on-shell WZ action while $1$ comes from the DBI part. Using $ \vol{(M_3)} =2^3\vol{(\mathbb{R}\text{P}^{3})} $ gives the formula in \eqref{eq:ABJMFs}.

\section{Probe details}\label{app:probedetails}

In this appendix, we give details for the calculation of the entanglement entropy of a spherical subregion using the probe brane method that does not require the knowledge of the backreacted bulk metric.

\subsection{Entropy from probe brane actions}\label{subapp:bannondetails}

At $ n=1 $, the hyperbolic black hole metric \eqref{eq:hypbh} is simply $ \text{AdS}_{d+1} $ in Rindler coordinates $(\tau,\zeta,v,\Omega)$ \eqref{eq:rindlerads} which are related to Poincar\'e coordinates $(r,t_E,\rho,\Omega)$ via
\begin{gather}
r = \frac{1}{R}\left(\zeta\cosh{v} + \sqrt{\zeta^{2}-1}\,\cos{\tau} \right), \quad t_E =  R\,\frac{ \sqrt{\zeta^{2}-1}\,\sin{\tau}}{\zeta\cosh{v} + \sqrt{\zeta^{2}-1}\,\cos{\tau}}\nonumber\\
\rho = R\,\frac{\zeta\,\sinh{v}}{\zeta\cosh{v} + \sqrt{\zeta^{2}-1}\,\cos{\tau}} \ .
\label{eq:RindlertoPoincare}
\end{gather}
The metric of $ \text{AdS}_{d+1} $ in Poincar\'e coordinates is given by
\begin{equation}
ds^{2} =  \frac{dr^{2}}{r^{2}} + r^{2}(dt_E^{2} + d\rho^{2} + \rho^{2}d\Omega^{2}_{d-2}) \ .
\end{equation}
In Poincar\'e coordinates, the RT surface is the hemisphere $r^{-2} + \rho^2 = R^2$ which corresponds to $\zeta = 1$ via \eqref{eq:RindlertoPoincare}.

We will denote the worldvolume of the brane $\xi_{n,0}$ in the three directions $(\tau,\zeta,v)$ of the hyperbolic black hole by $\mathcal{W}_n$. When $n = 1$, the embedding $\xi_{1,0}$ of the brane fills the region
\begin{equation}
r > r_\text{q}
\end{equation}
in Poincar\'e coordinates. In Rindler coordinates, this translates to
\begin{equation}
\tilde{r}(\tau,\zeta,v) \equiv \zeta\cosh{v} + \sqrt{\zeta^{2}-1}\,\cos{\tau}> \R  \ .
\end{equation}
More generally for $ n>1 $, the embedding $ \xi_{n,0} $ fills the region
\begin{equation}
\tilde{r}(\tau,\zeta,v) > C_n(\tau,\zeta,v)
\end{equation}
of the hyperbolic black hole where the location of the tip of the brane $ C_n $ is determined by $ \xi_{n,0} $ and satisfies $ C_1 = r_\text{q}R $. We will also impose two separate UV cut-offs $ \zeta < \Lambda_{\zeta} $ and $ v < \Lambda_{v} $. The worldvolume of the brane can thus be written as
\begin{equation}
\mathcal{W}_n = \{(\tau,\zeta,v)\,\lvert\,\tilde{r}(\tau,\zeta,v) > C_n(\tau,\zeta,v)\}\ .
\label{eq:Wn}
\end{equation}
where the coordinates $(\tau,\zeta,v)$ are in the domain
\begin{equation}
    0< \tau<2\pi,\quad \zeta_n<\zeta<\Lambda_\zeta,\quad 0 < v < \Lambda_{v}\ .
\end{equation}
As explained in Section \ref{sec:probeentropy}, the flavor correction to the entropy is given by
\begin{equation}
    S^{(1)}(R) = -\frac{d}{dn} I_{\text{D}p}^{\text{ren}}[\xi_{n,0},g_{n,0}]\bigg\lvert_{n=1} \ ,
     \label{eq:S1app}
\end{equation}
where the renormalized on-shell D$ p $-brane action takes the form
\begin{equation}
    I_{\text{D}p}^{\text{ren}} = \lim_{\Lambda_\zeta\rightarrow \infty}(I_{\text{D}p}^{\text{reg}} + I_{\text{D}p}^{\text{ct}} )
\end{equation}
and the counterterms are supported only on the $\zeta = \Lambda_\zeta$ surface and there are no counterterms on the $v = \Lambda_v$ surface. This means that we are only renormalizing the divergences coming from $\Lambda_\zeta\rightarrow \infty$ while the divergences in $\Lambda_v\rightarrow \infty$ will be related to the physical UV divergence when $\epsilon\rightarrow 0$ of entanglement entropy. We will relate $\Lambda_v$ to $\epsilon$ by writing slices of constant $\tau$ and $\zeta$ in terms of a FG coordinate $V = 2R\, e^{-v} $ as
\begin{equation}
    \zeta^2(dv^2+\sinh^2{v}\,d\Omega_{d-2}^2) = \frac{\zeta^2}{V^2}\biggl[dV^2+\biggl(1- \frac{V^2}{4R^2}\biggr)^2\,R^2d\Omega_{d-2}^2\biggr]\ .
    \label{eq:constantzetatau}
\end{equation}
This ensures that the metric on the conformal boundary $V\rightarrow 0$ ($v\rightarrow \infty$) of \eqref{eq:constantzetatau} coincides with the metric of the spherical entangling surface which is a sphere of radius $R$. We put the cut-off at $V = \epsilon$ so that
\begin{equation}
    e^{\Lambda_v} = \frac{2R}{\epsilon}\ .
    \label{eq:cutoffrelation}
\end{equation}
The regularized D$ p $-brane on-shell action and counterterms are take the form
\begin{align}
I_{\text{D}p}^{\text{reg}}[\xi_{n,0},g_{n,0}] &= \mathcal{N}_{\text{D}p}\int_{\mathcal{W}_n} d\tau d\zeta dv\,\mathcal{L}_n(\xi_{n,0},\partial_i\xi_{n,0}) \label{eq:Dpactionapp}\\
I_{\text{D}p}^{\text{ct}}[\xi_{n,0},\gamma] &=\mathcal{N}_{\text{D}p}\int_{0}^{2\pi} d\tau  \int_{0}^{\Lambda_{v}} dv\,\mathcal{L}_n^{\text{ct}}(\xi_{n,0},\partial_i\xi_{n,0})\biggr\lvert_{\zeta = \Lambda_\zeta} \ ,
\label{eq:Dpactionctapp}
\end{align}
where the index $i =\{ \tau,\zeta,v\}$ and $\gamma$ is the induced metric on the $\zeta = \Lambda_\zeta$ cut-off surface. In writing the integration domain of the counterterm, we used that for $\zeta = \Lambda_\zeta \rightarrow \infty$ the condition $\tilde{r}(\tau,\Lambda_\zeta,v) > C_n(\tau,\Lambda_\zeta,v)$ does not restrict the ranges of $(\tau,v)$.\footnote{This is certainly true at $n = 1$ in which case the condition is $\cosh{v}+\cos{\tau}> 0$ or equivalently $\cos{\tau} > -\cosh{v}$ which is satisfied for all $0 < \tau<2\pi$ and $0 < v < \infty$.}
\begin{figure}[t]
    \centering
    \includegraphics[width=0.75\textwidth]{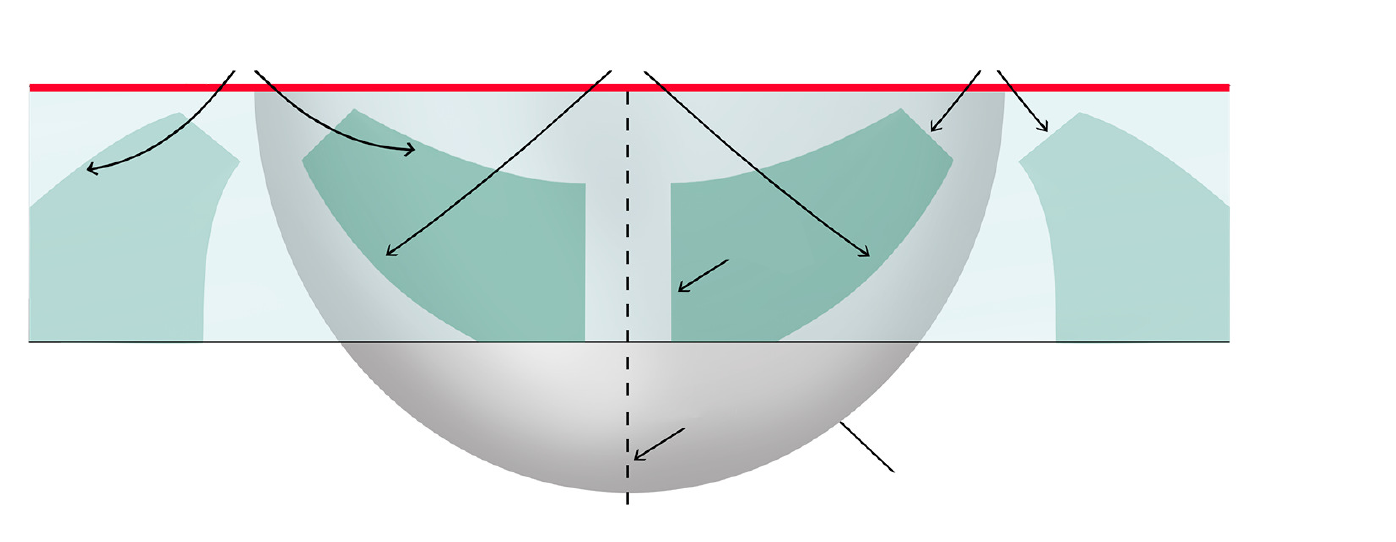}
    \put(-134,8){RT surface}
    \put(-174,28){$v=0$}
    \put(-35,45){$r=r_q$}
    \put(-163,67){$v=\delta_v$}
    \put(-106,112){$v=\Lambda_v$}
    \put(-194,112){$\zeta=1+\delta_\zeta$}
    \put(-278,112){$\zeta=\Lambda_\zeta$}
    \caption{Schematics of the integration boundaries presented by a cross-section of Fig. \ref{fig:cavity}. The integration regions are presented by green areas and their respective cut-offs are indicated by arrows.}
    \label{fig:IntegrationBoundaries-probe}
\end{figure}
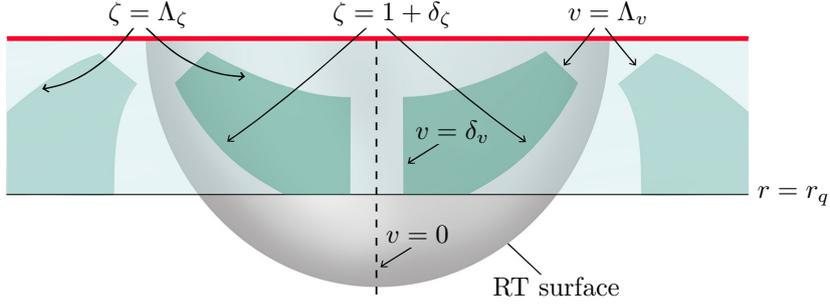

Let us now compute the entropy correction \eqref{eq:S1app}. The $n$-derivative in \eqref{eq:S1app} acts both on the integration domain $\mathcal{W}_n$ and on the integrand of the D-brane action \eqref{eq:Dpactionapp} as
\begin{equation}
\frac{d}{dn}I_{\text{D}p}^{\text{reg}} = \mathcal{N}_{\text{D}p}\int_{\partial \mathcal{W}_1\backslash \text{UV}} dS\,m_i\,\partial_n E^i_n\,\mathcal{L}_n +\, \mathcal{N}_{\text{D}p}\int_{\mathcal{W}_n} d\tau d\zeta dv\, \frac{d}{dn}\mathcal{L}_n \ ,
\label{totalnderivativen}
\end{equation}
which is known as the Reynolds transport theorem \cite{flanders1973differentiation}. In this expression,
\begin{equation}
    \tau = E_n^\tau(y^1,y^2)\ ,\quad \zeta = E_n^\zeta(y^1,y^2)\ ,\quad v= E_n^v(y^1,y^2)\ ,
\end{equation}
with $y = (y^1,y^2)\in \mathbb{R}^2$ is the embedding of $\partial \mathcal{W}_n$ into $\mathbb{R}^3$, $dS = d^2y\,\lVert\partial_{1}\vec{E}_n\times\partial_{2}\vec{E}_n\lVert$ with $\vec{E}_n = (E^\tau_n,E^\zeta_n,E^v_n)$ is the area element of $\partial \mathcal{W}_n$ in $\mathbb{R}^3$ and $m_i$ is its outward-pointing unit normal vector. In \eqref{totalnderivativen} we have used that the UV cut-off boundaries $\zeta = \Lambda_\zeta, v = \Lambda_v$ do not depend on $n$ and are hence not included in the boundary term of \eqref{totalnderivativen} which we denote as $\partial \mathcal{W}_1\backslash \text{UV}$. Equation \eqref{totalnderivativen} is the three-dimensional version of the Reynolds transport theorem proven in \cite{flanders1973differentiation} and it is written in a fixed coordinate system. In Appendix \ref{app:boundaryterms} we will also use the coordinate invariant version of the theorem (see equation \eqref{eq:generalvariation}).

In the second term of \eqref{totalnderivativen}, the variation $\tfrac{d}{dn}\mathcal{L}_n$ consists of two parts: the change of the embedding $ \partial_n \xi_{n,0} $ and the change due to the explicit $ n $-dependence of $ \mathcal{L}_n $ included in the blackening factor $f_n(\zeta)$. We get
\begin{equation}
\frac{d}{dn}\mathcal{L}_n = \partial_n\mathcal{L}_n +\left[ \frac{\partial \mathcal{L}_n}{\partial\xi_{n,0}} - \partial_i\left(\frac{\partial \mathcal{L}_n}{\partial(\partial_i \xi_{n,0})} \right)  \right](\partial_n \xi_{n,0})  + \partial_i\biggl(\frac{\partial \mathcal{L}_n}{\partial (\partial_i \xi_{n,0})}(\partial_n\xi_{n,0})\biggr) \ .
\label{variationlagr}
\end{equation}
After setting $ n = 1 $, the second term in \eqref{variationlagr} vanishes since $\xi_{1,0}$ satisfies the brane equations of motion. The second term in \eqref{totalnderivativen} becomes
\begin{equation}
    \int_{\mathcal{W}_n}d\zeta d\tau dv\, \frac{d}{dn}\mathcal{L}_n = \int_{\mathcal{W}_n} d\tau d\zeta dv\,(\partial_n\mathcal{L}_n)\,+ \int_{\partial \mathcal{W}_n}dS\,m_i\,\frac{\partial \mathcal{L}_n}{\partial (\partial_i \xi_{n,0})}(\partial_n\xi_{n,0})\ .
    \label{integrandvariation}
\end{equation}
Hence, the derivative \eqref{totalnderivativen} at $n=1$ can be written as
\begin{equation}
    \frac{d}{dn}I_{\text{D}p}^{\text{reg}}\bigg\lvert_{n = 1} = \mathcal{I} + \mathcal{B}^{(1)} + \mathcal{B}^{(2)}\ ,
\end{equation}
where the bulk and the boundary terms are given by
\begin{align}
    \mathcal{I} &= \mathcal{N}_{\text{D}p}\int_{\mathcal{W}_1} d\tau d\zeta dv\,\partial_n\mathcal{L}_n(\xi_{1,0},\partial_i\xi_{1,0})\lvert_{n=1}\\
    \mathcal{B}^{(1)} &=\mathcal{N}_{\text{D}p}\int_{\partial \mathcal{W}_1\backslash \text{UV}}dS\,\mathcal{L}_1\,m_i\,(\partial_n E^i_n)\lvert_{n=1}\label{eq:curlyB1}\\
    \mathcal{B}^{(2)} &=\mathcal{N}_{\text{D}p}\int_{\partial \mathcal{W}_1}dS\,m_i\,\frac{\partial \mathcal{L}_1}{\partial (\partial_i \xi_{1,0})}(\partial_n\xi_{n,0})\lvert_{n=1} \ .\label{eq:curlyB2}
\end{align}
The boundary $\partial \mathcal{W}_1$ consists of multiple components which give the following boundary terms (see Fig.~\ref{fig:IntegrationBoundaries-probe}).
\begin{enumerate}
\item Boundary terms in the $ \zeta $-direction at the RT surface $ \zeta = 1+\delta_\zeta $ (with $\delta_\zeta\rightarrow 0^{+}$), denoted by $ \mathcal{B}^{(1,2)}_{\text{RT}} $, and at infinity $ \zeta = \Lambda_{\zeta} $, denoted by $ \mathcal{B}^{(2)}_{\zeta = \infty} $.
\item Boundary terms in the $ v $-direction at $ v = \delta_v $ (with $\delta_v\rightarrow 0^{+}$), denoted by $ \mathcal{B}_{v = 0}^{(2)} $, and at $ v = \Lambda_{v} $, denoted by $ \mathcal{B}^{(2)}_{v = \infty} $.
\item Boundary terms at the tip of the brane $ r(\tau,\zeta,v)= r_{\text{q}} $ denoted by $ \mathcal{B}^{(1,2)}_{\text{tip}} $.
\end{enumerate}
The RT surface $\zeta = \zeta_1 = 1$ is one component and we can parametrize it using $(\tau,v)$ whose integration domain is given by $\tilde{r}(\tau,1,v) > \R $. This is equivalent to $\cosh{v} > \R  $ which does not constrain the integration range of $\tau$, but the lower limit of $v$-integration is determined to be
\begin{equation}
    v_0 = 
    \begin{cases}
        0\ ,\quad &\R   < 1\\
        \arcosh{\R}\ ,\quad &\R   \geq 1 
    \end{cases}\ .
    \label{v0}
\end{equation}
The boundary terms at $\zeta = 1$ are thus
\begin{align}
    \mathcal{B}_{\text{RT}}^{(1)} &= \mathcal{N}_{\text{D}p}\int_{0}^{2\pi} d\tau \int_{v_0}^{\Lambda_v} dv\,\frac{1}{d-1}\,\mathcal{L}_1(\xi_{1,0},\partial_i\xi_{1,0})\bigg\lvert_{\zeta = 1+\delta_\zeta}\\
\mathcal{B}_{\text{RT}}^{(2)}&=\mathcal{N}_{\text{D}p}\int_{0}^{2\pi} d\tau \int_{v_0}^{\Lambda_v} dv\,(-1)\,\frac{\partial \mathcal{L}_1}{\partial (\partial_\zeta \xi_{1,0})}(\partial_n\xi_{n,0})\lvert_{n=1}\bigg\lvert_{\zeta = 1+\delta_\zeta} \ ,
\end{align}
where we have used that on this surface $m_i = -\delta^\zeta_i$ and that
\begin{equation}
m_i\,\partial_nE^i_n\lvert_{n=1}\, = -\partial_n\zeta_n\lvert_{n=1}\, = \frac{1}{d-1} = 
\begin{dcases}
\frac{1}{2}\ ,\quad &d = 3\\
\frac{1}{3}\ ,\quad &d=4 
\end{dcases}\ .
\label{eq:zetander}
\end{equation}
The boundary term at $\zeta = \infty$ is explicitly
\begin{align}
\mathcal{B}_{\zeta = \infty}^{(2)} &= \mathcal{N}_{\text{D}p}\int_{0}^{2\pi}d\tau \int_{0}^{\Lambda_v} dv\,\frac{\partial \mathcal{L}_1}{\partial (\partial_\zeta \xi_{1,0})}(\partial_n\xi_{n,0})\lvert_{n=1}\biggr\lvert_{\zeta = \Lambda_\zeta}.
\end{align}
The boundary terms at $v = 0$ and $v = \infty$ are explicitly
\begin{align}
\mathcal{B}_{v = 0}^{(2)} &= \mathcal{N}_{\text{D}p}\int_{0}^{2\pi}d\tau \int_{1}^{\Lambda_\zeta} d\zeta\,\frac{\partial \mathcal{L}_1}{\partial (\partial_v \xi_{1,0})}(\partial_n\xi_{n,0})\lvert_{n=1}\biggr\lvert_{v = \delta_v}\\
\mathcal{B}_{v = \infty}^{(2)} &= \mathcal{N}_{\text{D}p}\int_{0}^{2\pi}d\tau \int_{1}^{\Lambda_\zeta} d\zeta\,\frac{\partial \mathcal{L}_1}{\partial (\partial_v \xi_{1,0})}(\partial_n\xi_{n,0})\lvert_{n=1}\biggr\lvert_{v = \Lambda_v}.
\end{align}
The boundary terms at the tip of the brane $ r= r_\text{q}$ are given by
\begin{align}
    \mathcal{B}^{(1)}_{\text{tip}} &=\mathcal{N}_{\text{D}p}\int_{r = r_{\text{q}}}dS\,\mathcal{L}_1\,m_i\,(\partial_n E^i_n)\lvert_{n=1}\\
    \mathcal{B}^{(2)}_{\text{tip}} &=\mathcal{N}_{\text{D}p}\int_{r = r_{\text{q}}}dS\,m_i\,\frac{\partial \mathcal{L}_1}{\partial (\partial_i \xi_{1,0})}(\partial_n\xi_{n,0})\lvert_{n=1}\ .
\end{align}
Hence, the boundary terms \eqref{eq:curlyB1}--\eqref{eq:curlyB2} decompose as
\begin{equation}
    \mathcal{B}^{(1)}  = \mathcal{B}_{\text{RT}}^{(1)}+\mathcal{B}_{\text{tip}}^{(1)}\ ,\quad \mathcal{B}^{(2)} =\mathcal{B}_{\text{RT}}^{(2)}+\mathcal{B}_{\text{tip}}^{(2)} +\mathcal{B}_{\zeta = \infty}^{(2)}+\mathcal{B}_{v = 0}^{(2)}+\mathcal{B}_{v = \infty}^{(2)}\ .
\end{equation}
The renormalized action also contains the $n$-derivative of the counterterms which we denote\footnote{Note they are not part of $\mathcal{B}^{(1)}$ \eqref{eq:curlyB1} even though we use the same superscript notation.} by
\begin{equation}
    \mathcal{B}^{(1)}_{\zeta = \infty} = \mathcal{N}_{\text{D}p}\int_{0}^{2\pi}d\tau \int_{0}^{\Lambda_v} dv\,\frac{d}{dn}\mathcal{L}_n^{\text{ct}}(\xi_{n,0},\partial_i\xi_{n,0})\lvert_{n=1}\biggr\lvert_{\zeta = \Lambda_\zeta}\ .
\end{equation}
The entanglement entropy is thus given by
\begin{equation}
S^{(1)}(R) = \mathcal{I} + \mathcal{B}_{\text{RT}}^{(1)}+\mathcal{B}_{\text{RT}}^{(2)}+ \mathcal{B}_{\text{tip}}^{(1)}+\mathcal{B}_{\text{tip}}^{(2)}+\mathcal{B}^{(1)}_{\zeta = \infty} +\mathcal{B}_{\zeta = \infty}^{(2)}+\mathcal{B}_{v = 0}^{(2)}+\mathcal{B}_{v = \infty}^{(2)} \ .
\end{equation}
The boundary terms at the tip of the brane arise due to changes in the embedding of the brane and in the coordinate location of the tip as determined by the embedding. Since the tip is not a physical boundary but the surface where the brane caps off smoothly, we expect $\mathcal{B}_{\text{tip}}^{(1,2)}$ to cancel each other
\begin{equation}    \mathcal{B}_{\text{tip}}^{(1)}+\mathcal{B}_{\text{tip}}^{(2)} = 0\,
    \label{eq:branetipcancellation}
\end{equation}
due to the regularity condition $\xi_{1,0}(r_{\text{q}})^{-1} = 0$. We prove the cancellation in Appendix \ref{app:boundaryterms} where we also give a coordinate independent treatment of the variation of the brane action. The cancellation \eqref{eq:branetipcancellation} is exactly analogous to the cancellation of the boundary term at the tip of the RT surface in computation of entanglement entropy using the RT formula in Section \ref{subsec:perturbativeRT}. In addition, we will see explicitly below that $\mathcal{B}_{v = 0}^{(2)}=\mathcal{B}_{v = \infty}^{(2)} = 0$ in both of our examples. Hence including only the non-zero terms, the entropy takes the form
\begin{equation}
S^{(1)}(R) = \mathcal{I} + \mathcal{B}_{\text{RT}}^{(1)}+\mathcal{B}_{\text{RT}}^{(2)} + \mathcal{B}_{\zeta = \infty}^{(1)}+\mathcal{B}_{\zeta = \infty}^{(2)}\ .
\end{equation}
We will see that $\mathcal{B}_{\text{RT}}^{(2)}$ is generically non-zero.

\subsection{Probe D7-branes}\label{subapp:probeD7}

In this appendix, we give details of the probe computation of flavor entanglement entropy presented in Section \ref{subsec:probeD7} in the case of a D7-brane on $\text{AdS}_5\times \text{S}^5$.

\paragraph{Holographic renormalization.}

Near the conformal boundary $\zeta\rightarrow \infty$, the probe brane embedding has the on-shell expansion \cite{Karch:2005ms}
\begin{equation}
\xi_{n,0}(\tau,\zeta,v) = \frac{\xi_-(\tau,v)}{\zeta} +\frac{\xi_+(\tau,v)}{\zeta^{3}} + \frac{\Theta}{\zeta^{3}}\log{\zeta} + \ldots \ .
\label{eq:D7asymptoticexpansion}
\end{equation}
As a result, the regularized on-shell D7-brane action \eqref{eq:Dpactionapp} with the Lagrangian \eqref{eq:D7lagrangiantext} has the expansion
\begin{align}
I_{\text{D7}}^{\text{reg}} =&\, \mathcal{N}_{\text{D7}}\int_{0}^{2\pi} d\tau \int_{v_0}^{\Lambda_{v}} dv\, \sinh^{2}{v} \\
&\times\left[\frac{1}{4}\,\Lambda_{\zeta}^{4}-\frac{1}{2}\,\xi_-^{2}\Lambda_{\zeta}^{2} -\frac{1}{2}\left(2\Theta\,\xi_-+\xi_-^{2}-(\partial_\tau\xi_-)^{2}-(\partial_v\xi_-)^{2}\right)\log{\Lambda_{\zeta}}+ \mathcal{O}(\Lambda_\zeta^0)\right], \quad \Lambda_{\zeta} \rightarrow \infty \ .\nonumber
\label{eq:regactiondivsD7}
\end{align}
The possible counterterms are \cite{Karch:2005ms}
\begin{gather}
I^{(1)} \propto \int \sqrt{\gamma}\ , \quad I^{(2)} \propto\int  \sqrt{\gamma}\,R_{\gamma}\ , \quad I^{(3)} \propto\int  \sqrt{\gamma}\,\xi^{2}\ , \quad I^{(4)} \propto\int  \sqrt{\gamma}\,R_{\gamma}\xi^{2}\log{\xi}\ ,\\
I^{(5)}\propto \int \sqrt{\gamma}\,\gamma^{ij}\,\partial_i \xi\,\partial_j\xi\log{\xi}\ , \quad I^{(6)} \propto\int  \sqrt{\gamma}\,\biggl(R_{\gamma ij}R^{ij}_{\gamma} - \frac{1}{3}R_{\gamma}\biggr)\log{\xi} \ , \quad I_{\text{fin}} \propto\int  \sqrt{\gamma}\,\xi^{4} \ ,
\end{gather}
where the integral is over the $\zeta = \Lambda_\zeta$ cut-off surface, $\gamma$ is the induced metric of the cut-off surface, $R_{\gamma ij}$ is the Ricci tensor, and $R_\gamma$ is the Ricci scalar built from $\gamma$. We can compute
\begin{equation}
R_{\gamma} = -\frac{6}{\zeta^{2}}\ ,\quad \gamma^{ij}\,\partial_i \xi\,\partial_j\xi = f_n(\zeta)^{-1}\,(\partial_\tau \xi)^{2} + \zeta^{-2}\,(\partial_v \xi)^{2}\ , \quad R_{\gamma ij}R^{ij}_{\gamma} - \frac{1}{3}R_{\gamma} = 0 \ .
\end{equation}
Thus, the counterterm Lagrangian $\mathcal{L}^{\text{ct}}_n$ in \eqref{eq:Dpactionapp} is explicitly
\begin{eqnarray}
\mathcal{L}^{\text{ct}}_n & = & \zeta^{3}\,f_n(\zeta)^{1\slash 2}\,\sinh^{2}{v}\\
& & \times\left[a_1+a_2\,\zeta^{-2}+a_3\,\xi^{2}+a_4\,\zeta^{-2}\xi^{2}\log{\zeta}+a_5\,\left(f_n(\zeta)^{-1}\,(\partial_\tau \xi)^{2} + \zeta^{-2}\,(\partial_v \xi)^{2}\right)\log{\zeta}+a_{\text{fin}}\,\xi^{4}\right] \ ,\nonumber
\label{eq:LnctD7}
\end{eqnarray}
where $a_1,\ldots,a_5$ and $a_{\text{fin}}$ are constants. The counterterm action in \eqref{eq:Dpactionapp} has the near boundary expansion
\begin{eqnarray}
I_{\text{D7}}^{\text{ct}} & = & \mathcal{N}_{\text{D7}}\int_{0}^{2\pi} d\tau \int_{0}^{\Lambda_{v}} dv\,\sinh^{2}{v}\\
& & \times\biggl[a_1\Lambda_{\zeta}^{4}+(b_1+a_3\,\xi_{-}^{2})\,\Lambda_{\zeta}^{2}+\left(2a_3\,\Theta\,\xi_-+a_{4}\,\xi_-^{2}+a_5\,[(\partial_\tau\xi_-)^{2}+(\partial_v\xi_-)^{2}]\right)\log{\Lambda_{\zeta}} + \mathcal{O}(1)\biggr]\ ,\nonumber
\end{eqnarray}
where $ b_1 = -\frac{1}{2}a_1+a_2 $. Hence, to cancel the divergences we must set
\begin{equation}
a_1 = -\frac{1}{4}\ , \quad a_3 = \frac{1}{2}\ , \quad a_4 = \frac{1}{2}\ , \quad a_5 = -\frac{1}{2} \ ,
\end{equation}
along with $ b_1 = 0 $ which gives
\begin{equation}
a_2 = -\frac{1}{8} \ .
\end{equation}
Hence, the counterterm action \eqref{eq:Dpactionapp} with the counterterm Lagrangian \eqref{eq:LnctD7} that cancels divergences of \eqref{eq:regactiondivsD7} is given by
\begin{eqnarray}
I_{\text{D7}}^{\text{ct}} & = & \mathcal{N}_{\text{D7}}\int_{0}^{2\pi} d\tau \int_{0}^{\Lambda_{v}} dv\,\sinh^{2}{v}\label{eq:IctD7exp}\nonumber\\
& & \times\biggl[-\frac{1}{4}\Lambda_{\zeta}^{4}+\frac{1}{2}\,\xi_{-}^{2}\Lambda_{\zeta}^{2}+\left(\Theta\,\xi_-+\frac{1}{2}\,\xi_-^{2}-\frac{1}{2}\,[(\partial_\tau\xi_-)^{2}+(\partial_v\xi_-)^{2}]\right)\log{\Lambda_{\zeta}} \\
& &\quad +\frac{1}{32}(3+4\zeta_n^4-4\zeta_n^2)-\frac{1}{4}\xi_-^2+a_{\text{fin}}\xi_-^4+\xi_-\xi_+\biggr]\ .\nonumber
\end{eqnarray}

\paragraph{Boundary terms in the entanglement entropy.}

Let us compute the boundary terms $\mathcal{B}_{\text{RT}}^{(1,2)}$ and $\mathcal{B}_{\zeta=\infty}^{(1,2)}$ by starting with the UV terms. For the first term $\mathcal{B}_{\zeta=\infty}^{(1)}$, we get from \eqref{eq:IctD7exp} that
\begin{align}
\mathcal{B}_{\zeta=\infty}^{(1)} &= \mathcal{N}_{\text{D7}}\int_{0}^{2\pi} d\tau \int_{0}^{\Lambda_{v}} dv\,\sinh^{2}{v}\,\biggl(-\frac{1}{12}+\xi_-\partial_n\xi_+\lvert_{n=1}\biggr) \\
&=\pi\,\mathcal{N}_{\text{D7}}\,\biggl(\frac{R^2}{\epsilon^2}-\log{\frac{2R}{\epsilon}}\biggr)\biggl(-\frac{1}{12}+\xi_-\partial_n\xi_+\lvert_{n=1}\biggr)\ ,
\label{eq:BUV1D7}
\end{align}
where we used that the source $\xi_-$ is independent of $n$ and that the $v$-integral evaluates to
\begin{equation}
\int_{0}^{\Lambda_{v}} dv\,\sinh^2{v} = \frac{1}{2}\biggl(\frac{R^2}{\epsilon^2}-\log{\frac{2R}{\epsilon}} + \mathcal{O}(\epsilon^2)\biggr)\ ,
\end{equation}
where we used the cut-off \eqref{eq:cutoffrelation}. The second UV term is explicitly
\begin{equation}
    \mathcal{B}_{\zeta=\infty}^{(2)}  = \mathcal{N}_{\text{D7}}\int_{0}^{2\pi} d\tau \int_{0}^{\Lambda_v} dv\,\frac{\zeta^3\sinh^2{v}\cos^3{(\xi_{1,0})}\,f_1(\zeta)\,(\partial_\zeta\xi_{1,0})\,(\partial_n\xi_{n,0})\lvert_{n=1}}{\sqrt{1+\zeta^{-2}\,(\partial_v\xi_{1,0})^{2} + f_1(\zeta)\,(\partial_\zeta \xi_{1,0})^{2}+f_1(\zeta)^{-1}\,(\partial_{\tau}\xi_{1,0})^{2}}}\biggr\lvert_{\zeta = \Lambda_\zeta}.
\end{equation}
Substituting the asymptotic expansion \eqref{eq:D7asymptoticexpansion} of the embedding gives
\begin{equation}
    \mathcal{B}_{\zeta=\infty}^{(2)}  = -\mathcal{N}_{\text{D7}}\int_{0}^{2\pi} d\tau \int_{0}^{\Lambda_v} dv\,\sinh^2{v}\,\xi_-\partial_n\xi_+\lvert_{n=1} + \mathcal{O}(\Lambda_{\zeta}^{-2})
\end{equation}
which becomes
\begin{equation}
    \mathcal{B}_{\zeta=\infty}^{(2)} = -\pi\,\mathcal{N}_{\text{D7}}\,\biggl(\frac{R^2}{\epsilon^2}-\log{\frac{2R}{\epsilon}}\biggr)\,\xi_-\partial_n\xi_+\lvert_{n=1}\ .
    \label{eq:BUV2D7}
\end{equation}
Hence, the sum of \eqref{eq:BUV1D7} and \eqref{eq:BUV2D7} is given by
\begin{equation}
    \mathcal{B}_{\zeta=\infty}^{(1)} + \mathcal{B}_{\zeta=\infty}^{(2)} = -\frac{\pi}{12}\,\mathcal{N}_{\text{D7}}\,\biggl(\frac{R^2}{\epsilon^2}-\log{\frac{2R}{\epsilon}}\biggr)\ ,
    \label{eq:BUVD712}
\end{equation}
where all dependence on $\partial_n\xi_+\lvert_{n=1}$ has canceled.

Let us then compute the boundary terms at $v = 0$ and $v = \infty$. From the equations of motion we find that the solution $ \xi_{n,0} $ near $ v = \infty $ behaves as
\begin{equation}
\xi_{n,0}(\tau,\zeta,v) =  \tilde{\xi}_-(\tau,\zeta)\,e^{-v} + \tilde{\xi}_+(\tau,\zeta)\,e^{-2v}+ \mathcal{O}(e^{-3v}),\quad v \rightarrow \infty \ .
\end{equation}
It follows that
\begin{equation}
\frac{\partial \mathcal{L}_1}{\partial (\partial_v \xi_{1,0})}(\partial_n\xi_{n,0})\lvert_{n=1}\, =-\frac{1}{4}\,\zeta\,\tilde{\xi}_-\partial_n\tilde{\xi}_+\lvert_{n=1}\,e^{-v}+\mathcal{O}(e^{-2v}),\quad v\rightarrow \infty\,,
\label{eq:D7Bvinfty}
\end{equation}
which implies that
\begin{equation}
\mathcal{B}_{v = \infty}^{(2)} = 0\ .
\end{equation}
Near $ v = 0 $ one can show that the solution $ \xi_{n,0} $ goes as
\begin{equation}
\xi_{n,0}(\tau,\zeta,v) = \xi_0^{(n)}(\tau,\zeta) +\xi_1^{(n)}(\tau,\zeta)\, v^{2} + \mathcal{O}(v^{2}),\quad v\rightarrow 0\,.
\end{equation}
This implies that
\begin{equation}
\frac{\partial \mathcal{L}_1}{\partial (\partial_v \xi_{1,0})} = \mathcal{O}(v^{3}),\quad v\rightarrow 0\ ,
\end{equation}
and since $ \partial_n\xi_{n,0}\lvert_{n=1}\, = \mathcal{O}(1) $ is finite at $ v = 0 $, we get
\begin{equation}
\mathcal{B}_{v = 0}^{(2)} = 0\ .
\end{equation}
Let us then compute the boundary terms on the RT surface. To compute them, we need the on-shell behavior of the brane embedding $\xi_{n,0}$ near $\zeta = \zeta_n$ which is given by
\begin{equation}
\xi_{n,0}(\tau,\zeta,v) = \xi^{(0)}_n(v) + \xi^{(1)}_n(\tau,v)\, \sqrt{\zeta-\zeta_n} + \mathcal{O}(\zeta-\zeta_n)\ ,\quad \zeta\rightarrow \zeta_n \ ,
\label{eq:nearhorizonxi}
\end{equation}
where $\xi^{(0)}_n(v)$ and $\xi^{(1)}_n(\tau,v)$ are fixed by the equations of motion, but we will not need their expressions for $n\neq 1$. It follows that
\begin{equation}
\mathcal{B}^{(1)}_{\text{RT}} = \frac{1}{3}\,\mathcal{N}_{\text{D7}}\int_{0}^{2\pi} d\tau \int_{v_0}^{\Lambda_v} dv\,\sinh^{2}{v}\cos^{3}{\xi^{(0)}_n}\,\sqrt{1+(\partial_v\xi^{(0)}_n)^{2}+\frac{1}{2}(\xi^{(1)}_n)^{2}+\frac{1}{2}(\partial_\tau\xi^{(1)}_n)^{2}} \ .
\label{eq:BIR11-D7}
\end{equation}
From the embedding \eqref{eq:D7xi10}, we see that at $n=1$,
\begin{equation}
\xi^{(0)}_1(v) = \arcsin{(\R  \sech{v})}\ , \quad \xi^{(1)}_1(\tau,v) = \frac{\sqrt{2}\,\R  \cos{\tau}\,\sech{v}}{\sqrt{\cosh^{2}{v}-\R^2}} \ ,
\label{eq:xi01D7}
\end{equation}
and $\zeta_1 = 1$. Substituting to \eqref{eq:BIR11-D7} gives
\begin{equation}
\mathcal{B}^{(1)}_{\text{RT}} = \frac{1}{3}\,\mathcal{N}_{\text{D7}}\int_{0}^{2\pi} d\tau \int_{v_0}^{\Lambda_v} dv\,\sinh^{2}{v}\,\left(1 - \R^2\sech^{2}{v}\right) \ .
\end{equation}
For $\R   < 1$, we get
\begin{equation}
\mathcal{B}^{(1)}_{\text{RT}} =\frac{\pi}{3}\,\mathcal{N}_{\text{D7}}\,\biggl[\frac{R^2}{\epsilon^2}-(1+2\R^2)\log{\frac{2R}{\epsilon}}+2\R^2+\mathcal{O}(\epsilon^2)\biggr] \ ,
\label{eq:BIR1finalsmallmass}
\end{equation}
where we have used (up to an additive constant)
\begin{equation}
\int dv\, \sinh^{2}{v}\,\left(1 + \R^2\sech^{2}{v}\right) = -\frac{1}{2}\,(1-\R^2)\,v+\frac{1}{4}\sinh{(2v)}+\R^{2}\tanh{v} \ .
\end{equation}
For $\R  >1$, the result is
\begin{align}
\mathcal{B}^{(1)}_{\text{RT}} =\frac{\pi}{3}\,\mathcal{N}_{\text{D7}}\,&\biggl[\frac{R^{2}}{\epsilon^{2}}-(1+2\R^2)\log{\frac{2R}{\epsilon}}+2\R^2-3\R  \sqrt{\R^2-1}\biggr.\nonumber\\
&\qquad\qquad\qquad\qquad\qquad\biggl.+(1+2\R^2)\arcosh{\R} + \mathcal{O}(\epsilon^{2})\biggr] \ .
\label{eq:BIR1final}
\end{align}
To compute $\mathcal{B}_{\text{RT}}^{(2)}$, we need
\begin{equation}
\frac{\partial \mathcal{L}_1}{\partial (\partial_\zeta \xi_{1,0})} = \frac{\sinh^{2}{v}\cos^{3}{\xi^{(0)}_1}\,\xi^{(1)}}{\sqrt{1+(\partial_v\xi^{(0)}_1)^{2}+\frac{1}{2}(\xi^{(1)}_1)^{2}+\frac{1}{2}(\partial_\tau\xi^{(1)}_1)^{2}}}\,\sqrt{\zeta - 1} + \mathcal{O}(\zeta-1)\ , \quad \zeta \rightarrow 1 \ .
\label{eq:Lder2}
\end{equation}
From the on-shell behavior \eqref{eq:nearhorizonxi} it follows that
\begin{equation}
\partial_n\xi_{n,0}\lvert_{n=1} = \frac{1}{6}\frac{\xi^{(1)}_1(\tau,v)}{\sqrt{\zeta-1}} + \mathcal{O}(\sqrt{\zeta-1})\ , \quad \zeta \rightarrow 1 \ ,
\label{eq:xinderlimitD7}
\end{equation}
where we used \eqref{eq:zetander}. Due to this divergence, the product of \eqref{eq:Lder2} and \eqref{eq:xinderlimitD7} in $ \mathcal{B}_{\text{RT}}^{(2)} $ is finite when $ \zeta \rightarrow 1 $. The result is
\begin{equation}
\mathcal{B}_{\text{RT}}^{(2)} = -\frac{1}{6}\,\mathcal{N}_{\text{D7}}\int_{0}^{2\pi} d\tau \int_{v_0}^{\Lambda_v} dv\,\frac{\sinh^{2}{v}\cos^{3}{\xi^{(0)}_1}\,(\xi^{(1)}_1)^{2}}{\sqrt{1+(\partial_v\xi^{(0)}_1)^{2}+\frac{1}{2}(\xi^{(1)}_1)^{2}+\frac{1}{2}(\partial_\tau\xi^{(1)}_1)^{2}}} \ .
\end{equation}
Substituting \eqref{eq:xi01D7} gives
\begin{equation}
\mathcal{B}_{\text{RT}}^{(2)} = -\frac{1}{3}\,\mathcal{N}_{\text{D7}}\int_{0}^{2\pi} d\tau \int_{v_0}^{\Lambda_v} dv\,\cos^{2}{\tau}\sinh^{2}{v}\sech^{4}{v}\,\ \R^2\left(1 - \R^2\sech^{2}{v}\right),
\end{equation}
and performing the integrals gives
\begin{equation}
\mathcal{B}_{\text{RT}}^{(2)} = -\frac{\pi}{3}\,\mathcal{N}_{\text{D7}}\,
\begin{dcases}
\frac{1}{3}\R^2-\frac{2}{15}\R^4\ ,\quad &\R  < 1 \\
\frac{1}{3}\R^2-\frac{2}{15}\R^4+ \frac{2}{15}\frac{\sqrt{\R^2-1}}{\R  }\,\left(1-2\R^2+\R^4\right)\ ,\quad &\R  \geq 1 
\end{dcases}
\label{eq:BIR2final2} \ .
\end{equation}
We see that $ \mathcal{B}_{\text{RT}}^{(2)} $ does not contain any UV divergences in contrast to $ \mathcal{B}_{\text{RT}}^{(1)} $.

The reason why $ \mathcal{B}_{\text{RT}}^{(2)} $ is finite and non-zero is the divergence in $\partial_n\xi_{n,0}\lvert_{n=1}$ at $\zeta = 1$. The boundary term $\mathcal{B}_{v = 0}^{(2)}$ is mathematically of the same form, but $\partial_n\xi_{n,0}\lvert_{n=1}$ does not have a divergence at $v = 0$ so that $\mathcal{B}_{v = 0}^{(2)} = 0$.

\paragraph{Bulk integral in the entropy.}

The bulk integral $\mathcal{I}$ appearing in the entropy is explicitly
\begin{align}
\mathcal{I} = \frac{1}{2}\,\mathcal{N}_{\text{D7}}\int_{0}^{2\pi} d\tau \int_{1}^{\Lambda_\zeta} d\zeta \int_{v_0}^{\Lambda_v} dv\, \zeta^{3}\sinh^{2}{v}\cos^{3}{(\xi_{1,0})}\,\frac{\partial_nf_n(\zeta)\lvert_{n=1}}{f_1(\zeta)^{2}} \nonumber\\
\times\frac{f_1(\zeta)^{2}\,(\partial_\zeta \xi_{1,0})^{2}-(\partial_{\tau}\xi_{1,0})^{2}}{\sqrt{1+\zeta^{-2}\,(\partial_v\xi_{1,0})^{2} + f_1(\zeta)\,(\partial_\zeta \xi_{1,0})^{2}+f_1(\zeta)^{-1}\,(\partial_{\tau}\xi_{1,0})^{2}}} \ ,
\label{eq:ID71}
\end{align}
where $ f_1(\zeta) = \zeta^{2}-1 $.

In order to compute the integral \eqref{eq:ID71}, we perform a change of coordinates from Rindler to Poincar\'e \eqref{eq:RindlertoPoincare}, and rescale the coordinates as: $\{ \tilde{t}_{E}=\frac{t_E}{R}, \tilde{\rho}=\frac{\rho}{R}, \tilde{z}=\frac{z}{R}\}$, where we have also introduced the coordinate $z\equiv\frac{1}{r}$. With this transformation, the integral has the form:
\begin{eqnarray}
    \mathcal{I}=\mathcal{N}_{\text{D7}}\int_{\epsilon/R}^{1/\R  }d\tilde{z} \int_{-\infty}^{\infty}d\tilde{t}_{E}\int_{0}^{+\infty}d\tilde{\rho} \frac{N}{D} \ ,
\end{eqnarray}
where
\begin{align}
N&=-16\R^2\tilde{z}\left(1-\R^2\tilde{z}^2\right) \tilde{\rho}^2\bigg[\tilde{t}_{E}^8+4\tilde{t}_{E}^6\left(1-\tilde{z}^2+\tilde{\rho}^2\right)+\left(\tilde{z}^4-(1-\tilde{\rho}^2)^2\right)^2-4\tilde{t}_{E}^2\big(\tilde{z}^6+\tilde{\rho}^2\nonumber\\
&\qquad+\tilde{\rho}^4-\tilde{\rho}^6-1+\tilde{z}^2\left(1-\tilde{\rho}^2\right)^2+3\tilde{z}^4(1+\tilde{\rho}^2)\big)+\tilde{t}_{E}^4\left(6-10\tilde{z}^4+4\tilde{\rho}^2+6\tilde{\rho}^4-8\tilde{z}^2(1+\tilde{\rho}^2)\right)\bigg]\nonumber\\
D&=3\bigg[\left((1+\tilde{t}_{E}^2+\tilde{z}^2)^2+2(\tilde{t}_{E}^2+\tilde{z}^2-1)\tilde{\rho}^2+\tilde{\rho}^4\right)\left(\tilde{t}_{E}^4+(\tilde{z}^2+\tilde{\rho}^2-1)^2+2\tilde{t}_{E}^2(1+\tilde{z}^2+\tilde{\rho}^2)\right)\bigg]^2 \ .
\end{align}
We first perform the integration in $\tilde{\rho}$. This could be done directly but it is simpler to resort to Cauchy's theorem. We therefore promote $\tilde{\rho}$ to a complex variable and we integrate along a closed contour delimited by the segment $\tilde{\rho} \in (-Q,+Q)$, and a quadrant of radius $Q$ in the upper half-plane connected to the previous segment. Sending $Q\rightarrow +\infty$ equals \textit{twice} the integral (recall that the integrand is even in both $\tilde{\rho}$ and $\tilde{t}_{E}$), since the contribution from the integral along the arc of the quadrant vanishes in this limit. By Cauchy's theorem, the integral boils down to the sum over residues at the pole locations. This first integral results in 
\begin{align}
\mathcal{I}&=\mathcal{N}_{\text{D7}}\int_{\frac{\epsilon}{R}}^{\frac{1}{\R  }}d\tilde{z} \int_{-\infty}^{+\infty} d\tilde{t}_{E} \frac{i \pi}{12\tilde{z}} \R^2(\R^2\tilde{z}^2-1)\bigg[\frac{-4+16\tilde{t}_{E}^2+6\tilde{z}^2-2i\tilde{t}_{E}(-7+3\tilde{t}_{E}^2+3\tilde{z}^2)}{\sqrt{-\tilde{z}^2-(i+\tilde{t}_{E})^2}}\nonumber \\
&\qquad-\frac{-4+16\tilde{t}_{E}^2+6\tilde{z}^2+2i\tilde{t}_{E}(-7+3\tilde{t}_{E}^2+3\tilde{z}^2)}{\sqrt{-\tilde{z}^2-(-i+\tilde{t}_{E})^2}}-\frac{2(10\tilde{t}_{E}^4-\tilde{z}^2+16\tilde{z}^2\tilde{t}_{E}^2+6\tilde{z}^4)}{(-\tilde{t}_{E}^2-\tilde{z}^2)^\frac{3}{2}} \bigg] \ .
\end{align}
Despite the explicit appearance of the $i$ factors, one can check that the previous expression is real. Next, we perform the integral in $\tilde{t}_{E}$. We can exploit the even parity of the integrand in $\tilde{t}_{E}$ to find
\begin{align}
\mathcal{I}=\mathcal{N}_{\text{D7}}\int_{\frac{\epsilon}{R}}^{\frac{1}{\R  }}d\tilde{z} \frac{\pi \R^2(\R^2\tilde{z}^2-1)}{6\tilde{z}}&\Theta(1-\tilde{z})\bigg[2\sqrt{1-\tilde{z}^2}(2\tilde{z}^2-1)\nonumber\\
& +\tilde{z}^2\bigg(\log(2-\tilde{z}^2+2\sqrt{1-\tilde{z}^2})-2\log{\tilde{z}}\bigg)\bigg] \ ,
\end{align}
where $\Theta$ is the Heaviside step function. Performing the remaining integral in $\tilde{z}$, for $\R  <1$ the result is
\begin{align}
\mathcal{I}_- &= \frac{\pi}{3}\,\mathcal{N}_{\text{D7}}\, \R^2\biggl(\log\frac{2R}{\epsilon}-\frac{13}{6}+\frac{1}{10}\,\R^2\biggr)\ ,
\label{eq:Ismallmass}
\end{align}
while for $\R  >1$ it is
\begin{align}
 \mathcal{I}_+ &= 
    \frac{\pi}{3}\,\mathcal{N}_{\text{D7}}\,  \biggl[\R^2\log{\frac{2R}{\epsilon}}-\frac{4}{15\R }\sqrt{\R^2-1}\bigg(1-\frac{97}{16}\R^2+\frac{3}{8}\R^4\bigg)-\frac{13}{6}\R^2+\frac{1}{10}\R^4\nonumber \\
     &\quad -\frac{1}{4}(1+\R^2)\arcosh{\R}\biggr]\ .
\label{eq:Ilargemass}
\end{align}

\subsection{Probe D6-branes}\label{subapp:probeD6}

In this appendix, we give details of the probe computation of flavor entanglement entropy presented in Section \ref{subsec:probeD6} in the case of a D6-brane on $\text{AdS}_4\times \mathbb{C}\text{P}^3$.

\paragraph{Holographic renormalization.}

We will add counterterms to remove divergences appearing when taking the limit $\Lambda_\zeta \rightarrow \infty$. The on-shell asymptotic behavior of the D6-brane embedding is given by
\begin{equation}
\xi_{n,0}(\tau,\zeta,v) = \frac{\pi}{2}+\frac{\xi_-(\tau,v)}{\zeta} + \frac{\xi_+(\tau,v)}{\zeta^{2}} + \mathcal{O}(\zeta^{-3})\ , \quad \zeta \rightarrow \infty \ .
\label{eq:D6asymptoticexpansion}
\end{equation}
The regularized on-shell D6-brane action \eqref{eq:Dpactionapp} with the Lagrangian \eqref{eq:D6lagrangiantext} has the expansion
\begin{equation}
I_{\text{D6}}^{\text{reg}} = \mathcal{N}_{\text{D6}}\int_{0}^{2\pi} d\tau \int_{v_0}^{\Lambda_{v}} dv\, \sinh{v}\,\left[\frac{1}{2}\,\Lambda_{\zeta}^{3}-\frac{1}{2}\,\xi_-^{2}\Lambda_{\zeta} + \mathcal{O}(\Lambda_\zeta^0)\right]\ , \quad \Lambda_{\zeta} \rightarrow \infty \ .
\label{eq:regactiondivs}
\end{equation}
The counterterms that cancel the $ \Lambda_\zeta \rightarrow \infty $ divergences are given by
\begin{equation}
I_{\text{ct}}^{(1)} \propto\int \sqrt{\gamma}\,\sin{\xi}\ , \quad I_{\text{ct}}^{(2)} \propto\int \sqrt{\gamma}\,\ell^{2}R_\gamma \ ,
\label{eq:invariants}
\end{equation}
where the integral is over the $\zeta = \Lambda_\zeta$ cut-off surface, $\gamma$ is the induced metric of the cut-off surface and $R_\gamma$ is the Ricci scalar built from $\gamma$. In our coordinates:
\begin{equation}
\sqrt{\gamma} = \ell^{3}\,\zeta^{2}\,f_n(\zeta)^{1\slash 2}\,\sinh{v}\, \sin{\xi}\ , \quad \ell^{2}R_{\gamma} = -\frac{2}{\zeta^{2}}\ .  
\end{equation}
so that the counterterm Lagrangian in \eqref{eq:Dpactionapp} takes the form
\begin{equation}
\mathcal{L}^{\text{ct}}_n = \zeta^{2}\,f_n(\zeta)^{1\slash 2}\,\sinh{v}\,\sin{\xi}\,\left(a_1\sin{\xi}+a_2\,\zeta^{-2}\right) \ ,
\end{equation}
where $a_1,a_2$ are constants. The counterterm action \eqref{eq:Dpactionapp} has the expansion
\begin{equation}
I_{\text{D6}}^{\text{ct}} = \mathcal{N}_{\text{D6}}\int_{0}^{2\pi} d\tau\int_{0}^{\Lambda_v} dv\,\sinh{v}\,\biggl[a_1\Lambda_{\zeta}^{3} +(b_1+b_2\,\xi_-^{2})\,\Lambda_{\zeta} -\frac{1}{2}\,a_1\,(\zeta_n^{3}-\zeta_n + 4\,\xi_-\xi_+) + \mathcal{O}(\Lambda^{-1}_{\zeta}) \biggr] \ ,
\end{equation}
where $ b_1 = -\frac{1}{2}a_1+a_2 $ and $ b_2 = -a_1 $. Thus, to cancel the divergences in \eqref{eq:regactiondivs}, we have to set
\begin{equation}
a_1 = -\frac{1}{2}\ , \quad a_2 = -\frac{1}{4}\ . 
\end{equation}
Hence, the counterterm action that cancels the divergences of \eqref{eq:regactiondivs} has the expansion
\begin{equation}
I^{\text{ct}}_{\text{D6}} = \mathcal{N}_{\text{D6}}\int_{0}^{2\pi} d\tau\int_{0}^{\Lambda_v} dv\,\sinh{v}\,\biggl[-\frac{1}{2}\,\Lambda^{3}_{\zeta}+\frac{1}{2}\,\xi_-^{2}\Lambda_{\zeta} +\frac{1}{4}\,(\zeta_n^{3}-\zeta_n)  +\xi_-\xi_++ \mathcal{O}(\Lambda^{-1}_{\zeta})\biggr]\ .
\label{eq:Ictexpansion}
\end{equation}

\paragraph{Boundary terms in the entropy.}

Let us compute the boundary terms $\mathcal{B}_{\text{RT}}^{(1,2)}$ and $\mathcal{B}_{\zeta=\infty}^{(1,2)}$ by starting with the UV terms. From \eqref{eq:Ictexpansion} it follows that
\begin{align}
\mathcal{B}_{\zeta=\infty}^{(1)} &= \mathcal{N}_{\text{D6}}\int_{0}^{2\pi} d\tau\int_{0}^{\Lambda_v} dv\,\sinh{v}\,\biggl[-\frac{1}{4}+\xi_-\partial_n\xi_+\lvert_{n=1}\, +\, \mathcal{O}(\Lambda^{-1}_{\zeta})\biggr] \\
&= 2\pi\,\mathcal{N}_{\text{D6}}\,\biggl(\frac{R}{\epsilon} - 1\biggr)\biggl(-\frac{1}{4}+\xi_-\partial_n\xi_+\lvert_{n=1}\biggr) \ ,
\label{eq:BUV1}
\end{align}
where we used the integral
\begin{equation}
    \int_{0}^{\Lambda_v} dv\,\sinh{v} = \frac{R}{\epsilon} - 1 + \mathcal{O}(\epsilon)\ .
\end{equation}
The second UV term is explicitly
\begin{equation}
    \mathcal{B}_{\zeta=\infty}^{(2)} = \mathcal{N}_{\text{D6}}\int_{0}^{2\pi} d\tau \int_{0}^{\Lambda_v} dv\,\frac{\zeta^2\sinh{v}\sin{(\xi_{1,0})}\,f_1(\zeta)\,(\partial_\zeta\xi_{1,0})\,(\partial_n\xi_{n,0})\lvert_{n=1}}{\sqrt{1+\zeta^{-2}\,(\partial_v\xi_{1,0})^{2} + f_1(\zeta)\,(\partial_\zeta \xi_{1,0})^{2}+f_1(\zeta)^{-1}\,(\partial_{\tau}\xi_{1,0})^{2}}}\biggr\lvert_{\zeta = \Lambda_\zeta}.
\end{equation}
Substituting the asymptotic expansion \eqref{eq:D6asymptoticexpansion} gives
\begin{equation}
    \mathcal{B}_{\zeta=\infty}^{(2)}  = -\mathcal{N}_{\text{D6}}\int_{0}^{2\pi} d\tau \int_{0}^{\Lambda_v} dv\,\sinh{v}\,\xi_-\partial_n\xi_+\lvert_{n=1} + \mathcal{O}(\Lambda_{\zeta}^{-2})\ ,
\end{equation}
which becomes
\begin{equation}
\mathcal{B}_{\zeta=\infty}^{(2)}  = -2\pi\,\mathcal{N}_{\text{D6}}\,\biggl(\frac{R}{\epsilon} - 1\biggr)\,\xi_-\partial_n\xi_+\lvert_{n=1} \ .
\label{eq:BUV2}
\end{equation}
The sum of \eqref{eq:BUV1} and \eqref{eq:BUV2} is thus given by
\begin{equation}
\mathcal{B}_{\zeta=\infty}^{(1)}+\mathcal{B}_{\zeta=\infty}^{(2)} = -\frac{\pi}{2}\,\mathcal{N}_{\text{D6}}\,\biggl(\frac{R}{\epsilon} - 1\biggr)\ ,
\end{equation}
where all dependence on $\partial_n\xi_+\lvert_{n=1}$ has canceled as in the case of the D7-brane.

The computation of the boundary terms at $v = 0$ and $v = \infty$ goes in the same way as in the D7-case. The on-shell behaviors of the embedding is given by
\begin{align}
    \xi_{n,0}(\tau,\zeta,v) &= \frac{\pi}{2}+ \tilde{\xi}_-(\tau,\zeta)\,e^{-v} + \tilde{\xi}_+(\tau,\zeta)\,e^{-2v} + \mathcal{O}(e^{-3v})\ ,\quad v \rightarrow \infty\\
    \xi_{n,0}(\tau,\zeta,v) &= \xi_0^{(n)}(\zeta) +\xi_1^{(n)}(\tau,\zeta)\, v^{2} + \mathcal{O}(v^{2})\ ,\quad v\rightarrow 0\,.
\end{align}
and this time
\begin{equation}
\frac{\partial \mathcal{L}_1}{\partial (\partial_v \xi_{1,0})}(\partial_n\xi_{n,0})\lvert_{n=1}\, =-\frac{1}{2}\,\tilde{\xi}_-\partial_n\tilde{\xi}_+\lvert_{n=1}\,e^{-2v}+\mathcal{O}(e^{-3v})\ ,\quad v\rightarrow \infty\,.
\label{eq:D6Bvinfty}
\end{equation}
These give $\mathcal{B}_{v = \infty}^{(2)} = 0$. In addition, $\partial_n\xi_{n,0}\lvert_{n = 1}$ is finite at $v = 0$ while
\begin{equation}
\frac{\partial \mathcal{L}_1}{\partial (\partial_v \xi_{1,0})} = \mathcal{O}(v^{2})\ ,\quad v\rightarrow 0 \ ,
\end{equation}
so that $\mathcal{B}^{(2)}_{v = 0} = 0$ as well.

Let us then compute the RT surface boundary terms. To compute them, we need the on-shell behavior of the probe D6-brane embedding $\xi_{n,0}$ near $\zeta = \zeta_n$ ,
\begin{equation}
\xi_{n,0}(\tau,\zeta,v) = \xi^{(0)}_n(v) + \xi^{(1)}_n(\tau,v)\, \sqrt{\zeta-\zeta_n} + \mathcal{O}(\zeta-\zeta_n) \ ,
\end{equation}
which is the same as in the D7-brane case. It follows that
\begin{equation}
\mathcal{B}^{(1)}_{\text{RT}} = \frac{3}{4}\mathcal{N}_{\text{D6}}\int_{0}^{2\pi} d\tau \int_{v_0}^{\Lambda_v} dv\,\sinh{v}\sin{\xi^{(0)}_n}\,\biggl[\sqrt{1+(\partial_v\xi^{(0)}_n)^{2}+\frac{1}{2}(\xi^{(1)}_n)^{2}+\frac{1}{2}(\partial_\tau\xi^{(1)}_n)^{2}}+\frac{1}{2}\sin{\xi^{(0)}_n}\biggr] \ .
\label{eq:BIR11}
\end{equation}
From the embedding \eqref{eq:D6xi10} we see that at $n=1$,
\begin{equation}
\xi^{(0)}_1(v) = \arccos{(\R  \sech{v})}\ , \quad \xi^{(1)}_1(\tau,v) = \frac{\sqrt{2}\,\R \cos{\tau}\,\sech{v}}{\sqrt{\cosh^{2}{v}-\R^2}}\ .
\label{eq:xi01D6}
\end{equation}
Substituting to \eqref{eq:BIR11} gives
\begin{equation}
\mathcal{B}^{(1)}_{\text{RT}} = \frac{3}{4}\,\mathcal{N}_{\text{D6}}\int_{0}^{2\pi} d\tau \int_{v_0}^{\Lambda_v} dv\,\sinh{v}\,\left(1 - \frac{1}{3}\R^2\sech^{2}{v}\right) \ .
\end{equation}
Performing the integrals gives
\begin{equation}
\mathcal{B}^{(1)}_{\text{RT}} =\frac{3\pi}{2}\,\mathcal{N}_{\text{D6}}\,
\begin{dcases}
\frac{R}{\epsilon}-1-\frac{1}{3}\R^2\ , \quad &\R  < 1 \\
\frac{R}{\epsilon}-\frac{4}{3}\R \ , \quad &\R  \geq 1 
\end{dcases}\ ,
\label{eq:BIR1finalD6}
\end{equation}
where we have used (up to an additive constant)
\begin{equation}
\int dv\, \sinh{v}\,\left(1 - \frac{1}{3}\R^2\sech^{2}{v}\right) = \cosh{v}+\frac{1}{3}\R^2\sech{v} \ .
\end{equation}
To compute $\mathcal{B}_{\text{RT}}^{(2)}$, we need
\begin{equation}
\frac{\partial \mathcal{L}_1}{\partial (\partial_\zeta \xi_{1,0})} = \frac{\sinh{v}\sin{\xi^{(0)}_1}\,\xi^{(1)}_1}{\sqrt{1+(\partial_v\xi^{(0)}_1)^{2}+\frac{1}{2}(\xi^{(1)}_1)^{2}+\frac{1}{2}(\partial_\tau\xi^{(1)}_1)^{2}}}\,\sqrt{\zeta - 1} + \mathcal{O}(\zeta-1)\ , \quad \zeta \rightarrow 1 \ .
\label{eq:Lder}
\end{equation}
From the on-shell behavior it follows that
\begin{equation}
\partial_n\xi_{n,0}\lvert_{n=1} = \frac{1}{4}\frac{\xi^{(1)}_1(\tau,v)}{\sqrt{\zeta-1}} + \mathcal{O}(\sqrt{\zeta-1})\ , \quad \zeta \rightarrow 1 \ ,\label{eq:xinderlimit}
\end{equation}
where we have used \eqref{eq:zetander}. Hence, the product of \eqref{eq:Lder} and \eqref{eq:xinderlimit} in $ \mathcal{B}_{\text{RT}}^{(2)} $ is finite when $ \zeta \rightarrow 1 $. The result is
\begin{equation}
\mathcal{B}_{\text{RT}}^{(2)} = -\frac{1}{4}\,\mathcal{N}_{\text{D6}}\int_{0}^{2\pi} d\tau \int_{v_0}^{\Lambda_v} dv\,\frac{\sinh{v}\sin{\xi^{(0)}_1}\,(\xi^{(1)}_1)^{2}}{\sqrt{1+(\partial_v\xi^{(0)}_1)^{2}+\frac{1}{2}(\xi^{(1)}_1)^{2}+\frac{1}{2}(\partial_\tau\xi^{(1)}_1)^{2}}} \ .
\end{equation}
Substituting \eqref{eq:xi01D6} gives
\begin{equation}
\mathcal{B}_{\text{RT}}^{(2)} = -\frac{1}{2}\,\mathcal{N}_{\text{D6}}\int_{0}^{2\pi} d\tau \int_{v_0}^{\Lambda_v} dv\,\cos^{2}{\tau}\sech^{2}{v}\sinh{v}\,\R^2\ ,
\end{equation}
and performing the integrals gives
\begin{equation}
\mathcal{B}_{\text{RT}}^{(2)} = -\frac{\pi}{6}\,\mathcal{N}_{\text{D6}}\,
\begin{dcases}
\R^2 \ ,\quad &\R < 1 \\
\frac{1}{\R } \ ,\quad &\R \geq 1 
\end{dcases} \ ,
\label{eq:BIR2finalD6}
\end{equation}
where we have used (up to an additive constant)
\begin{equation}
\int dv\,\sech^{2}{v}\sinh{v} = -\frac{1}{3}\sech^{3}{v} \ .
\end{equation}

\paragraph{Bulk integral in the entanglement entropy.}

The remaining bulk integral is explicitly
\begin{eqnarray}
\mathcal{I} & = & \frac{1}{2}\,\mathcal{N}_{\text{D6}}\int_{0}^{2\pi} d\tau \int_{1}^{\infty} d\zeta \int_{v_0}^{\infty} dv\, \zeta^{2}\sinh{v}\sin{(\xi_{1,0})}\,\frac{\partial_nf_n(\zeta)\lvert_{n=1}}{f_1(\zeta)^{2}} \nonumber\\
& & \qquad\times\frac{f_1(\zeta)^{2}\,(\partial_\zeta \xi_{1,0})^{2}-(\partial_{\tau}\xi_{1,0})^{2}}{\sqrt{1+\zeta^{-2}\,(\partial_v\xi_{1,0})^{2} + f_1(\zeta)\,(\partial_\zeta \xi_{1,0})^{2}+f_1(\zeta)^{-1}\,(\partial_{\tau}\xi_{1,0})^{2}}} \ ,
\label{eq:ID61}
\end{eqnarray}
where $ f_1(\zeta) = \zeta^{2}-1 $ and we have taken the limits $ \Lambda_{\zeta}\rightarrow \infty $, $ \Lambda_{v}\rightarrow \infty $ in the upper limits of the integrals, because the integrals are convergent on-shell.

We will now compute the bulk integral \eqref{eq:ID61}. For this, we will change variables to the rescaled coordinate system introduced in Subsection~\ref{subapp:probeD7}. In the new coordinate system, the integral reads:
\begin{align}
\mathcal{I}&=\mathcal{N}_{\text{D6}}\int_{\frac{\epsilon}{R}}^{\frac{1}{\R }}d\tilde{z}\int_{-\infty}^{\infty}d\tilde{t}_{E}\int_{0}^{\infty}d\tilde{\rho} \frac{N}{D}\ ,
    \label{eq:rescaledID6}
\end{align}
where
\begin{align}
    N &= 4R^2r_\text{q}^2\tilde{z}\tilde{\rho}\bigg[\tilde{t}_{E}^8+4\tilde{t}_{E}^6(1-\tilde{z}^2+\tilde{\rho}^2)+\left(\tilde{z}^4-(1-\tilde{\rho}^2)^2\right)^2+ 2\tilde{t}_{E}^4\big(3-5\tilde{z}^4+2\tilde{\rho}^2+3\tilde{\rho}^4\nonumber\\
    &\qquad\qquad-4\tilde{z}^2(1+\tilde{\rho}^2)\big)-4\tilde{t}_{E}^2\big(\tilde{z}^6-1+\tilde{\rho}^2+\tilde{\rho}^4-\tilde{\rho}^6+\tilde{z}^2(1-\tilde{\rho}^2)^2+3\tilde{z}^4(1+\tilde{\rho}^2)\big)\bigg]\\
D &= \bigg[(1+\tilde{t}_{E}^2+\tilde{z}^2)^2+2\tilde{\rho}^2(\tilde{t}_{E}^2+\tilde{z}^2-1)+\tilde{\rho}^4\bigg]^{\frac{3}{2}}\bigg[\tilde{t}_{E}^2+(\tilde{z}^2+\tilde{\rho}^2-1)^2+2\tilde{t}_{E}^2(1+\tilde{z}^2+\tilde{\rho}^2)\bigg]^{2}\nonumber \ .
\end{align}
By integrating first in $\tilde{\rho}$, we get
\begin{align}
\mathcal{I}&=\frac{\mathcal{N}_{\text{D6}}}{2}\int_{\frac{\epsilon}{R}}^{\frac{1}{\R }}d\tilde{z}\int_{-\infty}^{\infty}d\tilde{t}_{E}\R^2\bigg[\frac{\tilde{z}(\tilde{z}-1)}{\tilde{t}_{E}^2+(1-\tilde{z})^2}-\frac{\tilde{z}(\tilde{z}+1)}{\tilde{t}_{E}^2+(1+\tilde{z})^2}+\frac{2\tilde{z}}{\tilde{t}_{E}^2+\tilde{z}^2}\nonumber\\
&\qquad+(1-i\tilde{t}_{E})\log(-\frac{\tilde{t}_{E}+i\tilde{z}}{i+\tilde{t}_{E}})-(1+i\tilde{t}_{E})\log\frac{\tilde{t}_{E}(-i+\tilde{t}_{E})+\tilde{z}(\tilde{z}+1)}{(i-\tilde{t}_{E})(\tilde{t}_{E}+i(\tilde{z}-1))}\\
&\qquad+(1+i\tilde{t}_{E})\log\frac{\tilde{t}_{E}-i\tilde{z}}{i-\tilde{t}_{E}}+(-1+i\tilde{t}_{E})\log\frac{\tilde{t}_{E}(-i-\tilde{t}_{E})-\tilde{z}(\tilde{z}+1)}{(i+\tilde{t}_{E})(\tilde{t}_{E}-i(\tilde{z}-1))}\bigg]\ .\nonumber
\end{align}
This expression is straightforward to integrate with respect to $\tilde{t}_{E}$ giving us:
\begin{align}
\mathcal{I}&=\mathcal{N}_{\text{D6}}\int_{\frac{\epsilon}{R}}^{\frac{1}{\R }}d\tilde{z}\R^2 \bigg[\tilde{z}\bigg(\arctan\frac{\tilde{t}_{E}}{\tilde{z}-1}-\arctan\frac{\tilde{t}_{E}}{\tilde{z}+1}\bigg)+2\bigg(\arctan\frac{\tilde{t}_{E}}{\tilde{z}}+\tilde{z}^2\arctan\frac{\tilde{z}}{\tilde{t}_{E}}\bigg)\nonumber\\
&\qquad+\frac{2-i\tilde{t}_{E}}{2}\tilde{t}_{E}
\log\frac{-\tilde{t}_{E}-i\tilde{z}}{i+\tilde{t}_{E}}-\frac{2+i\tilde{t}_{E}}{2}\tilde{t}_{E}\log\frac{\tilde{t}_{E}(-i+\tilde{t}_{E})+\tilde{z}(\tilde{z}+1)}{(i-\tilde{t}_{E})(\tilde{t}_{E}+i(\tilde{z}-1))}\nonumber\\
&\qquad+(1-\tilde{z}^2)\bigg(\arctan\frac{\tilde{z}-1}{\tilde{t}_{E}}+\arctan\frac{\tilde{z}+1}{\tilde{t}_{E}}\bigg)+\frac{2+i\tilde{t}_{E}}{2}\tilde{t}_{E}\log\frac{\tilde{t}_{E}-i\tilde{z}}{i-\tilde{t}_{E}}\nonumber \\
&\qquad+\frac{i\tilde{t}_{E}-2}{2}\tilde{t}_{E}\log\frac{\tilde{t}_{E}(i+\tilde{t}_{E})+\tilde{z}(\tilde{z}+1)}{(-i-\tilde{t}_{E})(\tilde{t}_{E}-i(\tilde{z}-1))} \bigg]\Bigg\lvert_{0}^{+\infty}
\end{align}
when evaluated at the endpoints of integration, the previous expression acquires a simple form:
\be
\mathcal{I}=\mathcal{N}_{\text{D6}}\int_{\frac{\epsilon}{R}}^\frac{1}{\R }d\tilde{z} \pi \R^2(1-\tilde{z}-\tilde{z}^2)\Theta(1-\tilde{z})
\ee
which can be immediately integrated giving us
\begin{equation}
\mathcal{I} = \pi\,\mathcal{N}_{\text{D6}}\,\times
\begin{dcases}
\frac{1}{6}\,\R^2\ , \quad &\R  < 1 \\
\R -\frac{1}{3\R}-\frac{1}{2}\ , \quad &\R  \geq 1 
\end{dcases} \ .
\label{eq:IfinalD6app}
\end{equation}

\section{Boundary terms and diffeomorphism invariance}\label{app:boundaryterms}

In this appendix, we will reformulate the analysis of probe D$p$-brane actions and their variations in a coordinate invariant manner. We carefully analyze certain boundary terms arising in the variation and relate them to diffeomorphism Ward identities. We show that regularity of the brane embedding implies cancellation of boundary terms at the tip of the brane.

\paragraph{The brane action.} The D$p$-brane action takes the form
\begin{equation}
I_{\mathcal{W}}[g,\xi] = \int_{\mathcal{W}\times Y} d^{p+1}x\,\sqrt{\hat{g}} \ ,
\end{equation}
where $\hat{g}$ is the pull-back of the metric $g$ onto the brane, $\mathcal{W}$ describes the worldvolume of the brane in the three directions of the hyperbolic black hole spanning a constant-$\Omega$ slice and $Y$ describes the remaining directions of the worldvolume $Y = S^{d-2}\times M_q$. Let $x^i$ with $i = 1,2,3$ denote the coordinates on the constant-$\Omega$ slice of the hyperbolic black hole and $\tilde{x}^\alpha$ with $\alpha = 1,\ldots, p-2$ the remaining worldvolume coordinates. In Appendix \ref{subapp:bannondetails}, we used $x^i = (\tau,\zeta,v)$, but here we will consider the slice in a general coordinate system. 

The brane embedding depends only on the hyperbolic black hole coordinates $\xi = \xi(x) = \xi(x^i)$ so the non-zero components of the induced metric of the brane are
\begin{equation}
\hat{g}_{ij}(x) = g_{ij}(x) +(\partial_i\xi)(\partial_j\xi)\ ,\quad \hat{g}_{\alpha\beta}(x,\tilde{x}) = g_{\alpha\beta}(\xi(x),\tilde{x}) \ ,
\end{equation}
where we used that in our examples $g_{\xi\xi} = 1$. Using the determinant formula
\begin{equation}
\det{\hat{g}_{ij}} = [1 + g^{ij}\,(\partial_i\xi)(\partial_j\xi)]\,\det{g_{ij}}
\end{equation}
the brane action becomes
\begin{equation}
I_{\mathcal{W}}[g,\xi] = \int_{\mathcal{W}} d^{3}x\sqrt{g}\,\mathcal{L} \ ,
\label{eq:Dpbraneactionv2}
\end{equation}
where the Lagrangian is
\begin{equation}
\mathcal{L}(\xi,\partial_i\xi,g^{ij}) = K(\xi)\,\sqrt{1 + g^{ij}\,(\partial_i\xi)(\partial_j\xi)}
\label{eq:absD7lagr}
\end{equation}
and we have defined
\begin{equation}
K(\xi(x)) = \int_{Y} d^{p-2}\tilde{x}\,\sqrt{\det{g_{\alpha\beta}(\xi(x),\tilde{x})}}\;.
\end{equation}
One can check that this formula gives the D7- and D6-brane actions \eqref{eq:D7lagrangiantext} and \eqref{eq:D6lagrangiantext}. Notice that in this appendix, we will separate the metric determinant $\sqrt{g}$ from the Lagrangian density $\mathcal{L}$ in \eqref{eq:Dpbraneactionv2}. We do this to make the diffeomorphism invariance (under three-dimensional diffeomorphisms acting in the $x^i$ directions) of the action manifest.

The boundary $\partial \mathcal{W}$ of the integration domain consists of multiple components with one being the tip $\mathcal{T}$ where the brane ends. From this three-dimensional point of view, the tip is a boundary, while from the higher-dimensional point of view, it is simply the surface where the brane smoothly caps off. Let $x = W(y)$ be the embedding of $\mathcal{T}$ in $\mathbb{R}^3$ where $ y = (y^1,y^2) \in \mathbb{R}^2$ are the two coordinates of $\mathcal{T}$. The embedding is determined by the condition
\begin{equation}
    \xi(W(y)) = 
    \begin{dcases}
        0 \ ,\quad &\text{D7-brane}\\
        \frac{\pi}{2} \ ,\quad &\text{D6-brane}\\
    \end{dcases}
    \label{eq:branetipembedding}
\end{equation}
and the integration domain $\mathcal{W}$ is given by points $x\in \mathbb{R}^3$ for which $0<\xi(x)<\pi\slash 2$ (for both D7- and D6-branes).

\paragraph{Variation of the action.} Consider the variation of the action \eqref{eq:Dpbraneactionv2} with respect to the metric $g_{ij}$, the brane embedding $\xi(x)$ and the integration domain $\mathcal{W}$. The variation of the integration domain produces additional boundary terms in accordance with the higher-dimensional generalization of the Leibniz integral rule. We can compute these boundary terms using the general theorem of \cite{flanders1973differentiation} which applies to integrations of differential forms over higher-dimensional domains (see also \cite{Reddiger:2019lqh} and references therein). In our case, the differential form being integrated is the volume form times the Lagrangian density $\mathcal{L}$. The result is
\begin{equation}
\delta I_{\mathcal{W}}[g,\xi] = \int_{\mathcal{W}} d^{3}x\,\delta(\sqrt{g}\, \mathcal{L})+\int_{\partial\mathcal{W}} d^{2}y\sqrt{h}\,n_i\,\delta E^{i}\,\mathcal{L} \ ,
\label{eq:generalvariation}
\end{equation}
where $\sqrt{h}$ is the square root of the determinant of the induced metric, $n_i$ is the outward-pointing unit normal vector normalized in the metric $g$ (it satisfies $g^{ij}\,n_i\,n_j = 1$) and $\delta E^i$ is the variation of the embedding $x^i = E^i(y)$ of $\partial \mathcal{W}$. The variation of the bulk term is given by
\begin{align}
\delta (\sqrt{g}\,\mathcal{L}) = \frac{1}{2}\,\sqrt{g}\,T_{ij}\,\delta g^{ij}+\sqrt{g}\,\biggr[\frac{\partial \mathcal{L}}{\partial \xi} - \nabla_i\biggl(\frac{\partial \mathcal{L}}{\partial (\nabla_i\xi)}\biggr)\biggl]\,\delta \xi + \sqrt{g}\,\nabla_i\biggl(\frac{\partial \mathcal{L}}{\partial (\nabla_i\xi)}\,\delta \xi\biggr)\ ,
\end{align}
where we have defined the effective three-dimensional stress tensor of the brane
\begin{equation}
    T_{ij}\equiv \frac{2}{\sqrt{g}}\frac{\partial (\sqrt{g}\,\mathcal{L})}{\partial g^{ij}} \ .
\end{equation}
Substituting to \eqref{eq:generalvariation} gives
\bea
\delta I_{\mathcal{W}}[g,\xi] &=& \int_{\mathcal{W}} d^{3}x\,\sqrt{g}\,\biggl\{\frac{1}{2}\,T_{ij}\,\delta g^{ij}+\biggr[\frac{\partial \mathcal{L}}{\partial \xi} - \nabla_i\biggl(\frac{\partial \mathcal{L}}{\partial (\nabla_i\xi)}\biggr)\biggl]\,\delta \xi\biggr\}\nonumber\\
&& +\int_{\partial\mathcal{W}} d^{2}y\sqrt{h}\,\biggl(n_i\,\frac{\partial \mathcal{L}}{\partial (\partial_i\xi)}\,\delta \xi+n_i\,\delta E^{i}\,\mathcal{L}\biggr) \ .
\label{eq:braneactionvarmaster}
\eea
We will now apply this equation to derive equations of motion for $\xi$ and diffeomorphism Ward identities.

\paragraph{Diffeomorphism Ward identity.} Since the brane action is diffeomorphism invariant the stress tensor $T_{ij}$ satisfies certain Ward identities. Under the three-dimensional diffeomorphism $F(x)$, the fields transform as
\begin{equation}
\widetilde{g}_{ij}(x) = \frac{\partial F^{k}}{\partial x^{i}}\frac{\partial F^{l}}{\partial x^{j}}\,g_{kl}(F(x))\ ,\quad \widetilde{\xi}(x) = \xi(F(x)) \ .
\end{equation}
The brane action satisfies\footnote{In fact, the integration domain $\widetilde{W}$ determined by $\widetilde{\xi}$ coincides with $ F(\mathcal{W})$ so that \eqref{eq:diffeoinvariance} can be written nicely as $I_{\widetilde{\mathcal{W}}}[\widetilde{g},\widetilde{\xi}] = I_{\mathcal{W}}[g,\xi]$.}
\begin{equation}
    I_{F(\mathcal{W})}[\widetilde{g},\widetilde{\xi}] = I_{\mathcal{W}}[g,\xi]\ ,
    \label{eq:diffeoinvariance}
\end{equation}
which is the statement of diffeomorphism invariance. We will consider infinitesimal diffeomorphisms
\begin{equation}
    F^i(x) = x^i + \delta F^i(x)
\end{equation}
and write \eqref{eq:diffeoinvariance} as $\delta I_{\mathcal{W}}[g,\xi] = 0 $ where the variation is given by
\begin{equation}
   \delta E^i(y) = -\delta F^i(E(y))\ ,\quad \delta g^{ij} = -\nabla^{i}\delta F^j - \nabla^j \delta F^i\ ,\quad \delta\xi = \delta F^{i}\,\partial_i \xi\ ,
\end{equation}
where we have defined
\begin{gather}
\nabla_i\delta F_{j} = \partial_i\delta F_j-\Gamma^{k}_{ij}\,\delta F_k\ ,\quad \Gamma^{k}_{ij} = \frac{1}{2}\,g^{kl}\,(\partial_ig_{jl}+\partial_jg_{il}-\partial_lg_{ij})\ .
\end{gather}
Substituting these variations to \eqref{eq:braneactionvarmaster} and integrating the $T_{ij}\,\nabla^i\delta F^j$ term by parts gives
\begin{align}
0 = \delta I_{\mathcal{W}}[g,\xi]&=\int_{\mathcal{W}} d^{3}x\sqrt{g}\,\delta F^{j}\,\biggl\{\nabla^iT_{ij}+\biggr[\frac{\partial \mathcal{L}}{\partial \xi} - \nabla_i\biggl(\frac{\partial \mathcal{L}}{\partial (\nabla_i\xi)}\biggr)\biggl]\,(\partial_j \xi)\biggr\} \nonumber\\
&\qquad +\int_{\partial\mathcal{W}} d^{2}y\sqrt{h}\,\delta F^j\,\biggl[-n^{i}\,T_{ij}+n_i\,\biggl(\frac{\partial \mathcal{L}}{\partial (\partial_i\xi)}\,(\partial_j\xi)-\delta^i_j\mathcal{L}\biggr)\biggr]\ .
\end{align}
Hence, diffeomorphism invariance \eqref{eq:diffeoinvariance} implies that the stress tensor $T_{ij}$ satisfies the equations
\begin{equation}
    \nabla^{i}T_{ij} = -\biggr[\frac{\partial \mathcal{L}}{\partial \xi} - \nabla_i\biggl(\frac{\partial \mathcal{L}}{\partial (\nabla_i\xi)}\biggr)\biggl]\,(\partial_j\xi)\ ,\quad n^{i}\,T_{ij}\lvert_{\partial\mathcal{W}}\,=n_i\,\biggl(\frac{\partial \mathcal{L}}{\partial (\partial_i\xi)}\,(\partial_j\xi)-\delta^i_j\mathcal{L}\biggr)\bigg\lvert_{\partial\mathcal{W}}\ .
\end{equation}
One can check explicitly that these equations are satisfied for the Lagrangian \eqref{eq:absD7lagr}. As a result, when $\xi$ satisfies the equations of motion, the stress tensor is conserved
\begin{equation}
    \nabla^{i}T_{ij}  =0 \ .
\end{equation}
On the tip of the brane $\mathcal{T}$ the outward-pointing unit normal $P^i= g^{ij}P_j$ is given by
\begin{equation}
P_i = \frac{\partial_i \xi}{\sqrt{g^{kl}\,(\partial_k\xi)(\partial_l \xi)}}\bigg\lvert_{\mathcal{T}} \ ,
\label{eq:Pidef}
\end{equation}
and it satisfies $g^{ij}\,P_iP_j = 1$. Hence the second equation on $\mathcal{T}$ can be written as
\begin{equation}
    P^{i}P^jT_{ij}\lvert_{\mathcal{T}}\, =\biggl(\frac{\partial \mathcal{L}}{\partial (\partial_i\xi)}\,(\partial_i\xi)-\mathcal{L}\biggr)\bigg\lvert_{\mathcal{T}}\ .
\end{equation}
Unlike in the conservation equation, the right-hand side here does not vanish on-shell. Instead, the equation
\begin{equation}
    P^{i}P^jT_{ij}\lvert_{\mathcal{T}}\, = 0
    \label{eq:genregularity}
\end{equation}
is understood as a boundary condition for $\xi$. In fact, it is equivalent to the regularity of the brane embedding at the point where the brane caps off in the higher-dimensional description. Substituting the Lagrangian \eqref{eq:absD7lagr} gives explicitly
\begin{equation}
    P^{i}P^jT_{ij}\lvert_{\mathcal{T}}\, = -\frac{K(\xi)}{\sqrt{1 + g^{ij}\,(\partial_i\xi)(\partial_j\xi)}}\bigg\lvert_{\mathcal{T}}\,.
\end{equation}
We can see that \eqref{eq:genregularity} is equivalent to
\begin{equation}
    g^{ij}\,(\partial_i\xi)(\partial_j\xi)\lvert_{\mathcal{T}}\, = \infty\ .
\end{equation}
We can see how the regularity condition fixes the brane embedding in the AdS background in terms of the location of the tip. The embedding is a function of $r$ only $\xi = \xi(r)$ and the tip $\mathcal{T}$ is located at $r = r_{\text{q}}$. The regularity condition amounts to
\begin{equation}
    \xi'(r_{\text{q}}) = \infty\,.
    \label{eq:regrcoordinate}
\end{equation}
For example in the D7-brane case, the solution of the equation of motion has two integration constants which are fixed by the Dirichlet boundary conditions $\xi(r_{\text{q}}) = 0$, $\xi(\infty) = \frac{\pi}{2}$ and the regularity condition \eqref{eq:regrcoordinate} in terms of the single parameter $r_{\text{q}}$. The result is the embedding \eqref{eq:D7n1}.

This discussion also applies to the diffeomorphism invariant area functional \eqref{eq:firstEE} appearing in the RT computation of the entanglement entropy in Section \ref{subsec:perturbativeRT}. The diffeomorphism invariant Ward identity relates the regularity condition for the embedding of the RT surface, the third condition in \eqref{eq:rhoboundaryconditions}, to the vanishing of the normal directed component of the stress tensor of the area functional at the tip.

\paragraph{Vanishing of boundary terms.} Let us consider a general variation of $\xi$ under which also the embedding $W(y)$ of $\mathcal{T}$ changes. This is what one considers in the replica computation of the entanglement entropy $\delta = \partial_n$. The variation of the embedding of the tip $ \mathcal{T} $ is obtained from
\begin{equation}
\xi(W(y))= 0 \ .
\end{equation}
Taking the variation of \eqref{eq:branetipembedding} gives
\begin{equation}
\delta \xi(W(y)) + (\partial_i\xi)(W(y))\,\delta W^{i}(y) = 0
\end{equation}
so that we get
\begin{equation}
\delta W^{i}(y) = -\frac{g^{ij}\,(\partial_j\xi)}{g^{kl}\,(\partial_k\xi)(\partial_l \xi)}\,\delta\xi\bigg\lvert_{x = W(y)} =  -\frac{P^{i}}{\sqrt{g^{kl}\,(\partial_k\xi)(\partial_l \xi)}}\,\delta\xi\bigg\lvert_{\mathcal{T}} \ ,
\label{eq:deltaW}
\end{equation}
where $P^i = g^{ij}P_j$ is the outward-pointing unit normal vector of $\mathcal{T}$ defined by \eqref{eq:Pidef}. Using the general formula \eqref{eq:braneactionvarmaster}, the variation of the action becomes
\bea
\delta I_{\mathcal{W}}[g,\xi] &=& \int_{\mathcal{W}} d^{3}x\,\sqrt{g}\,\biggl\{\frac{1}{2}\,T_{ij}\,\delta g^{ij}+\biggr[\frac{\partial \mathcal{L}}{\partial \xi} - \nabla_i\biggl(\frac{\partial \mathcal{L}}{\partial (\nabla_i\xi)}\biggr)\biggl]\,\delta \xi\biggr\}\label{eq:braneactionvar3} \\
&&+\int_{\mathcal{T}} d^{2}y\sqrt{h}\,\biggl(\frac{\partial \mathcal{L}}{\partial (\partial_i\xi)}\,(\partial_i\xi)-\mathcal{L}\biggr)\frac{\delta\xi}{\sqrt{g^{kl}\,(\partial_k\xi)(\partial_l \xi)}}\nonumber \\
&&+\int_{\partial\mathcal{W}\,\backslash \mathcal{T}} d^{2}y\sqrt{h}\,\biggl(n_i\,\frac{\partial \mathcal{L}}{\partial (\partial_i\xi)}-n_i\,\delta E^{i}\mathcal{L}\biggr)\,\delta \xi \ .\nonumber
\eea
The boundary term on $ \mathcal{T} $ can be identified with the sum $ \mathcal{B}_{\text{IR}}^{(1)}+\mathcal{B}_{\text{IR}}^{(2)} $ of terms defined in Appendix~\ref{subapp:bannondetails} and it vanishes upon imposing the regularity condition \eqref{eq:genregularity}. Assuming $ \xi $ satisfies the Euler--Lagrange equation, we get
\begin{align}
\delta I_{\mathcal{W}}[g,\xi] &= \int_{\mathcal{W}} d^{3}x\,\sqrt{g}\,\frac{1}{2}\,T_{ij}\,\delta g^{ij} +\int_{\partial\mathcal{W}\,\backslash \mathcal{T}} d^{2}y\sqrt{h}\,\biggl(n_i\,\frac{\partial \mathcal{L}}{\partial (\partial_i\xi)}-n_i\,\delta E^{i}\mathcal{L}\biggr)\,\delta \xi \ .\nonumber
\end{align}
The first bulk integral here is equal to $\mathcal{I}$ while the remaining boundary terms include $ \mathcal{B}_{\text{RT}}^{(1,2)} $, $ \mathcal{B}_{\zeta=\infty}^{(1,2)} $, $ \mathcal{B}_{v = 0}^{(2)} $, and $ \mathcal{B}_{v = \infty}^{(2)} $.

\bibliographystyle{JHEP}
\bibliography{flavorEE}

\end{document}